\documentclass[12pt]{article}   
\usepackage{amsmath,amsthm,latexsym,amssymb,amsfonts,epsfig}% AMS-style for Hanno
\allowdisplaybreaks
\makeatletter
\@addtoreset{equation}{section}
\makeatother

\newtheorem{Theorem}{Theorem}[section]

\newtheorem{Lemma}{Lemma}[section]

\newcommand{\vla}[0]{\xrightarrow{\hspace{5ex}}}
\newcommand{\evec}[1]{\vec{#1}}% euclidean vectors
\newcommand{\norm}[1]{\left\lVert #1 \right\rVert}    % norm 
\newcommand{\betr}[1]{\left\lvert #1 \right\rvert}    % betrag
\newcommand{\scpr}[2]{\langle#1,#2\rangle} % scalar product
\newcommand{\comm}[2]{\left[#1,#2\right]} % commutator
\newcommand{\lzwo}{L^2} % ein L2-raum 
\newcommand{\aver}[1]{\big\langle\big\langle#1\big\rangle\big\rangle}  % the average
\newcommand{\expec}[1]{\langle#1\rangle}

\newcommand{\qqquad}{\qquad\qquad}

\newcommand{\blank}{\,\cdot\,}
\newcommand{\R}{\mathbb{R}}              % die reellen zahlen
\newcommand{\Z}{\mathbb{Z}}              % die ganzen zahlen
\newcommand{\C}{\mathbb{C}}              % die komplexen zahlen
\DeclareMathSymbol{\comp}{\mathrel}{AMSa}{"62} % kompakte tm
\DeclareMathOperator{\diver}{div}
\DeclareMathOperator{\rotat}{rot}

\DeclareMathOperator{\tr}{Tr}
\DeclareMathOperator{\sutwo}{SU(2)}

\DeclareMathOperator{\uone}{U(1)}

\def\be#1\ee{\begin{equation}#1\end{equation}}
\def\ba#1\ea{\begin{eqnarray}#1\end{eqnarray}}
\def\Nl{{\mathchoice
{\setbox0=\hbox{$\displaystyle\rm N$}\hbox{\hbox to0pt
{\kern0.4\wd0\vrule height0.9\ht0\hss}\box0}}
{\setbox0=\hbox{$\textstyle\rm N$}\hbox{\hbox to0pt
{\kern0.4\wd0\vrule height0.9\ht0\hss}\box0}}
{\setbox0=\hbox{$\scriptstyle\rm N$}\hbox{\hbox to0pt
{\kern0.4\wd0\vrule height0.9\ht0\hss}\box0}}
{\setbox0=\hbox{$\scriptscriptstyle\rm N$}\hbox{\hbox to0pt
{\kern0.4\wd0\vrule height0.9\ht0\hss}\box0}}}}
\def\Zl{{\mathchoice
{\setbox0=\hbox{$\displaystyle\rm Z$}\hbox{\hbox to0pt
{\kern0.4\wd0\vrule height0.9\ht0\hss}\box0}}
{\setbox0=\hbox{$\textstyle\rm Z$}\hbox{\hbox to0pt
{\kern0.4\wd0\vrule height0.9\ht0\hss}\box0}}
{\setbox0=\hbox{$\scriptstyle\rm Z$}\hbox{\hbox to0pt
{\kern0.4\wd0\vrule height0.9\ht0\hss}\box0}}
{\setbox0=\hbox{$\scriptscriptstyle\rm Z$}\hbox{\hbox to0pt
{\kern0.4\wd0\vrule height0.9\ht0\hss}\box0}}}}
\def\Ql{{\mathchoice
{\setbox0=\hbox{$\displaystyle\rm Q$}\hbox{\hbox to0pt
{\kern0.4\wd0\vrule height0.9\ht0\hss}\box0}}
{\setbox0=\hbox{$\textstyle\rm Q$}\hbox{\hbox to0pt
{\kern0.4\wd0\vrule height0.9\ht0\hss}\box0}}
{\setbox0=\hbox{$\scriptstyle\rm Q$}\hbox{\hbox to0pt
{\kern0.4\wd0\vrule height0.9\ht0\hss}\box0}}
{\setbox0=\hbox{$\scriptscriptstyle\rm Q$}\hbox{\hbox to0pt
{\kern0.4\wd0\vrule height0.9\ht0\hss}\box0}}}}
\def\Rl{{\mathchoice
{\setbox0=\hbox{$\displaystyle\rm R$}\hbox{\hbox to0pt
{\kern0.4\wd0\vrule height0.9\ht0\hss}\box0}}
{\setbox0=\hbox{$\textstyle\rm R$}\hbox{\hbox to0pt
{\kern0.4\wd0\vrule height0.9\ht0\hss}\box0}}
{\setbox0=\hbox{$\scriptstyle\rm R$}\hbox{\hbox to0pt
{\kern0.4\wd0\vrule height0.9\ht0\hss}\box0}}
{\setbox0=\hbox{$\scriptscriptstyle\rm R$}\hbox{\hbox to0pt
{\kern0.4\wd0\vrule height0.9\ht0\hss}\box0}}}}
\def\Cl{{\mathchoice
{\setbox0=\hbox{$\displaystyle\rm C$}\hbox{\hbox to0pt
{\kern0.4\wd0\vrule height0.9\ht0\hss}\box0}}
{\setbox0=\hbox{$\textstyle\rm C$}\hbox{\hbox to0pt
{\kern0.4\wd0\vrule height0.9\ht0\hss}\box0}}
{\setbox0=\hbox{$\scriptstyle\rm C$}\hbox{\hbox to0pt
{\kern0.4\wd0\vrule height0.9\ht0\hss}\box0}}
{\setbox0=\hbox{$\scriptscriptstyle\rm C$}\hbox{\hbox to0pt
{\kern0.4\wd0\vrule height0.9\ht0\hss}\box0}}}}
\def\Hl{{\mathchoice
{\setbox0=\hbox{$\displaystyle\rm H$}\hbox{\hbox to0pt
{\kern0.4\wd0\vrule height0.9\ht0\hss}\box0}}
{\setbox0=\hbox{$\textstyle\rm H$}\hbox{\hbox to0pt
{\kern0.4\wd0\vrule height0.9\ht0\hss}\box0}}
{\setbox0=\hbox{$\scriptstyle\rm H$}\hbox{\hbox to0pt
{\kern0.4\wd0\vrule height0.9\ht0\hss}\box0}}
{\setbox0=\hbox{$\scriptscriptstyle\rm H$}\hbox{\hbox to0pt
{\kern0.4\wd0\vrule height0.9\ht0\hss}\box0}}}}
\def\Ol{{\mathchoice
{\setbox0=\hbox{$\displaystyle\rm O$}\hbox{\hbox to0pt
{\kern0.4\wd0\vrule height0.9\ht0\hss}\box0}}
{\setbox0=\hbox{$\textstyle\rm O$}\hbox{\hbox to0pt
{\kern0.4\wd0\vrule height0.9\ht0\hss}\box0}}
{\setbox0=\hbox{$\scriptstyle\rm O$}\hbox{\hbox to0pt
{\kern0.4\wd0\vrule height0.9\ht0\hss}\box0}}
{\setbox0=\hbox{$\scriptscriptstyle\rm O$}\hbox{\hbox to0pt
{\kern0.4\wd0\vrule height0.9\ht0\hss}\box0}}}}
\begin{document}
\title{Towards the QFT on Curved Spacetime Limit of QGR.\\ 
II: A Concrete Implementation}
\author{
H. Sahlmann\thanks{sahlmann@aei-potsdam.mpg.de}, 
T. Thiemann\thanks{thiemann@aei-potsdam.mpg.de} \\
       MPI f. Gravitationsphysik, Albert-Einstein-Institut, \\
           Am M\"uhlenberg 1, 14476 Golm near Potsdam, Germany}
\date{{\small PACS No. 04.60, Preprint AEI-2002-050}}

\maketitle 

\begin{abstract}
The present paper is the companion of \cite{ST01} in which we 
proposed a scheme 
that tries to derive the Quantum Field Theory (QFT) on Curved Spacetimes
(CST) limit from background independent Quantum General Relativity (QGR).
The constructions of \cite{ST01} make heavy use of the notion of
\textit{semiclassical states for QGR}. In the present paper, we employ  
the complexifier coherent states for QGR recently proposed 
by Thiemann and Winkler as semiclassical states, and thus  
fill the general formulas obtained in \cite{ST01} with life.

We demonstrate how one can, under some simplifying assumptions, 
explicitely compute expectation values of the operators relevant for
the gravity-matter Hamiltonians of \cite{ST01} in the complexifier
coherent states.  
These expectation values give rise to effective matter Hamiltonians
on the background on which the gravitational coherent state is peaked 
and thus induce approximate notions of $n-$particle states and 
matter propagation on fluctuating spacetimes. We display the details
for the scalar and the electromagnetic field. 

The effective theories exhibit two types of corrections as compared to 
the the ordinary QFT on CST. The first is due to the quantum
fluctuations of the gravitational field, the second arises from the 
fact that background independence forces both geometry and matter to 
propagate on a spacetime of the form $\Rl\times\gamma$ where $\gamma$ is a 
(random) graph.

Finally we obtain explicit 
numerical predictions for non-standard dispersion relations for the
scalar and the electromagnetic field. They should, 
however, not be taken too seriously, due to the many ambiguities in 
our scheme, the analysis of the physical significance of which 
has only begun.  
We show however, that one can classify these ambiguities at least in 
broad terms. 
\end{abstract}
%------------------------------------------------------------------------
\section{Introduction}
\label{s1}
%------------------------------------------------------------------------
Canonical, non-perturbative Quantum General Relativity (QGR) has by now
reached the status of a serious candidate for a quantum theory of the 
gravitational field:
First of all, the formulation of the theory is mathematically rigorous. 
Although there are 
no further inputs other than the fundamental principles of four-dimensional, 
Lorentzian General Relativity and quantum theory, the  
theory predicts that there is a built in {\it fundamental discreteness} 
at Planck scale distances and therefore an UV cut-off precisely due to its
diffeomorphism invariance (background independence). Next, while most of the 
results have so far been obtained using the canonical operator language,
also a path integral formulation (``spin foams'') is currently 
constructed. Furthermore, as a first physical application, a rigorous, 
microscopical derivation of the Bekenstein-Hawking entropy -- area law has 
been established. 
The reader interested in all the technical details of QGR and its 
present status is 
referred to the exhaustive review article \cite{Thiemann:2001yy} and 
references therein, and to \cite{Rovelli:1998yv} for a less technical 
overview. For a comparison with other
approaches to quantum gravity see
\cite{Horowitz:2000sh,Rovelli:1997qj,Rovelli:1999hz}.\\ 

A topic that has recently attracted much attention is to explore the
regime of QGR where the quantized gravitational field behaves ``almost
classical'', i.e. approximately like a given classical solution to the 
field equations. Only if such a regime exists, one can really claim
that QGR is a viable candidate theory for quantum gravity.  
Consequently, efforts have been made to identify so called 
\textit{semiclassical states} in the Hilbert space of QGR, states 
that reproduce a given classical geometry in terms of their
expectation values and in which the quantum mechanical fluctuations are 
small \cite{Ashtekar:1992tm,Bombelli:2000ua,Thiemann:2000bw,Thiemann:2000ca,Thiemann:2000bx}.
Also, it has been investigated how gravitons emerge as carriers of the 
gravitational interaction in the semiclassical regime of the theory
\cite{IR1,IR2,Ashtekar:1991mz}. 
The recent investigation of Madhavan and others \cite{V1,V2,V3,Ashtekar:2001xp} on the
relation between the Fock representations used in conventional quantum 
field theories and the one in QGR further illuminate the relation
between QGR and a perturbative treatment based on gravitons. 

In \cite{ST01} we developed and discussed a general scheme how one can 
define a theory of quantum matter coupled to quantum gravity in the
setting of QGR and investigate its semiclassical limit.  
In the present paper we concretize the results of \cite{ST01} by
employing a specific proposal
\cite{Thiemann:2000bw,Thiemann:2000ca,Thiemann:2000bx} 
for semiclassical states for QGR. 
As the present paper relies on the general approach as well as on
specific results of \cite{ST01}, it should be read together with the
latter. Especially the discussion of the conceptual issues arising in
the present context is much more completely covered in \cite{ST01}. 
Also, it should be stressed that the cautionary remarks concerning
our results made there apply even more to the present paper: The
analysis of the semiclassical regime of QGR in general, 
as well as that of the coherent states 
\cite{Thiemann:2000bw,Thiemann:2000ca,Thiemann:2000bx} for QGR
specifically has only begun recently, and so \textit{the main purpose of our
work is exploratory}. 

In the present paper, we roughly proceed in three steps: 
Firstly, we review the coherent state family introduced in 
\cite{Thiemann:2000bw,Thiemann:2000ca,Thiemann:2000bx,Sahlmann:2001nv} and fix the 
parameters in its definition in such a way that best semiclassical
behavior is obtained for the observables relevant to our
considerations.  
Then, under some simplifying assumptions, we compute the expectation values
in the coherent states for the operators relevant for setting up the
effective QFT for the matter fields according to \cite{ST01}.  
Finally, we use the resulting effective theory to approximately
compute the quantum gravity corrections to the dispersion relations
for the scalar and the electromagnetic field.  

Let us consider these steps in more detail:\\

In \cite{Thiemann:2000bw,Thiemann:2000ca,Thiemann:2000bx,Sahlmann:2001nv}, a
promising family of semiclassical states have been constructed and
analyzed. Each member
of this family is labelled by a (random) graph $\gamma$ and a point
$m\in {\cal M}$ in the gravitational phase space. Other states derived 
by the complexifier method \cite{Thiemann:2002vj} could be used as well
but for the exploratory purposes of this series of papers it is sufficient 
to stick to those simplest ones.

Three scales enter the definition of the coherent states and are of
considerable importance for their semiclassical properties. These
scales are the {\it microscopic Planck scale} $\ell_P$, the 
{\it mesoscopic graph scale} $\epsilon$ which represents the average length of an edge of $\gamma$ 
as measured by the three metric determined by $m$ and a {\it macroscopic
curvature scale} $L$ which characterizes the scale at which matter (and 
thus geometry) vary. 
While $\ell_P,L$ are determined by the input $m$,
the scale $\epsilon$ is a priori a free parameter. We fix it by asking
that a natural family of observables be well approximated by our coherent 
states which leads quite generically to a geometric mean type of 
behavior, concretely $\epsilon\propto \ell_P^\alpha L^{1-\alpha}$ where
$0<\alpha<\frac{1}{2}$. In contrast to the weave proposal \cite{Ashtekar:1992tm}
the graph scale is larger than the Planck scale due to the fact that we 
do not only approximate the three geometry but also the extrinsic 
curvature which forces the coherent state to depend on all possible 
spin representations of SU(2) and not only the defining (or any 
other single) one.

The analysis of \cite{Sahlmann:2001nv} revealed that the coherent states proposed 
do not approximate well coordinate dependent observables like the holonomy 
or the electric flux operator. However, we discovered that operators which 
classically correspond to integrals of scalar densities of weight one are 
extremely well approximated. This class of observables 
contains
Hamiltonian constraints and all spatially diffeomorphism 
invariant quantities which suffice to separate the points of the 
diffeomorphism invariant phase space. The intuitive reason for this is 
the following point which has been stressed for years, among others, 
especially by Rovelli \cite{Rovelli:1991ph, Rovelli:1991pi}:
{\it Matter can be located only where Geometry is excited!}
Classically this follows from Einstein's equations. In the quantum theory
it is reflected by the fact that matter and geometry degrees of freedom 
are necessarily located on the same graph 
\cite{Thiemann:1998rq,Thiemann:1998rt}. Imagine now constructing a 
diffeomorphism invariant area operator $\widehat{\mbox{Ar}}$. In contrast 
to its companion $\widehat{\mbox{Ar}}(S)$ well studied in the literature
it does not depend on an externally prescribed coordinate surface, rather
in order to model the measurement of the are of the desk table on which
you are working right now one would construct a coherent state of the 
combined matter and geometry Hilbert space which is peaked on flat space
and, say, on an electromagnetic field which is zero everywhere except
for a region in the vicinity of the table. This way the dynamics 
automatically forces the surface to be adapted to the graph on which the 
coherent state depends.\\

In a next step, we compute coherent state expectation values for 
the gravitational degrees of freedom that appear in the matter--geometry 
Hamiltonians. This computation, although straightforward in principle,
turns out to be quite tedious in practice. To keep the computational 
effort on a tolerable level and maintain some clarity of presentation, 
we simplify things by doing the calculation only for the Abelian
(I\"on\"u-Wigner) limit $\uone^3$ of $\sutwo$ as gauge group. 
The computations done in \cite{Thiemann:2000ca,Thiemann:2000bx}
exemplify that this replacement does not change the results
qualitatively, and therefore seems acceptable for the exploratory 
purposes of the
present paper. A calculation in full generality should only be carried 
out after other issues have been settled, and will probably necessitate
the use of computers.\\  

Ground-breaking work on the phenomenology of QGR has been done in 
\cite{Gambini:1998it,Alfaro:1999wd,Alfaro:2001rb,Alfaro:2001gk}.  
In these works, corrections to the standard dispersion relations 
for matter fields due to QGR have been obtained. 
Since we are dealing 
with a theory for matter coupled to QGR in \cite{ST01} and the present
work, it is an important question whether these results can be 
confirmed in the present setting. Therefore,
as a final step, we formulate effective matter Hamiltonians on a graph
based on the expectation values obtained before. 
The resulting theory is that of fields propagation on a random
graph. It bears a remarkable
similarity to models considered in lattice gauge theory
\cite{Christ:1982zq,Christ:1982ck,Christ:1982ci}, and there is also a
close analogy to the propagation of phonons in amorphous solids. 
As we have discussed at length in \cite{ST01}, the resulting
dynamics for the matter fields is very complicated, and analytic
results in the literature on lattice gauge theory and on amorphous solids 
are few. (To say the least. See however \cite{Tanguy:2002} for a
beautiful numerical study of some two dimensional models from
condensed matter physics.)
Already a simplified one dimensional system (whose
definition along with some results was sketched in \cite{ST01} and will 
be covered more completely in \cite{ST03}) shows many of 
the complications (optical and acoustic branches, fuzziness of
dispersion relations at high energies etc.) that are to be expected
for the dynamics of fields propagating in a QGR background. 
Therefore, to compute dispersion relations for the models obtained, 
we have to rely on an approximation scheme denoted ``graph averages'', 
geared to the description of the dynamics in the limit where the
energy of the fields is low (or, equivalently, their wavelength
large). This approximation scheme leads to precise numerical values 
for all correction coefficients in the dispersion relations, once we 
have fixed a random process that generates our sample graph. The
results we obtain are similar to those of
\cite{Gambini:1998it,Alfaro:2001rb} in many respects, but differ in
the scaling of the corrections. 

The validity of the approximation scheme we use has been discussed in 
\cite{ST01} but certainly merits future investigation. 
In any case, the resulting formulas can probably effectively handled
in full generality only by a computer.\\

Let us finish with a brief description of the contents of the sections 
to follow:

The next section contains a short review of the construction of the 
complexifier coherent states
\cite{Thiemann:2000bw,Thiemann:2000ca,Thiemann:2000bx}.  

In section \ref{se3} we analyze the relation between the different
scales that enter the definition of the coherent states, and their
semiclassical properties. Relying on this analysis we fix the
parameters of the coherent states for the rest of the paper.

Section \ref{s4} is the longest of the present paper. We show how
expectation values in the coherent states can be computed and do the
concrete calculations for the operators occurring in the Hamiltonians 
for scalar, the electromagnetic and fermionic fields coupled to gravity. 

Section \ref{s5} deals with the computation of dispersion
relations. We implement the procedure outlined in \cite{ST01} and
compare our results to the ones in the literature. 

Finally, in section \ref{s6} we summarize what we have tried to do and 
what could be achieved with present technology. We conclude with a list 
of the open conceptual and technical questions that this work has left us 
with.
%------------------------------------------------------------------------
\section{Complexifier Coherent States}
\label{s2}
%------------------------------------------------------------------------
The purpose of the present section is to review the construction and 
basic properties of the coherent states for QGR
\cite{Thiemann:2000bw,Thiemann:2000ca,Thiemann:2000bx}. For an
introduction to the formalism of QGR as a whole we refer the reader to 
\cite{Rovelli:1998yv,Thiemann:2001yy}, or to the brief introduction in 
\cite{ST01}.\\ 

As already pointed out in the introduction, the task of constructing 
\textit{semiclassical states} for QGR has received much attention 
\cite{Ashtekar:1992tm,Bombelli:2000ua,Thiemann:2000bw,Thiemann:2000ca,Thiemann:2000bx}.
Semiclassical states are states, so far in the kinematical Hilbert space 
of QGR, that approximate a specific classical geometry in the sense that
expectation values of observables in such a state are close to the
respective classical values and the quantum mechanical fluctuations
are small. These requirements can certainly not be met for
\textit{all} possible observables, so the definition of a
semiclassical state also involves specification of the class of
observables that are well approximated. 
  
In the present work, we will use the gauge theory coherent states (GCS 
for short) constructed in \cite{Thiemann:2000bw} and subsequently analyzed in
\cite{Thiemann:2000ca,Thiemann:2000bx,Sahlmann:2001nv}. These states are only one 
example of a large class of semiclassical states, called the 
\textit{complexifier coherent states}. We
refer to \cite{Thiemann:2002vj} for an investigation of this class of
states as well as a discussion of the relationship to
\cite{Bombelli:2000ua,V1,V2,V3,Ashtekar:2001xp}.
 
The main mathematical tool used in the construction of the GCS is a
generalization due to Hall \cite{Hall1994,Hall1997} of the well 
known coherent states for the harmonic oscillator.  
The basic observation underlying this generalization is that the
harmonic oscillator coherent states can be obtained as analytic
continuation of the heat kernel on $\R^n$: 
\begin{equation*}
 \psi^t_z(x)=e^{-t\Delta}\delta_{x'}(x)\Big\rvert_{x'\longrightarrow
   z},\qquad x\in \R^n, z\in \C^n, 
\end{equation*}
the Laplacian $\Delta$ in the above formula playing the role of 
a \textit{complexifier}. 

It has been shown in \cite{Hall1994} that 
coherent states on a connected compact Lie group $G$ can analogously be
defined as analytic continuations of the heat kernel
\begin{equation}
\label{eq2.2}
 \psi^t_g(h)=e^{-t\Delta_G}\delta^{(G)}_{h'}(h)\Big\rvert_{h'\longrightarrow u}
\end{equation}
to an element $u$ of the complexification $G^\C$ of $G$. 
These states have nice mathematical properties. Among other things,
they are minimal uncertainty states for a certain pair of operators
and they form an overcomplete set in the Hilbert space over $G$ derived 
from the Haar measure.  

The case of
this construction relevant for the definition of GCS is $G=\sutwo$. Its
complexification is given by SL(2,$\C$) and can be parametrized as
\begin{equation}
  \label{eq2.1}
  u=\exp\left[i\tau_jp^j/2\right]h,\qqquad
  p^k \in \R^3, h \in \sutwo  
\end{equation}
where $i\tau_k, k=1,2,3$ denote the Pauli matrices. 

A crucial question in view of applications to 
the construction of semiclassical states for QGR is whether the states 
\eqref{eq2.2} obey peakedness properties analogous to that of the
harmonic oscillator coherent states. 
In \cite{Thiemann:2000ca} it was shown that this is indeed 
the case: For $u$ given by $p,h$ via the parametrization
\eqref{eq2.1}, the following holds: 
\begin{itemize}
\item $\psi^t_u$ is exponentially (Gaussian) peaked with respect to the 
multiplication operator $\hat{h}$ on the group at the point $h$. The
width of the peak is approximately given by $\sqrt{t}$ 
\item $\psi^t_u$ is Gaussian peaked with respect to the invariant 
  vector-fields at a point $p/t$ in the associated momentum
  representation. The width of the peak is approximately given by
  $1/\sqrt{t}$. 
\end{itemize}
For a more precise formulation of these statements we refer to
\cite{Thiemann:2000ca}. 

In QGR, the configuration degrees of freedom are represented by the
holonomies along edges $e$ of a graph $\gamma$ embedded in $\Sigma$.   
To use the coherent states on $\sutwo$ for the construction of
semiclassical states for QGR, momentum observables, that are associated to a 
graph in a similar way as the holonomies have to be defined. 
This was done in \cite{Thiemann:2000bv}. 
The construction can be summarized as follows
(for the many details we refer the reader to the original work):  
To each graph $\gamma$ fix once and for all a dual 2-complex
$P_\gamma$, i.e. roughly speaking a set of surfaces $(S_e)_{e\in 
  E(\gamma)}$ which  intersect each other in common boundaries at most 
and such that the edge $e$ of $\gamma$ intersects only $S_e$ and that this 
intersection is transversal. The surfaces $S_e$ shall be given an orientation
according to the orientations of the edges $e$, i.e. the pairing between 
the orientation two form on $S_e$ with the tangent vector field on $e$
at the  intersection point should be positive.
Also to each point $p$ lying in a surface $S_e$ fix an analytic path $\rho(p)$
connecting the intersection point $S_e\cap e$ with $p$ and denote the
part of $e$ from $e(0)$ to $S_e\cap e$ by $e^{\text{in}}$.

With the help of these structures, we can now define the quantity
\begin{equation}
\label{eq1.20}
p^e_j(A,E)= -\frac{1}{2a^2}\tr\left[\tau_j 
h_{e^{\text{in}}}\left(\int_{S_e}
h_{\rho(p)}E^a(p)h_{\rho(p)}^{-1}\epsilon_{abc}\,dS^{bc}(p)\right)
h^{-1}_{e^{\text{in}}}\right].
\end{equation}
where $a$ is a length scale introduced to make $p^e_j$ dimensionless
and whose relation with $t$ is $t=\ell_P^2/a^2$. 
The key feature of this new variable $p^e_j$ is that 
\begin{equation}
\label{eq1.15}
\left\{p^e_j,h_{e'}\right\}=
\frac{\kappa^2}{a^2}\delta_{e,e'}\frac{\tau_j}{2}h_{e'},\qquad
\left\{p^e_i,p^{e'}_j\right\}=-
\frac{\kappa^2}{a^2} \delta_{e,e'}\epsilon_{ijk} p^e_k
\end{equation}
where $\epsilon_{ijk}$ are the structure constants of $\sutwo$. Therefore, 
if
$h_e$ is represented by the multiplication operator $\widehat{h}$ on
the cylindrical subspace corresponding to $e$, $p^e_j$ can be
represented by the right invariant vector-field $i t X^j$
acting on the cylindrical subspace corresponding to $e$.

Having the momentum variables $p^e_j$ at disposal, 
the construction of the GCS can now be finished. It needs three inputs: 
\begin{itemize}
\item A point $(A^{(0)},E^{(0)})$ in the classical phase space that
  should be approximated. 
\item A graph $\gamma$ and a corresponding dual polyhedronal
  decomposition $P_\gamma$ of $\Sigma$, and the associated path system 
  $\Pi_\gamma$. 
\item The parameter $t$ or, equivalently, the length scale $a$. 
\end{itemize}
For each edge $e$ of the graph $\gamma$, one can now compute the
holonomy $h_e^{(0)}$ in the classical connection $A^{(0)}$ and the
classical quantities $p^{(0)e}_j$ depending on $A^{(0)},E^{(0)}$ as 
expressed in \eqref{eq1.20}. 
The gauge coherent state for QGR is then defined as 
\begin{equation*}
  \psi^t_{(A^{(0)},E^{(0)})}(h_{e_1},\ldots,h_{e_N})\doteq 
\prod_{n=1}^N\psi^t_{g_{e_n}(A_0,E_0)}(h_{e_n})
\end{equation*}
where $e_1,\ldots,e_N$ represent the edges of the graph $\gamma$
and $g_e(A_0,E_0)=\exp(p^{(0)e}_j\tau_j/2)h_e^{(0)}$.  

The states thus defined inherit the peakedness properties of the
coherent states \eqref{eq2.2} in an obvious way with respect to the
elementary observables $\widehat{h}_{e_1},\ldots,\widehat{h}_{e_N}$ and 
$\widehat{p}^{e_1}_j,\ldots,\widehat{p}^{e_N}_j$. For more complicated
observables, a more detailed consideration has to be given. This is
the topic of the next section. 
We will see that this analysis fixes the parameter $t$ as well as the average
edge length of the graph $G$, thus reducing the freedom in the
construction of the GCS considerably. 
%-----------------------------------------------------------------------
\section{Observables and Scales}
\label{se3}
%-----------------------------------------------------------------------
In the previous section we saw that the complexifier coherent states 
$\psi_{m,\gamma}$
that will be used in this paper (see \cite{Thiemann:2002vj} for 
generalizations) depend on a point $m\in {\cal M}$ and a triple 
$(\gamma,P_\gamma,\Pi_\gamma)$ where $\gamma$ is a graph, $P_\gamma$ a 
polyhedronal decomposition of $\Sigma$ dual to $\gamma$ and $\Pi_\gamma$
is an associated path system. The states $\pi_{m,\gamma}$ are linear 
combinations of spin network states over $\gamma$ (and all of its 
subgraphs) with coefficients which depend on $m,P_\gamma,\Pi_\gamma$.
We are interested in the question which kind of operators $\hat{O}$ are 
approximated 
well by these states, that is, for which holds that expectation values are
close to the classical value and for which the fluctuations are small.

By construction, they approximate very well the holonomy operators 
$\hat{h}_e$ and the electric flux operators $\hat{E}_j(S_e)$ where 
$e$ runs through the set of edges of $\gamma$ and $S_e$ is a face in the 
polyhedronal decomposition dual to $e$. But how about more general
operators such as $\hat{h}_p,\widehat{\mbox{Ar}(S)}$ where $p$ is an
arbitrary path and $S$ an arbitrary surface? First of all, unless 
$p$ is a composition of edges of $\gamma$ we have 
$<\psi_{m,\gamma},\hat{h}_e\psi_{m,\gamma}>=0$ due to the orthogonality 
of spin-network states. Secondly, the expectation values of the area 
operator suffer from the ``staircase problem" \cite{Sahlmann:2001nv} which says that
unless $S$ is composed of the $S_e$ then its expectation value will be 
off the correct value. 

The first reaction is: The states are not good,
they must be improved. One such improvement could be by averaging
over an ensemble of graphs \cite{Bombelli:2000ua} but as shown in 
\cite{Thiemann:2002vj} this still does not improve the holonomy expectation 
values. Thus, one could think that one should construct semiclassical 
states of a completely different type, maybe going to a new representation 
of the canonical commutation relations \cite{V1,V2,V3}. However, this 
is not easy if the present formulation of QGR is to be kept as shown in 
\cite{Thiemann:2002vj}. It therfore seems that we are in trouble. 

There is a second possibility however: Maybe we are just trying to 
approximate the wrong observables? Notice that it is a {\it physical
input} which observables should be well approximated, certainly we do
not expect all classical quantities to be approximated well in the quantum 
theory. This is even true for simple finite dimensional systems such as 
the harmonic oscillator: The energy itself is well approximated but not 
its exponential. In our case, traces of holonomy operators and area 
operators are certainly natural candidates for operators to be well
approximated because they are gauge invariant, suffice to separate the 
points of the gravitational phase space and are simple functions of the 
basic operators that the whole quantization is based on, namely
holonomy and electric flux operators. Is it possible that there are 
observables which are better suited for our semiclassical considerations?

A first hint of how such observables should look like comes from the 
observation that the volume operator $\widehat{\mbox{Vol}}(R)$ for a 
coordinate region $R$ does not suffer from the staircase problem.
A detailed analysis shows that this happens because the region $R$ 
corresponds to a {\it three-dimensional submanifold} of $\Sigma$ rather 
than one -- or two dimensional ones. We therefore are led to the proposal
that one should not look at holonomy and area operators but rather at
quantities that classically come from three-dimensional integrals.
There are classical observables of that kind that one can construct and 
which separate the points of the gauge invariant gravitational phase space 
as well: Let $\omega_a$ be a one form, say of rapid decrease, and consider
\ba \label{3.1}
Q(\omega) &:=&
\int_\Sigma d^3x \frac{E^a_j E^b_j}{\sqrt{\det(q)}}\omega_a\omega_b
\\ \label{3.2}
M(\omega) &:=&
\int_\Sigma d^3x \frac{B^a_j B^b_j}{\sqrt{\det(q)}}\omega_a\omega_b
\ea
where $E^a_j E^b_j=:\det(q) q^{ab}$ and $B^a_j=
\frac{1}{2}\epsilon^{abc} F_{bc}^j$ and where $F$ is the curvature of the 
connection $A$. Notice that both (\ref{3.1}), (\ref{3.2}) are of the 
type of operators that can be quantized with the methods of 
\cite{Thiemann:1998rt} in a background independent fashion since they are 
integrals of scalar densities. Moreover, they suffice to separate the 
points of the gauge invariant phase space as one can see by suitably
restricting the support of $\omega_a$ and by the polarization identity 
for quadratic forms.

The crucial fact about these quantities is now as follows: When we 
quantize them along the lines of \cite{Thiemann:1998rt} they become 
diffeomorphism covariant, densely defined, closed operators on the 
kinematical QGR Hilbert space ${\cal H}_0$ of the following structure
\be \label{3.3}
\hat{O}(\omega)T_s=\sum_{v\in V(s(\gamma))} \;\;
\sum_{v\in \partial 
e,\partial e';e,e' \in E(\gamma(s))} \;\;\omega(e)\omega(e') 
\hat{O}_{v;e,e'}T_s
=:\sum_{v\in V(s(\gamma))} \hat{O}_\gamma(\omega,v) T_s
\ee
where $T_s$ is a spin-network state with underlying graph $\gamma(s)$ and 
$V(\gamma),E(\gamma)$ denote the sets of vertices and edges of a graph
respectively and $\omega(e)=\int_e \omega$. The fact that an action 
only at vertices takes place in (\ref{3.3}) is due to the appearance of 
the volume operator which enters the stage due to the factor of 
$1/\sqrt{\det(q)}$ in (\ref{3.1}), (\ref{3.2}) which is required by 
background independence and the requirement that only density one valued 
quantities can be quantized in a background independent way 
\cite{Thiemann:1998rt}.
The operator $\hat{O}_{v;e,e'}$ is a polynomial formed out of holonomy 
operators {\it along the edges of $\gamma(s)$} and powers of the volume 
operator restricted to an arbitrarily small neighborhood of the vertex $v$.
Now the coherent states are constructed precisely in such a way that
the holonomy operators along the edges of $\gamma(s)$ are well 
approximated and, as we will explicitly prove in this work, they also
approximate very well the volume operator of 
\cite{Rovelli:1995ge,Ashtekar:1997fb} {\it 
at least if the graph is six-valent, e.g. of cubic topology}. (For 
other graph topologies the 
prefactor $\frac{1}{8\cdot 3!}$, which enters the square roots that 
defines
the volume operator, would presumably need to be adapted to the vertex 
valence,
it should be larger (smaller) for valences smaller (larger) than six).

Thus, due to the Ehrenfest properties proved in \cite{Thiemann:2000bx}
we conclude that at least for coherent states based on graphs with
cubic topologies the operators (\ref{3.1}), (\ref{3.2}) are approximated 
well (with small fluctuations) provided the expectation values of 
(\ref{3.3}) define a Riemann sum approximation of the classical
integrals (\ref{3.1}), (\ref{3.2}). This is, however, the case by the 
very construction of such operators as outlined in \cite{Thiemann:1998rt}.
Thus, the mechanism responsible for the fact that no such problems as for 
the area and holonomy operators arise is due to the fact that for 
operators
coming from volume integrals the elementary electric flux and holonomy
operators involved are automatically those adapted to the graph in 
question.

That only cubic graphs should give rise to the correct classical limit
might be disturbing at first but it is on the other hand not too 
surprising: The volume operator at a given vertex $v$ is a square root 
of an operator which
in turn is a sum of basic operators, one for each unordered triple of 
distinct edges 
incident at $v$ in \cite{Rovelli:1995ge} and in \cite{Ashtekar:1997fb} one considers only 
those triples which have linearly independent tangents at $v$.
Each of these basic operators is a third order homogeneous polynomial in 
electric flux operators. With respect to our coherent states, each 
polynomial gives a 
contribution of the same order of magnitude. If $n$ is the valence of 
the vertex of $v$ then there will be altogether $N(n)$ terms where 
$N(n)=n[n-1][n-2]$ for the operator in \cite{Rovelli:1995ge} while for the 
operator of \cite{Ashtekar:1997fb} this number is smaller whenever there are 
triples of edges with co-planar tangents at $v$. The smallest valence 
for which the volume operator does not vanish is $n=3$ in which case
$N(3)\le 6$. Since each term corresponds to the volume of the cell of the 
polyhedronal decomposition dual to $\gamma$, the factor $1/48$ dividing 
the sum over triples is too large. Now $N(4)\le 24$ is still too small
while $N(5)\le 120$ is already definitely too large for the volume 
operator of \cite{Rovelli:1995ge}. For the cubic topology we have, however, 
precisely $N(6)=48$ for \cite{Ashtekar:1997fb} because the only triples that 
contribute are formed by 
those spanning the eight octants defined by the coordinate system defined 
by the tangents of the six edges at $v$. For graphs of higher 
valence, unless there are sufficiently many coplanar triples, the 
\cite{Ashtekar:1997fb} volume 
operator also over-counts the classical volume. Notice that none of these 
statements proves that one operator is proved over the other, it just 
means that our coherent states do not approximate both equally well. Only 
if one would know that our states are ``the correct choice'', could one 
distinguish between the two kinds of volume operators on physical grounds.

Intuitively, it is actually not too bad that only graphs of low valence 
should give rise to the correct classical limit. After all, one would not
try to approximate a classical integral by Riemann sums in terms of graphs 
with vertices of arbitrarily high topology. Such graphs should describe 
quantum states without classical correspondence. It is also natural that
cubic graphs are somehow distinguished because the classical integral 
is locally defined by a Cartesian coordinate system.

Having convinced ourselves that the coherent states of the previous 
section actually do make sense at least for operators of the kind
(\ref{3.1}) and (\ref{3.2}) we turn to the question how the scale 
$\epsilon$ should be chosen. In order to quantize the classical integral
\be \label{3.4}
O(\omega)=\int_\Sigma d^3x O^{ab} \omega_a\omega_b
\ee
the procedure adopted in \cite{Thiemann:1998rt} was to define the 
operator on the spin-network basis. Thus, let $\gamma$ be a graph
and $v\mapsto R_v$ a partition of $\Sigma$ where $v$ runs through
$V(\gamma)$. Let $\epsilon_v^3$ be the coordinate volume of $R_v$.
Then we have 
\be \label{3.5}
O(\omega)=\sum_{v\in V(\gamma)}
\int_{R_v} d^3x O^{ab} \omega_a\omega_b
\approx \sum_v \epsilon_v^3  O^{ab}(v) \omega_a(v) 
\omega_b(v)=:O_\gamma(\omega)
\ee
where in the last step we have replaced the integral by a Riemann sum.
The quantization of the term at $v$ in the sum in (\ref{3.5}) gives
rise to the operator
\be \label{3.5a} 
\hat{O}_\gamma(\omega,v)=
\sum_{v\in \partial e,\partial e';e,e' \in E(\gamma(s))} 
\omega(e)\omega(e') 
\hat{O}_{v;e,e'}
\ee
in (\ref{3.3}) and by construction its expectation value in a coherent 
state $\psi_{\gamma,m}$ gives back $O_\gamma(\omega,v)_{|m}$ to zeroth 
order in 
$\hbar$. Thus, apart from quantum corrections for the expectation value,
which we will call a {\it normal ordering error}, already the quantity 
$O(\omega,m)-O_\gamma(\omega_m)$ is in general non-zero. This 
{\it classical error} error will decrease with $\epsilon$. With 
the Euler-MacLaurin error estimation methods of \cite{Sahlmann:2001nv} one can prove 
an estimate of the form
\be \label{3.6}
|O(\omega,m)-O_\gamma(\omega,m)|\le [\frac{\epsilon}{L}]^\beta O(\omega,m)
\ee
where $\beta\ge 2$ and $L$ is the average size of the the quantity
$[O^{ab} \omega_a \omega_b]^{\prime\prime}/[O^{ab} \omega_a 
\omega_b]$ where the double prime denotes second derivatives.
Thus, $L$ captures information about the gravitational curvature as well
as the curvature of $\omega$. The size of $\beta$ depends strongly
on the randomness of the graph in question and also would change 
if one would average over graphs.

More precisely,
if we are interested in diffeomorphism invariant quantities (\ref{3.1}),
(\ref{3.2}) such as the matter Hamiltonians that we wish to 
approximate in the following sections, then we should set, e.g., 
$\omega_a=\phi_{,a}$ where 
$\phi$ is a scalar field or we should consider integrands of the form 
$q_{ab}E^a E^b/\sqrt{\det(q)}$ where $E^a$ is the Maxwell electric 
field. To see what the matter and geometry scales involved are, consider 
the time -- time
component of the Einstein equations for electromagnetic waves 
with vector potential $A=A_0 e^{i(|k|t-kx)}$. If $q^2$ is the electric 
charge, then the matter energy density is of the order $A_0^2 k^2/q^2$.
If $R$ denotes the curvature radius of the curvature tensor then 
we get from Einstein's equations $R^{-2}\approx (\ell_P A_0 k)^2/\alpha$
where $\alpha=\hbar q^2$ is the Feinstruktur constant. Thus,
if we introduce the wave length by $k=1/\lambda$ then 
$R^{-2}\approx (10 A_0 \ell_P)^2 \lambda^{-2}\ll \lambda^{-2}$ 
at least for weak electromagnetic waves $A_0\ll 10^{32} \mbox{cm}^{-1}$.
Thus, $L$ should, for the applications of this paper to be thought of 
being very close to the matter wave length $\lambda$ and $R$ is large,
so that the geometry is almost flat.

Let us now consider fluctuations. Since the quantities 
$M(\omega),Q(\omega)$ have different physical units, in order to compare
their fluctuations we should compare their relative fluctuations which
are dimension-free quantities. More precisely, we consider the 
expectation value in the coherent state $\psi_{\gamma,m}$ of the relative 
deviation squared $[\hat{O}/O(m)-1]^2$ between the operator $\hat{O}$
and its expected classical value $O(m)$ which is a proper measure 
for the total deviation of the operator from the classical quantity due
to 1) the fluctuation of the gravitational field and 2) its discrete
nature which 
forces us to work with graphs rather than continuous integrals. 
If we denote by $<.>_{m,\gamma}$ the expectation value in the 
coherent state $\psi_{m,\gamma}$ and if there is no normal ordering error 
then we arrive at 
\be \label{3.7}
<[\frac{\hat{O}}{O(m)}-1]^2>_{m,\gamma}\approx
<[\frac{\hat{O}_\gamma}{O_\gamma(m)}-1]^2>_{m,\gamma}+
[\frac{\hat{O}_\gamma(m)}{O(m)}-1]^2
\ee
The second term in (\ref{3.7}) is of order $(\epsilon/L)^{2\beta}$ as 
derived above. Now we see that for the quantity $M(\omega)$ the first
term is divergent for flat data because $M(\omega)=0$ while for 
$O=Q(\omega)$ there is no such problem. This is like comparing the 
relative fluctuations of $\hat{x},\hat{p}$ for the harmonic oscillator 
at the phase space point $(x,p)=(0,1)$ which of course makes little 
sense. To deal with this problem we chose the following strategy: 
We compare the relative fluctuations 
at {\it generic points in phase space} where we find a relation between
the scale $a$ of the coherent state and the scale $L$ and {\it then 
extend this relation to all points in $\cal M$}. This strategy is 
certainly ad hoc but we do not see any other possibility at this point 
to fix the size of $a$ by a more physical requirement. 

Accepting this we will consider non-flat data in which case generically
$L\approx R$ is closer to the curvature scale.
If we assume that the operators 
$\hat{O}_\gamma(\omega,v)$ in (\ref{3.3}) are much weaker correlated for 
distinct $v$ than for coinciding $v$ (as it turns out to be the case) then
we obtain 
\be \label{3.8}
<\hat{O}_\gamma^2>_{m,\gamma}-(<\hat{O}_\gamma>_{m,\gamma})^2
\approx \sum_v 
[<\hat{O}_{v,\gamma}^2>_{m,\gamma}-(<\hat{O}_{v,\gamma}>_{m,\gamma})^2]
\ee
Restricted to $\gamma$, the operator $\hat{O}_{v,\gamma}$ is a 
homogeneous polynomial 
of some rational power of the operators $P^e_j\approx E_j(S_e)/a^2$ 
which are of 
order $E_0 \epsilon^2/a^2$ where $E_0$ is some average value of
$E^a_j$ and $a$ is the coherent state scale introduced in the previous 
section. It is also a polynomial of some integral power of the operator 
$h_\alpha-h_\alpha^{-1}$ which is of order $B_0\epsilon^2$ where $B_0$
is some average value of $B^a_j$ and is approximately given by 
$E_0 L^{-2}$. As shown in \cite{Sahlmann:2001nv}, the fluctuations for the respective
vertices $v$ is effectively given by exchanging $O_{\gamma,v}(m)$ by
$t \partial^2 O_{\gamma,v}(m)/[\partial P(S_e)]^2
\approx t O_{\gamma,v}/P(S_e)^2$ for the electric
fluctuations and by 
$t \partial^2 h_\alpha^2  O_{\gamma,v}(m)/[\partial h_\alpha]^2
\approx t O_{\gamma,v}/[h_\alpha-h_\alpha^{-1}]^2$ 
for the magnetic ones where $\alpha$ is some loop incident at $v$. 
Inserting $P(S_e)\approx E_0\epsilon^2/a^2$ and 
$h_\alpha-h_\alpha\approx E_0 \epsilon^2/R^2\approx E_0 \epsilon^2/L^2$.
equating (\ref{3.8}) for $O=M(\omega),O=Q(\omega)$ respectively 
immediately leads to $a\approx L$. 

While the derivation of this result is maybe not entirely convincing,
it is actually the only choice from a classical point of view: Since 
$L$ is the only classical scale available and the complexifier 
generator $C$ for our coherent states, from which
the scale $a$ derives, is a classical object, the scale $L$ is the only 
classical one in the problem that should be used in order to make 
$C/\hbar$ dimensionfree.

Coming back to flat space $m=(A_a^j,E^a_j)=(0,\delta^a_j)$ we want to fix 
$\epsilon$ by requiring that 
the relative fluctuation (\ref{3.7}) for $O=Q(\omega)$ is minimized.
This leads to the condition that (notice $E_0=1$)  
\be \label{3.9}
t\frac{\epsilon^3}{\mbox{Vol}(\mbox{supp}(\omega))} 
\frac{1}{[\epsilon^2/a^2]^2}+[\epsilon^2/L^2]^\beta
\ee
be minimized where $a:=L$. The fluctuation contribution depends on the 
volume of the support of $\omega$. Since we want to resolve regions 
with our graph of the linear size bigger or equal than $L$ (think
of $L$ as the smallest wavelength to be resolved for our applications)
we obtain that (\ref{3.9}) is certainly dominated by
\be 
t\frac{L}{\epsilon}+[\epsilon^2/L^2]^\beta
\ee
This function has a unique minimum at 
\be \label{3.10}
\epsilon=\ell_P^\alpha L^{1-\alpha},\;\;\alpha=\frac{1}{\beta+\frac{1}{2}}
\le \frac{2}{5}<\frac{1}{2}
\ee
In \cite{Sahlmann:2001nv} we chose  
$\mbox{Vol}(\mbox{supp}(\omega))\ge\epsilon^3$ and different 
observables, adapted to the graph in question, in order to have the 
lattice degrees of freedom well approximated and led to $\alpha \approx 
1/6$. However, it is clear that this choice would lead to boundary effects 
if the support of $\omega$ is not adapted to the graph in question which
would be unnatural. Such boundary effects
are avoided by $\mbox{Vol}(\mbox{supp}(\omega))\ge L^3$
and go at most as the quotient between the volume of a shell of 
thickness $\epsilon$ at the 
boundary of a region of volume $L^3$ and its volume, that is, as 
$\epsilon/L$. This drives the 
lattice scale $\epsilon$ closer to the Planck scale. Notice that in 
any case $\ell_P\ll \epsilon \ll L$.

This concludes the present section. The relations $a:=L$ and 
(\ref{3.10}) will be our working proposal.
%----------------------------------------------------------------------
\section{Coherent States Expectation Values}
\label{s4}
%----------------------------------------------------------------------
The purpose of this section is to present the calculation of the
expectation values of the various terms occurring in the Hamiltonians 
of section 4 in \cite{ST01} in the coherent states for QGR discussed
in the preceeding sections. 
In the first part we will explain the simplifying assumptions used for the
computation and introduce the necessary notation. Section \ref{sea.2} 
is devoted to the computation of the expectation values of the volume
operator $\widehat{V}_v$ and the operator
\begin{equation}
  \label{eq2.12}
  \hat{Q}^j_e(v,r)=\frac{1}{4r}\mbox{tr}(\tau_j h_e[h_e^{-1},(\hat{V}_v)^r]),
\end{equation}
as they are the basic building blocks of the Hamiltonians obtained in
\cite{ST01}. 
In section \ref{sea.3} the results are used to give the  
expectation values of the geometric operators occurring in the
Hamiltonians for the scalar and the electromagnetic field. 
%----------------------------------------------------------------------
\subsection{Implementation of the Simplifying Assumptions}
%----------------------------------------------------------------------
\label{sea.1}

%----------------------------------------------------------------------
{\bf The cubic lattice:}\\
%----------------------------------------------------------------------
For reasons already explained in our companion paper,
the first simplification that we will make concerns the random graphs: 
In the following we will exclusively work with states based on graphs
of cubic topology. This simplifies both the notation and the c-number
coefficients occurring in the Hamiltonians. 
In a graph of cubic topology, each vertex is six-valent with three
edges ingoing and
three outgoing. We denote the outgoing edges by $e_I$,
$I=1,2,3$ and choose an ordering, such that the tangents of
$e_1,e_2,e_3$ form a right handed triple wrt. the given orientation of 
$\Sigma$. The vertices can be labeled by elements $v$ of $\Z^3$. We
write $e^+_I(v):=e_I(v),\;e^-_I(v):=e_I(v-I)^{-1}$ where $n-I$ denotes
the point in $\Zl^3$ translated one unit along the negative $I$ axis.
In keeping with that convention, we associate to 
$e^-_I(v)$ the dual surface $S_{e_I(v-I)}$ with its orientation
\textit{reversed}. 
%----------------------------------------------------------------------
\\\\{\bf Replacing $\sutwo$ by $\uone^3$:}\\
%----------------------------------------------------------------------
We substitute $SU(2)$ by $U(1)^3$ in our computation because the results
of \cite{Thiemann:2000ca,Thiemann:2000bx} reveal that the qualitative features are untouched 
so nothing conceptually new is learned when doing the much harder 
non-Abelian computation. For the exploratory purposes of this paper it
is thus sufficient to stick with the Abelian group. Consequently we will 
replace $\widehat{Q}$ as well as the volume operator 
itselve by appropriate $\uone^3$ counterparts. For $\uone^3$ each edge
is not labelled by a single, non-negative, half-integral spin degree of 
freedom but rather by three integers $n_j\in \Z,\;j=1,2,3$ and we have
three kinds of holonomies $h_e^j$. The generators $\tau_j$ of $\uone^3$
are simply $i$ (imaginary unit). The canonical commutation relations
on $\lzwo(\uone ^3,d^3\mu_H)$ are replaced by
\begin{align*} 
{[}\widehat{h}^j,\widehat{h}^k]&= 0 \\
{[}\widehat{p}_j,\widehat{h}^k]&= i t \delta_j^k\widehat{h}^j\\
{[}\widehat{p}_j,\widehat{p}_k]&= 0
\end{align*} 
(cf. \eqref{eq1.15}) with adjointness relations 
$(\widehat{h}^j)^\dagger=(\widehat{h}^j)^{-1}$, 
$(\widehat{p}_j)^\dagger=\widehat{p}_j$.
It follows that \eqref{eq2.12} gets replaced by
\begin{equation*}
\widehat{Q}_e^j(v,r)= \frac{i}{4r} \widehat{h}^j_e
\comm{(\widehat{h}^j_e)^{-1}}{\widehat{V}_v^r}, 
\end{equation*}
Finally the expression for the volume operator 
in our companion paper is replaced 
by
\begin{equation*}
\widehat{V}_{\gamma,v}=l_p^3\sqrt{|\epsilon^{jkl}
{[}\frac{\widehat{Y}^{e^+_1(v)}_j-\widehat{Y}^{e^-_1(v)}_j}{2}]\;
{[}\frac{\widehat{Y}^{e^+_2(v)}_k-\widehat{Y}^{e^-_2(v)}_k}{2}]\;
{[}\frac{\widehat{Y}^{e^+_3(v)}_l-\widehat{Y}^{e^-_3(v)}_l}{2}]|}
\end{equation*}
with $\widehat{Y}^e_j=ih^j\partial/\partial h^j$.

The $\uone ^3$ coherent states over any graph $\gamma$ are given by 
(see \cite{Thiemann:2000ca})
\begin{equation*}
\psi^t_{\gamma,m}=\otimes_{e\in E(\gamma)}\otimes_{j=1}^3\psi^t_{g^j_e(m)}
\end{equation*}
where 
\begin{equation*}
\psi^t_g=\sum_{n\in\Z} e^{-t n^2/2} (gh^{-1})^n
\end{equation*}
and $g^j_e(m)=e^{p^e_j(m)}h_e^j(m)\in\C -\{0\}=\uone ^\C$. Here $m$
is a point in the gravitational phase space and 
\begin{align*}
h_e^j(m)&\doteq {\cal P}\exp(i\int_e A^j) \nonumber\\
p^e_j(m)&\doteq \frac{1}{a^2}\int_{S_e} (\ast E)_j
\end{align*}
that is, due to the Abelian nature of our simplified gauge group the 
path system in $S_e$ is no longer needed.
 
As is obvious from the explicit form of the Hamiltonians, our
calculation can be done vertex by vertex since there is no
inter-gravitational interaction between the associated operators. 
We can therefore 
concentrate on a single vertex for the remainder of this section
and drop the label $v$ in what follows. 

For the sake of the computation to follow, we introduce the
shorthands
\begin{equation*}
h_{J\sigma j}\doteq h^j_{e^\sigma_J},\qquad
p_{J\sigma j}\doteq p^{e^\sigma_J}_j,\qquad
g_{J\sigma j}\doteq e^{p_{J\sigma j}}h_{J\sigma j}
\end{equation*}
and similarly the operators 
$\widehat{Y}_{J\sigma j}\doteq\widehat{Y}_j^{e^{\sigma}_J}$.  
Let us finally define 
\begin{equation}
\label{trans}
\widehat{\sqcup}\doteq \frac{1}{a^3}\widehat{V},\qquad
\widehat{q}_{J\sigma j}(r)\doteq 
\frac{r}{2ita^{3r}}\widehat{Q}^j_{e^\sigma_J}(r).
\end{equation}
Note that $\widehat{q}$ is essentially selfadjoint. 

The huge advantage of $\uone ^3$ 
over $\sutwo$ is that the ``spin-network functions'' 
\begin{equation*}
T_{\{n_{J\sigma j}\}}(\{h_{J\sigma j}\})=
\prod_{J\sigma j} h_{J\sigma j}^{-n_{J\sigma j}} 
\end{equation*}
are simultaneous eigenfunctions of all the $\widehat{Y}_{J\sigma j}$ with 
respective eigenvalue $n_{J\sigma j}$. Even better, the operator
$\widehat{q}_{J_0\sigma_0 j_0}(r)$ is also diagonal with eigenvalue
\begin{equation*}
\lambda^r_{J_0\sigma_0 j_0}(\{n_{J\sigma j}\})=
2\frac{\lambda^r(\{n_{J\sigma j}\})-
\lambda^r(\{n_{J\sigma j}+\delta_{(J_0\sigma_0 j_0),(J\sigma j)}\})}{t},
\end{equation*}
where 
\begin{equation*}
\lambda^r(\{n_{J\sigma j}\})
=t^{3r/2} \left(\sqrt{|\epsilon^{jkl}
{[}\frac{n_{1,+,j}-n_{1,-,j}}{2}]\;
{[}\frac{n_{2,+,k}-n_{2,-,k}}{2}]\;
{[}\frac{n_{3,+,l}-n_{3,-,l}}{2}]|}\right)^r.
\end{equation*}
%----------------------------------------------------------------------
\subsection{The Expectation Values of $\widehat{q}$}
%----------------------------------------------------------------------
\label{sea.2}
Now we will explicitly calculate the expectation values of powers of 
the operators 
$\widehat{q}$ and $\widehat{\sqcup}$ .
The gravitational parts of the matter Hamiltonians constructed in \cite{ST01}
are all sums and products
of these operators which act only on the edges of a specific vertex,
therefore we can restrict consideration to a single vertex and consequently 
to a part 
\begin{equation*}
\psi^t_{\{g_{J\sigma j}\}}(\{h_{J\sigma j}\})
\doteq \prod_{J\sigma j} \psi^t_{g_{J\sigma j}}(h_{J\sigma j}) 
\end{equation*}
of the coherent state which just contains the factors corresponding to
the edges of a single vertex.\newline
What we are looking for is the expectation value of an arbitrary
polynomial of the $\widehat{q}$:
\begin{align} \label{fluc4.18}
\expec{\,\cdot\,}
&\doteq
\frac{\scpr{\psi^t_{\{g_{J\sigma j}\}}}{
\prod_{k=1}^N\widehat{q}_{J_k\sigma_k j_k}(r_k)
\psi^t_{\{g_{J\sigma j}\}}}}{||\psi^t_{\{g_{J\sigma j}\}}||^2}
\nonumber\\
&=
\frac{
\sum_{\{n_{J\sigma j}\}} e^{-t\sum_{J,\sigma,j} n_{J\sigma j}^2}
e^{2\sum_{J\sigma j} p_{J\sigma j} n_{J\sigma j}}
\prod_{k=1}^N \lambda^r_{J_k\sigma_k j_k}(\{n_{J\sigma j}\})
}
{\prod_{J,\sigma,j}||\psi^t_{g_{J\sigma j}}||^2}
\end{align}
where (see \cite{Thiemann:2000ca})
\begin{equation}\label{fluc4.19}
||\psi^t_g||^2=\sqrt{\frac{\pi}{t}} e^{p^2/t}[1+K_t(p)],\;
g=e^p e^{i\varphi},\;|K_t(p)|\le K_t=O(t^\infty).
\end{equation}
As in \cite{Thiemann:2000ca}, in order to extract 
useful information out of the 
formula (\ref{fluc4.18}) it is of outmost importance to perform a 
Poisson transformation on it because we are interested in tiny values
of $t$ for which (\ref{fluc4.18}) converges rather slowly while the transformed 
series converges rapidly since then $t$ gets replaced by $1/t$. 
To that end, let us introduce $T\doteq \sqrt{t},x_{J\sigma j}\doteq T n_{J\sigma j}$,
whereupon
\begin{equation}\label{fluc4.20}
\expec{\,\cdot\,}=
\frac{
\sum_{\{x_{J\sigma j}\}} e^{-\sum_{J,\sigma,j} x_{J\sigma j}^2}
e^{2\sum_{J\sigma j} x_{J\sigma j} p_{J\sigma j}/T}
\prod_{k=1}^N \lambda^r_{J_k\sigma_k j_k}(\{x_{J\sigma j}\})
}
{\prod_{J,\sigma,j}||\psi^t_{g_{J\sigma j}}||^2}
\end{equation}
where 
\begin{align} \label{fluc4.21}
 \lambda^r_{J_0\sigma_0 j_0}(\{x_{J\sigma j}\})&=
2\frac{\lambda^r(\{x_{J\sigma j}\})-
\lambda^r(\{x_{J\sigma j}+T\delta_{(J_0\sigma_0 j_0),(J\sigma j)}\})}{t}
\nonumber\\
 \lambda^r(\{x_{J\sigma j}\})
&=t^{3r/4} \sqrt{|\epsilon^{jkl}
{[}\frac{x_{1,+,j}-x_{1,-,j}}{2}]\;
{[}\frac{x_{2,+,k}-x_{2,-,k}}{2}]\;
{[}\frac{x_{3,+,l}-x_{3,-,l}}{2}]|}^r
\end{align}
Then Poisson's theorem gives
\begin{equation}\label{fluc4.22}
\expec{\,\cdot\,}=
\frac{\frac{1}{T^{18}}
\sum_{\{n_{J\sigma j}\}} \int_{\R^{18}} d^{18}x 
e^{\sum_{J,\sigma,j} [-x_{J\sigma j}^2
+2 x_{J\sigma j} (p_{J\sigma j}-i\pi n_{J\sigma j})/T]}
\prod_{k=1}^N \lambda^r_{J_k\sigma_k j_k}(\{x_{J\sigma j}\})
}{\prod_{J,\sigma,j}||\psi^t_{g_{J\sigma j}}||^2}
\end{equation}
An observation that reduces the eighteen dimensional integral to a
nine dimensional one is that the integrand in (\ref{fluc4.22}) only depends
on $x_{Jj}\doteq x^-_{Jj}\doteq [x_{J,+,j}-x_{J,-,j}]/2$ and not on
$x^+_{Jj}\doteq [x_{J,+,j}+x_{J,-,j}]/2$. 
Consider also the analogous quantities
$p^\pm_{Jj}\doteq [p_{J,+,j}\pm p_{J,-,j}]/2,\;
n^\pm_{Jj}\doteq [n_{J,+,j}\pm n_{J,-,j}]/2$ and let
$p_{Jm}\doteq p^-_{Jj},\;n_{Jm}\doteq p^-_{Jj}$. 
Switching to the coordinates $x^\pm_{Jj}$, noticing that 
$|\det(\partial\{x_{J\sigma j}\}/
\partial\{x^+_{Jj},x^-_{Jj}\}|=2^9$ we obtain
\begin{align} \label{fluc4.23}
\expec{\,\cdot\,}&=
\frac{(\frac{2}{t})^9
\sum_{\{n_{J\sigma j}\}} 
[\int_{\R^9} d^9 x^+ e^{2\sum_{Jj} [-(x^+_{Jj})^2
+2 x^+_{Jj} (p^+_{Jj}-i\pi n^+_{Jj})/T]}]}
{\prod_{J,\sigma,j}||\psi^t_{g_{J\sigma j}}||^2}
\times\nonumber\\
&\qqquad\times
\left[\int_{\R^9} d^9 x e^{2\sum_{Jj} [-x_{Jj}^2
+2 x_{Jj} (p_{Jj}-i\pi n_{Jj})/T]}
\prod_{k=1}^N \lambda^r_{J_k\sigma_k j_k}(\{x_{Jj}\})\right]
\end{align}
where 
\begin{align} \label{fluc4.24}
\lambda^r_{J_0\sigma_0 j_0}(\{x_{Jj}\})=&
2\frac{\lambda^r(\{x_{J j}\})-
\lambda^r(\{x_{Jj}+T\delta_{(J_0 j_0),(J j)}/2\})}{t}
=:\lambda^r_{J_0 j_0}(\{x_{J j}\})
\nonumber\\
 \lambda^r(\{x_{Jj}\})
&=t^{3r/4} \left(|\det(\{x_{Jj}\}\right)^{r/2}
\end{align}
actually {\it no longer depends on $\sigma_0$}!
The integral over $x^+_{Jj}$ in (\ref{fluc4.24}) can be immediately performed 
by using a contour argument with the result
\begin{equation}\label{fluc4.25}
\expec{\,\cdot\,}=
\frac{(\frac{\sqrt{2\pi}}{t})^9
\sum_{\{n_{J\sigma j}\}} e^{\frac{2}{t}\sum_{Jj}(p^+_{Jj}-i n^+_{Jj})^2} 
[\int_{\R^9} d^9 x e^{2\sum_{Jj} [-x_{Jj}^2
+2 x_{Jj} (p_{Jj}-i\pi n_{Jj})/T]}
\prod_{k=1}^N \lambda^r_{J_k\sigma_k j_k}(\{x_{Jj}\})]
}
{\prod_{J,\sigma,j}||\psi^t_{g_{J\sigma j}}||^2}
\end{equation}
Finally, using (\ref{fluc4.19}) we can further simplify to
\begin{align} \label{fluc4.26}
\expec{\,\cdot\,}&=
\frac{\sqrt{\frac{2}{\pi}}^9}{[(1-K_t)^{18},(1+K_t)^{18}]}
\sum_{\{n_{J\sigma j}\}} 
e^{\frac{2}{t}\sum_{Jj}[(p^+_{Jj}-i \pi n^+_{Jj})^2-(p^+)_{Jj}^2-p_{Jj}^2]} 
\times\nonumber\\
&\times \int_{\R^9} d^9 x e^{2\sum_{Jj} [-x_{Jj}^2
+2 x_{Jj} (p_{Jj}-i\pi n_{Jj})/T]}
\prod_{k=1}^N \lambda^r_{J_k\sigma_k j_k}(\{x_{Jj}\})
\end{align}
where the notation for the denominator means that its value ranges
at most in the interval indicated. Its precise value will be irrelevant
for what follows since its departure from unity is $O(\infty)$.
\\[1ex]
%-----------------------------------------------------------
{\bf Only the $n_{J\sigma,j}=0$ terms matter:}\\[1ex]
%-----------------------------------------------------------
The remaining integral in (\ref{fluc4.26}) cannot be computed in closed form
so that we must confine ourselves to a judicious estimate. We wish to
show that the only term in the infinite sum of (\ref{fluc4.26}) which 
contributes
corrections to the classical result of finite order in $t$ is the one with 
$n_{J\sigma j}=0$ for all $J,\sigma, j$. In order to do that, we must 
demonstrate that all the other terms can be estimated in such a way 
that the series of their estimates converges to an $O(t^\infty)$ number.
This would be easy if we could complete the square in the exponent of the
integrand but since for $r/2$ not being an even 
positive integer the function $\lambda^r$ is not analytic in $\C^9$
we cannot immediately use a contour argument in order to estimate the
remaining integral. In order to proceed and to complete the square 
anyway we expand the product 
$\prod_{k=1}^N \lambda^r_{J_k\sigma_k j_k}(\{x_{Jj}\})$ into
monomials of the form 
$\prod_{k=1}^N \frac{\lambda^r(\{x_{Jj}+c^k_{Jj}\})}{t}$ with 
$c^k_{Jj}=T\delta_{J_k j_k,J j}/2$ or $c^k_{Jj}=0$ and estimate the
integrals over the latter. We trivially have 
\begin{equation}\label{fluc4.27}
\lambda^r(\{x_{Jj}+c^k_{Jj}\})=
t^{3r/4} ([\det(\{x_{Jj}+c^k_{Jj}\})]^2)^{r/4}
=t^{3r/4} \exp\left(\frac{r}{4}\ln([\det(\{x_{Jj}+c^k_{Jj}\})]^2)\right)
\end{equation}
where we must use the branch of the logarithm with $\ln(z)=\ln(|z|)+i\varphi$
for any complex number $z=|z|e^{i\varphi}$ with $\varphi\in [0,2\pi)$.
With this branch understood, in the form (\ref{fluc4.27}) the 
integrand of (\ref{fluc4.26}) becomes univalent on the entire complex
manifold $\C^9$ except at the points where $\det(\{x_{Jj}+c^k_{Jj}\})=0$. 
Now a laborious contour argument can be given tho the extent that we
can move the path of integration away from the real hyperplane in
$\C^9$ without changing the result. Therefore we can indeed complete
the square in the exponent. 

It remains to estimate (\ref{fluc4.26}) from above. Isolating the term
with $n_{J\sigma,j}=0$ for all $J,\sigma,j$ we have 
\begin{align} \label{fluc4.34}
& \betr{\expec{\,\cdot\,}-
\frac{\sqrt{\frac{2}{\pi}}^9}{[(1-K_t)^{18},(1+K_t)^{18}]}
\int_{\R^9} d^9 x e^{-2\sum_{Jj} x_{Jj}^2}
\prod_{k=1}^N \lambda^r_{J_k\sigma_k j_k}(\{x_{Jj}+p_{Jj}/T\})}
\nonumber\\
&=
\bigg|\frac{\sqrt{\frac{2}{\pi}}^9}{[(1-K_t)^{18},(1+K_t)^{18}]}
\sum_{\{n_{J\sigma j}\}\not=\{0\}} 
e^{\frac{2}{t}\sum_{Jj}[(p^+_{Jj}-i \pi n^+_{Jj})^2
+(p_{Jj}-i\pi n_{Jj})^2-(p^+)_{Jj}^2-p_{Jj}^2]} 
\times\nonumber\\
&\qqquad\times \int_{\R^9} d^9 x e^{-2\sum_{Jj} x_{Jj}^2}
\prod_{k=1}^N \lambda^r_{J_k\sigma_k j_k}(\{x_{Jj}+(p_{Jj}-i\pi 
n_{Jj})/T\})|
\nonumber\\
&\le
(\frac{2}{t})^N\bigg|\frac{\sqrt{\frac{2}{\pi}}^9}{(1-K_t)^{18}}
\sum_{\{n_{J\sigma j}\}\not=\{0\}} 
e^{-\frac{\pi^2}{t}\sum_{J\sigma j} n_{J\sigma j}^2}\int_{\R^9} d^9 x e^{-2\sum_{Jj} x_{Jj}^2}
\times\nonumber\\
&\qqquad\times 
\prod_{k=1}^N 
[e^{\frac{r}{2}\ln(|\det(\{T x_{Jj}+(p_{Jj}-i\pi n_{Jj})\})|)}+
e^{\frac{r}{2}
\ln(|\det(\{T x_{Jj}+t \delta_{(J j),(J_k j_k)}+(p_{Jj}-i\pi 
n_{Jj})\})|)}]\bigg|.
\end{align}
Let $w_{Jj}$ be a matrix of complex numbers and define the norm
$\norm{w}^2\doteq \sum_{Jj} |w_{Jj}|^2$ so that in particular
$||w_1+w_2||\le ||w_1||+||w_2||$ and $|w_{Jj}|\le \norm{w}$ for all $J,j$.
Now $\det(\{w_{Jj}\})$ is a linear combination of six monomials of the 
form $w_{J_1 j_1} w_{J_2 j_2} w_{J_3 j_3}$ so that 
$|\det(\{w_{Jj}\})|\le 6\norm{w}^3$. In particular,
$|\det(\{T x_{Jj}+(p_{Jj}-i\pi n_{Jj})\})|\le 6(T\norm{x}+\norm{p}+\pi \norm{n})^3$
and 
$|\det(\{T x_{Jj}+t \delta_{(J j),(J_k j_k)}/2+(p_{Jj}-i\pi 
n_{Jj})\})|\le 6(T\norm{x}+t+\norm{p}+\pi \norm{n})^3$. Invoking this result 
into (\ref{fluc4.34}) we find
\begin{align} \label{fluc4.35}
&\le
(\frac{4}{t})^N\bigg|\frac{\sqrt{\frac{2}{\pi}}^9}{(1-K_t)^{18}}
\sum_{\{n_{J\sigma j}\}\not=\{0\}} 
e^{-\frac{\pi^2}{t}\sum_{J\sigma j} n_{J\sigma j}^2}
\int_{\R^9} d^9 x e^{-2\norm{x}^2}
e^{\frac{Nr}{2}\ln(6[T\norm{x}+t+\norm{p}+\pi \norm{n}]^3)}\bigg|
\nonumber\\
&\le
(\frac{4\;6^{r/2}}{t})^N\bigg|\frac{\sqrt{\frac{2}{\pi}}^9}{(1-K_t)^{18}}
\sum_{\{n_{J\sigma j}\}\not=\{0\}} 
e^{-\frac{\pi^2}{t}\sum_{J\sigma j} n_{J\sigma j}^2}\times\nonumber\\ 
&\qqquad\times\int_{\R^9} d^9 x e^{-2\norm{x}^2}
[\frac{1}{4}+t \norm{x}^2+t+\norm{p}+\pi \norm{n}]^{[\frac{3Nr}{2}]+1}\bigg|
\end{align}
where $[3Nr/2]$ is the Gauss bracket of a real number (largest integer
smaller than or equal to $3Nr/2$) and in the last step we have 
used the elementary estimate $x\le x^2+1/4$ valid for any real number $x$.
The integral in the last line of (\ref{fluc4.35}) can be evaluated exactly
by invoking the binomial theorem. Consider the integrals of the form
\begin{equation}\label{fluc4.36}
I_k\doteq \sqrt{\frac{2}{\pi}}^m\int_{\R^m} d^m x
e^{-2\norm{x}^2} \norm{x}^{2k}
\end{equation}
for any positive integer $m$. Switching to polar coordinates one easily 
proves the recursion formula
\begin{equation}\label{fluc4.37}
I_k=\frac{m+2(k-1)}{4} I_{k-1}
\end{equation}
and since $I_0=1$ we find
\begin{align} \label{fluc4.38}
I_k&=\frac{(\frac{m}{2}+k-1)!}{2^k (\frac{m}{2})!}\mbox{ if }m\mbox{ even}
\nonumber\\
I_k&=\frac{(m-1+2k)!\;(\frac{m-1}{2})!}{8^k (m-1)! (\frac{m-1}{2}+k)!}
\mbox{ if }m\mbox{ odd}
\end{align}
Using the elementary estimate 
$e (n/e)^n\le n!\le e ((n+1)/e)^{n+1}$ we find for $0\le k\le n$
and $n\ge 2$ that 
\begin{align} \label{fluc4.39}
I_k\le e(\frac{m+2n}{2e})^{m/2} (\frac{m+2n}{4e})^{k}\doteq
C_{m,n}(\frac{m+2n}{4e})^{k}
\mbox{ if }m\mbox{ even}
\nonumber\\
I_k\le \frac{m-1}{2e}\frac{(\frac{m-1}{2})!}{(m-1)!} 
(\frac{m+2n}{m-1})^m (\frac{m+2n}{4(m-1)})^k
=:C_{m,n} (\frac{m+2n}{4(m-1)})^k
\mbox{ if }m\mbox{ odd}
\end{align}
In our case $m=9$ and $n=[\frac{3Nr}{2}]+1$. Thus, we can finish the 
estimate of (\ref{fluc4.35}) with
\begin{align} \label{fluc4.40}
& |<.>-
\frac{\sqrt{\frac{2}{\pi}}^9}{[(1-K_t)^{18},(1+K_t)^{18}]}
\int_{\R^9} d^9 x e^{-2\sum_{Jj} x_{Jj}^2}
\prod_{k=1}^N \lambda^r_{J_k\sigma_k j_k}(\{x_{Jj}+p_{Jj}/T\})|
\nonumber\\
&\le
\frac{(\frac{4\;6^{r/2}}{t})^N C_{9,[\frac{3Nr}{2}]+1}}{(1-K_t)^{18}}
\sum_{\{n_{J\sigma j}\}\not=\{0\}} 
e^{-\frac{\pi^2}{t}\sum_{J\sigma j} n_{J\sigma j}^2}
\times\nonumber\\
&\times 
[\frac{1}{4}+t\frac{9+2([\frac{3Nr}{2}]+1)}{32}+t+\norm{p}+\pi 
\norm{n}]^{[\frac{3Nr}{2}]+1} 
\end{align}
which is obviously of order $O(t^\infty)$. We can give a bound independent
of $p$ since in our applications $\norm{p}$ can be bounded by a constant of the 
order of $t^\alpha$. 

Let us summarize our findings in the form of a theorem.
\begin{Theorem} \label{th4.1}
Let $\norm{p(v)}^2\doteq \sum_{Jj} p_{Jj}(v)^2$.
Suppose that there 
exists a positive constant $K$ such that  
$\sup_{v\in V(\gamma),m\in{\cal M}}\norm{p(v)}=:\norm{p}\le K$ is uniformly 
bounded. Then for small $t$
\begin{equation} \label{fluc4.41}
\expec{\,\cdot\,}=\frac{\sqrt{\frac{2}{\pi}}^9}{[(1-K_t)^{18},(1+K_t)^{18}]}
\int_{\R^9} d^9 x e^{-2\sum_{Jj} x_{Jj}^2}
\prod_{k=1}^N \lambda^r_{J_k\sigma_k j_k}(\{x_{Jj}+p_{Jj}/T\})+O(t^\infty)
\end{equation}
independently of $m\in{\cal M},
v\in V(\gamma)$.
\end{Theorem}
%-------------------------------------------------------------------
{\bf Expansion of the remaining integral:}\\[1ex]
%-------------------------------------------------------------------
It remains to compute the power expansion (in $T$) of the remaining
integral in (\ref{fluc4.41}) and to show that at each order the remainder 
is smaller than the given order. We will see that only even powers 
of $T$ contribute so that this expansion is actually an expansion 
in $t$. The basic reason is that the expansion of the integrand 
in powers of $T$ is at the same time an expansion in powers of $x_{Jj}$
as is obvious from the explicit form of the functions $\lambda^r(\{x_{Jj}\})$.
These powers of $x_{Jj}$ are integrated against the Gaussian
$e^{-2\norm{x}^2}$ which is an even function under the reflection 
$x_{Jj}\to -x_{Jj}$ whence the integral for odd powers (an odd function under
reflection) must vanish. We will not be able to show that the integral
in (\ref{fluc4.41}), which certainly converges for any $p_{Jj},t$
(just set $\norm{n}=0$ in above estimate), can be 
expanded into an infinite series in powers of $t$, rather our estimates
will be only good enough in order to show that there is a maximal order
$n_0$ (which becomes infinite as $t\to 0$) in the sense that 
the remainder at order $n$ is smaller than the given order for 
all $n\le n_0$.
We will use rather coarse estimates which could possibly be much improved 
in order to raise the value of $n_0$ derived here but for all practical 
purposes the analysis 
described below will be sufficient since $n_0$ is anyway a rather large 
positive integer.

Consider once more the function $\lambda^r_{J\sigma j}(x+p/T)$:
Let us introduce $q\doteq p t^{-\alpha}$ which is of order unity and 
$s=t^{1/2-\alpha}$. Then 
\begin{equation}
\label{fluc4.42}
\lambda^r_{J\sigma j}(x+p/T)
=2|\det(p)|^{r/2}\frac{|\det(1+q^{-1}xs)|^{r/2}-|\det(1+q^{-1}xs
+q^{-1}\delta_{J j}sT/2)|^{r/2}}{t} 
\end{equation}
Now for any matrix $A$ we have $\det(1+A)=1+\tr (A)
+\frac{1}{2}[(\tr (A))^2-\tr (A^2)]+\det(A)=:1+z'_A$ and so 
$\det(1+A)^2=1+2z'_A+(z'_A)^2=:1+z_A=:y_A\ge 0$. Let
$y\doteq 1+z_{q^{-1}xs}$ and $y_1\doteq 1+z_{q^{-1}[xs+\sigma\delta_{Jj}sT]}$.
Then (\ref{fluc4.42}) becomes
\begin{equation}\label{fluc4.43}
\lambda^r_{J\sigma j}(x+p/T)|
=\frac{2|\det(p)|^{r/2}}{t}[y^{r/4}-y_1^{r/4}]
\end{equation}
and we should expand $y^{r/4},y_1^{r/4}$ around $y=y_1=1$. We now invoke
our knowledge that $0<r\le 1$ is a rational number, so we find positive 
integers $M>L>0$ without common prime factor such that $r/4=L/M$. Let us 
define recursively
\begin{align} \label{fluc4.44}
f^{(0)}_{L/M}(y)&\doteq  y^{L/M}, \nonumber\\
f^{(n+1)}_{L/M}(y) &\doteq  \frac{f^{(n)}_{L/M}(y)-f^{(n)}_{L/M}(1)}{y-1}.
\end{align}
It follows from this definition that
\begin{equation}\label{fluc4.45}
f^{(0)}_{L/M}(y)=\sum_{k=0}^n f^{(k)}_{L/M}(1) [y-1]^k+
f^{(n+1)}_{L/M}(y)[y-1]^{n+1}.
\end{equation}
\begin{Lemma} \label{la4.1}
We have 
\begin{equation}\label{fluc4.46}
f^{(k)}_{L/M}(1)=(L/M,k)
\end{equation}
where
\begin{equation*}
(L/M,k)\doteq \frac{(L/M)(L/M-1)\ldots(L/M-k+1)}{k !}
=(-1)^{k+1}\frac{L}{M}\frac{M-L}{2M}\frac{2M-L}{3M}\ldots\frac{(k-1)M-L}{kM}
\end{equation*}
and the following recursion holds for all $n\ge 1$
\begin{equation}\label{fluc4.47}
f^{(n+1)}_{L/M}(y)=
\frac{\sum_{k=1}^{L-1} f^{(n)}_{k/M}(y)-\sum_{l=1}^n f^{(l)}_{L/M}(1)
\sum_{k=1}^{M-1} f^{(n-l+1)}_{k/M}(y)}{\sum_{k=0}^{M-1} f^{(0)}_{k/M}(y)}.
\end{equation}
\end{Lemma}
The proof of the lemma consists in a straightforward Taylor expansion
(first part) and an induction (second part) and will not be reproduced 
here. 

The motivation for the derivation of this recursion is that it allows
us to estimate $|f^{(n+1)}_{L/M}(y)|$ once we have an estimate for all
the $|f^{(l)}_{k/M}(y)|$ with $0\le k\le M-1,0\le l \le n$.
\begin{Lemma} \label{la4.2}
For all $0<L\le M,\;n\ge 0$ we have 
\begin{equation}\label{fluc4.51}
|f^{(n)}_{L/M}(y)|\le (1+y)(\beta M)^n
\end{equation}
where $\beta>1$ is any positive number satisfying $\beta\ge 
1+\frac{\beta}{\beta-1}$, e.g. $\beta=3$.
\end{Lemma}
This lemma can be proven by induction, using the results of the
previous one. 

Using the expansion (\ref{fluc4.45}) and the fact that $y$ is a polynomial
in the $x_{Jj}$ it is possible evaluate the Gaussian integrals over 
the first $n$ terms the last one of which is obviously at least of order 
$s^n$. We would like to know at which order $n_0$ the remaining term
in (\ref{fluc4.45}) is no longer of order at least $s^{n_0+1}$. 

To that end recall that $y=1+2z+z^2$ where $z=\tr (A)+
\frac{1}{2}[(\tr (A))^2-\tr (A^2)]+\det(A)$ and 
$A_{jk}=s\sum_J (q^{-1})_{Jj}x_{Jk}$. We now have the following basic 
estimates 
\begin{align*} 
|\tr (A)|&=s|\sum_{Jj} q^{-1}_{Jj}x_{Jj}|\le s||q^{-1}||\;\norm{x}
\nonumber\\
|(q^{-1}x)_{jk}|&=|\sum_J q^{-1}_{Jj} x_{Jk}|
\le \sqrt{\sum_J [q^{-1}_{Jj}]^2} \sqrt{\sum_J [x_{Jk}]^2}
\nonumber\\
|\tr (A^2)|&= s^2|\sum_{jk} (q^{-1}x)_{jk}(q^{-1}x)_{kj}|
\le s^2|\sum_{jk} |(q^{-1}x)_{jk}|\;|(q^{-1}x)_{kj}|
\nonumber\\ 
&\le 
s^2[\sum_j \sqrt{\sum_J [q^{-1}_{Jj}]^2} \sqrt{\sum_J [x_{Jj}]^2}]
[\sum_k \sqrt{\sum_J [q^{-1}_{Jk}]^2} \sqrt{\sum_J [x_{Jk}]^2}]
\nonumber\\
&\le
s^2[\sqrt{\sum_j \sqrt{\sum_J [q^{-1}_{Jj}]^2}^2}
\sqrt{\sum_j \sqrt{\sum_J [x_{Jj}]^2}^2}]^2
\nonumber\\
&\le s^2 ||q^{-1}||^2\;\norm{x}^2
\nonumber\\
|\det(A)|&\le 6 s^3||q^{-1}x||^3 \le 6 s^3||q^{-1}||^3\;\norm{x}^3
\end{align*}
where in the first line we have made use of the Cauchy-Schwarz inequality
for the inner product $<x,x'>=\sum_{Jj} x_{Jj} x'_{Jj}$, in the second for 
the inner product $<x,x'>=\sum_J x_J x'_J$, in the fourth line for the inner
product $<x,x'>=\sum_j x_j x'_j$ and finally in the last line we have 
used the estimate derived between equations (\ref{fluc4.34}) and (\ref{fluc4.35}).
These estimates imply that
\begin{align*}
|z|&\le s\norm{q^{-1}}\;\norm{x}+ s^2\norm{q^{-1}}^2
\;\norm{x}^2+6|\det(q^{-1})|\; \norm{x}^3=:u(\norm{x}),\\
|y-1|&\le 2u+u^2=:P(\norm{x})
\end{align*}
and $P(\norm{x})$ is a polynomial of sixth order in $\norm{x}$.

We are now ready to estimate the Gaussian integral over the remainder:
\begin{align} \label{fluc4.55}
&E_n\doteq \betr{\sqrt{\frac{2}{\pi}}^9\int_{\R^9} d^9x e^{-2\norm{x}^2} 
f^{(n+1)}_{L/M}(y)[y-1]^{n+1}}
\nonumber\\
&\qqquad\le
\sqrt{\frac{2}{\pi}}^9 (3M)^{n+1} \int_{\R^9} d^9x e^{-2\norm{x}^2} 
[(P(\norm{x}))^{n+2}+2(P(\norm{x}))^{n+1}].
\end{align}
Consider an arbitrary polynomial in $\norm{x}$ of the form 
$$
P(x)=\sum_{k=0}^l a_k \norm{x}^k.
$$
By the multinomial theorem
$$
(P(x))^n=\sum_{n_0+..+n_l=n} \frac{n!}{(n_0 !)..(n_l)!}[\prod_{k=0}^l 
a_k^{n_k}]\;\norm{x}^{\sum_{k=0}^l k n_k}.
$$
Let us consider Gaussian integrals of the form
$$
\sqrt{\frac{2}{\pi}}^m\int_{\R^m} d^m x e^{-2\norm{x}^2} \norm{x}^n
=V_{m-1}\sqrt{\frac{2}{\pi}}^m
\int_0^\infty dr e^{-2r^2} r^{n+m-1}
=: V_{m-1}\sqrt{\frac{2}{\pi}}^m J_{n+m-1}
$$
where $V_m=2\pi^{m/2}/\Gamma(m/2)$ is the volume of $S^{m}$.
Now
\begin{align} \label{fluc4.56}
J_n &=\frac{\sqrt{2\pi}}{4} 2^{-3n/2} \frac{n!}{\frac{n}{2}!}
\mbox{ for }n\mbox{ even},
\nonumber\\
J_n &=\frac{1}{4} 2^{-(n-1)/2} (\frac{n-1}{2}!)
\mbox{ for }n\mbox{ odd},
\end{align}
and one immediately checks that 
\begin{equation*}
J_n\le \frac{\sqrt{2\pi}}{4}
\frac{[\frac{n}{2}]!}{2^{[\frac{n}{2}]}}
\end{equation*}
where $[.]$ again denotes the Gauss bracket. Using the above used estimate
for the factorial $n!\le e (\frac{(n+1)}{e})^{n+1}$ we may further estimate
\begin{equation*}
J_n\le 
\frac{e\sqrt{2\pi}}{4}
\frac{(\frac{n+1}{2e})^{\frac{n+1}{2}}}{2^{\frac{n-1}{2}}}
=\frac{e\sqrt{2\pi}}{4} 2^{-n}
(\frac{n+1}{e})^{\frac{n+1}{2}}
\end{equation*}
where we used $\frac{n-1}{2}\le[\frac{n}{2}]\le \frac{n}{2}$. Finally,
if $n\le n_M$ then
\begin{equation}\label{fluc4.58}
J_n\le 
\frac{e\sqrt{2\pi}}{4} 2^{-n}
(\frac{n_M+1}{e})^{\frac{n+1}{2}}.
\end{equation}
Combining these results we obtain the final estimate
\begin{align} \label{fluc4.59}
&\sqrt{\frac{2}{\pi}}^m\int_{\R^m} d^m x e^{-2\norm{x}^2} P(x)^n
=V_{m-1}\sqrt{\frac{2}{\pi}}^m \frac{e\sqrt{2\pi}}{4}
\sum_{n_0+..+n_l=n} \frac{n!}{(n_0 !)..(n_l)!}[\prod_{k=0}^l a_k^{n_k}]
J_{\sum_{k=0}^l k n_k+m-1}
\nonumber\\
&\le 
V_{m-1}\sqrt{\frac{2}{\pi}}^m \frac{e\sqrt{2\pi}}{4}
\sum_{n_0+..+n_l=n} \frac{n!}{(n_0 !)..(n_l)!}[\prod_{k=0}^l a_k^{n_k}]
2^{-(\sum_{k=0}^l k n_k+m-1)}
(\frac{m+ln}{e})^{\frac{\sum_{k=0}^l k n_k+m-1+1}{2}}
\nonumber\\
&= 
V_{m-1}\sqrt{\frac{2}{\pi}}^m \frac{e\sqrt{2\pi}}{2}
(\frac{m+ln}{4e})^{\frac{m}{2}}
\sum_{n_0+..+n_l=n} \frac{n!}{(n_0 !)..(n_l)!}
[\prod_{k=0}^l (a_k \sqrt{\frac{m+ln}{4e}}^k)^{n_k}]
\nonumber\\
&= 
V_{m-1}\sqrt{\frac{2}{\pi}}^m \frac{e\sqrt{2\pi}}{2}
(\frac{m+ln}{4e})^{\frac{m}{2}}
[\sum_{k=0}^l a_k \sqrt{\frac{1+ln}{4e}}^k]^n
\nonumber\\
&=: K_{m,l} (\frac{m+ln}{4e})^{\frac{m}{2}} 
P(\sqrt{\frac{m+ln}{4e}})
\end{align}
since $\sum_{k=0}^l k n_k\le ln=n_M-m$ for any configuration of the 
$n_k$ subject to the constraint $n_0+..+n_l=n$. 

In our case we have $m=9,l=6$ and thus we can bound the remainder
\eqref{fluc4.55} from above:
\begin{align} \label{fluc4.60}
E_n \le K_{9,6} (3M)^{n+1}\bigg[
(\frac{9+6(n+2)}{4e})^{\frac{9}{2}}&
(P(\sqrt{\frac{9+6(n+2)}{4e}})^{n+2}\\
&\qqquad+2(\frac{9+6(n+1)}{4e})^{\frac{9}{2}} (P(\sqrt{\frac{1+6(n+1)}{4e}})^{n+1}\bigg].\nonumber
\end{align}
For small $n$ the error $E_n$ is the number $s^{n+1}$ times a constant of 
order unity. For large $n$, however, the error becomes comparable 
to the order of accuracy (in powers of $s$) that we are interested in 
itself. The value $n=n_0$ from where onwards it does not make sense 
any longer to compute corrections can be estimated from the condition 
\begin{equation}\label{fluc4.61}
E_{n+1}/E_n\ge 1.
\end{equation}
Due to the complicated structure of (\ref{fluc4.60}) the precise value 
of $n_0$ cannot be computed analytically but its order of magnitude 
can be obtained under the self-consistency assumption that $n_0$ 
is quite large so that the change of 
$P(\sqrt{(9+6(n_0+2))/(4e)})$ as we change $n_0$ by $1$ is much
smaller than its value.  
A tedious but straightforward estimate shows that under this
assumption
\begin{equation}\label{fluc4.66}
n_0=\frac{4e (\frac{\tau_0(M)}{s ||q^{-1}||})^2-9}{6}-3
\end{equation}
where $\tau_0(M)$ is of order unity. 
Thus $n_0$ is a very large number if $||q^{-1}||$ is of order unity and 
$s$ is tiny. Moreover,
\begin{equation}\label{fluc4.67}
\delta P=2(u+1)(1+2\tau+18\tau^2)\delta\tau
= 6(u+1)u\delta\tau/\tau\le 6P\frac{\delta\tau}{\tau}.
\end{equation}
But under the change $\delta n=1$
\begin{equation}\label{fluc4.68}
\delta\tau\approx\frac{d\tau}{d n}\delta n=\frac{\tau}{9(9+2n)}
\end{equation}
whence
\begin{equation}\label{fluc4.69}
(\frac{\delta P}{P})_{n=n_0}\le \frac{2}{3(9+2n_0)}\ll 1
\end{equation}
as desired since $n_0$ is a large number.\\
\\
Let us now finally go back to our desired expectation value
(\ref{fluc4.41}) which we would like to compute up to some order
$n<n_0$ in $s$. Let again 
$y\doteq 1+z_{q^{-1}xs}=1+z$ and $y_{J\sigma j}\doteq 
1+z_{q^{-1}[xs+\delta_{Jj}sT/2]}=1+z_{J\sigma j}$ with 
$z_A=(z'_A)^2+2 z'_A,\;
z'_A=\tr (A)+\frac{1}{2}[(\tr (A))^2-\tr (A^2)]+\det(A)$
for any matrix $A$ and recall our convention $r/4=L/M$. Thus 
(\ref{fluc4.43}) becomes up to order $n$
\begin{align} \label{fluc4.70}
\lambda^r_{J\sigma j}(x+p/T)
&=\frac{2|\det(p)|^{2L/M}}{t}[y^{L/M}-y_{J\sigma j}^{L/M}]\\
&=\frac{2|\det(q)|^{2L/M}t^{6L/M\alpha}}{t}\{
[(y-y_{J\sigma j})\sum_{k=1}^n f^{(k)}_{L/M}(1)
\sum_{l=0}^{k-1}(y-1)^l (y_{J\sigma j}-1)^{k-1-l}]
\nonumber\\
&\qqquad+[f^{(n+1)}_{L/M}(y) (y-1)^{n+1}-f^{(n+1)}_{L/M}(y_{J\sigma j}) 
(y_{J\sigma j}-1)^{n+1}]\}
\nonumber
\end{align}
In order to compute (\ref{fluc4.41}) up to order $n$ with respect to $s$
we have to 
plug the expansions (\ref{fluc4.70}) into formula (\ref{fluc4.41}) and to collect 
all the contributions up to order $s^n$. The integral over 
the remainder is then still smaller as long as $n<n_0$ as shown above.
In the present work we are interested 
only in the leading order 
(classical limit) and next to leading order (first quantum correction)
as well as in an estimate of the error at the next to leading order.

A laborious but straightforward power counting reveals that 
\begin{equation}\label{fluc4.75}
\lambda^r_{J\sigma j}=\frac{sT}{t}(1+sx+(sx)^2+O(sT))
\end{equation}
where the notation just  
means that $\lambda^r_{J\sigma j}$ is a polynomial in $x_{Jj}$ of order two
where the monoms of order $0,1,2$ come with a power of $s$ of the
order indicated {\it or higher}.
We thus see that
if we wish to keep only terms up to order $(sT/t)^N$ and $(sT/t)^N s^2$ 
in $\prod_{k=1}^N \lambda^r_{J_k\sigma_k j_k}(x+p/T)$ it will be 
sufficient to do the following: For the term of order $(sT/t)^N$ keep only
the terms proportional to $x^0$ in each of the factors of the 
form (\ref{fluc4.75}) and for term of order $(sT/t)^N s^2$ keep either 
1. only the terms proportional to $x^2$ in one of the factors of the form 
(\ref{fluc4.75}) and only the terms of order $x^0$ in the others or 2. 
only the terms proportional to $x^1$ in two of the factors of the form 
(\ref{fluc4.75}) and only the terms of order $x^0$ in the others. Clearly 
terms of order $(sT/t)^N s$ do not survive since they are linear in $x$
and integrate to zero against the Gaussian. 

In estimating the error that we make notice that there are two errors, 
one coming from neglecting all higher orders in (\ref{fluc4.75}) and one from 
the remainder in the expansion (\ref{fluc4.70}). As for the first error, 
notice that all Gaussian integrals are of order unity so that
the indicated powers of $t$ correctly display the error (compared 
to $(sT/t)^N s^2$) as of higher order in $s$. As for the second error 
we can use the expansion (\ref{fluc4.70}) up to 
some order $n'>2$ until $s^{n'+1}\ll sT s^2$ in view of the estimate
(\ref{fluc4.60}). The minimal value of $n'$ depends on the value of $\alpha$.
For instance, if $\alpha=1/6$ as indicated by \cite{Sahlmann:2001nv} then 
$s=t^{1/3}$ so that $s^{n'-2}=t^{(n'-2)/3}\ll T=t^{1/2}$ means 
$n'>2+3/2$ so the minimal value would be $n'=4$ in this case. This value 
is well below $n_0\gg 1$ so that the error is indeed of higher order 
in $s$ as compared to $(sT/t)^N s^2$.

With these things said we can now actually compute the first contributing
correction to the classical limit. We will not bother with the higher 
order corrections since we just showed that they can be bounded by 
terms of sub-leading order as compared to $(sT/t)^N s^2$. In particular, 
we will replace the $O(t^\infty)$ corrections by zero in (\ref{fluc4.41}).
We then have  
\begin{align} \label{fluc4.76}
\expec{\,\cdot\,} &=
\sqrt{\frac{2}{\pi}}^9
\int_{\R^9} d^9 x e^{-2\norm{x}^2}
\bigg\{ [\prod_{k=1}^N \lambda^r_{J_k\sigma_k j_k}(x+p/T)_{|x^0}]
\nonumber\\
&\qqquad+ [\sum_{l=1}^N 
\lambda^r_{J_l\sigma_l j_l}(x+p/T)_{|x^2}] 
\prod_{k\not=l} 
\lambda^r_{J_k\sigma_k j_k}(x+p/T)_{|x^0}] 
\nonumber\\
&\qqquad + [\sum_{1\le l< m\le N}^N 
\lambda^r_{J_l\sigma_l j_l}(x+p/T)_{|x^1}] 
\lambda^r_{J_m\sigma_m j_m}(x+p/T)_{|x^1}] 
\prod_{k\not=l,m} 
\lambda^r_{J_k\sigma_k j_k}(x+p/T)_{|x^0}] \bigg\}\\
&\qqquad +O(t^{(N[3r/2-1]\alpha} sT)
\end{align}
where the restrictions mean the ones to the appropriate powers of
$x$ as derived above. It remains to explicitly compute these restrictions
and to do the Gaussian integrals. According to what we have said above 
we write
\begin{align} \label{fluc4.77}
\lambda^r_{J\sigma j}(x+p/T)
&= O(t^{[3r/2-1]\alpha} sT)+
2|\det(q)|^{r/2} t^{[3r/2-1]\alpha}\{
[f^{(1)}_{r/4}(1) (\frac{y-y_{J\sigma j}}{sT})_{|x^0}]
\nonumber\\
&+ [f^{(1)}_{r/4}(1) (\frac{y-y_{J\sigma j}}{sT})_{|x^1}]
+f^{(2)}_{r/4}(1) (\frac{y-y_{J\sigma j}}{sT})_{|x^0}
((y-1)_{|x^1}+ (y_{J\sigma j}-1)_{|x^1})]
\nonumber\\
&+ [f^{(1)}_{r/4}(1) (\frac{y-y_{J\sigma j}}{sT})_{|x^2}]
+f^{(2)}_{r/4}(1) (\frac{y-y_{J\sigma j}}{sT})_{|x^0}
((y-1)_{|x^2}+ (y_{J\sigma j}-1)_{|x^2})
\nonumber\\
&+f^{(3)}_{r/4}(1) (\frac{y-y_{J\sigma j}}{sT})_{|x^0}
(((y-1)_{|x^1})^2+((y_{J\sigma j}-1)_{|x^1})^2+(y-1)_{|x^1}
(y_{J\sigma j}-1)_{|x^1})]\}
\end{align}
And furthermore 
\begin{align} \label{fluc4.78}
y-1 &= 2s q^{-1}_{Mm} x_{Mm}+s^2(2 q^{-1}_{Mm} q^{-1}_{Nn}
-q^{-1}_{Mn} q^{-1}_{Nm}) x_{Mm} x_{Nn} +O(s^3)
\nonumber\\
&=: s C^{Mm} x_{Mm} + s^2 C^{Mm,Nn} x_{Mm} x_{Nn} +O(s^3)
\nonumber\\
y_{J\sigma j}-1 
&= 2s\tr (q^{-1}x)+s^2[2\tr (q^{-1}x)^2 
-\tr (q^{-1}x q^{-1} x)]+O(sT)
\nonumber\\
&=: s C^{Mm} x_{Mm} + s^2 C^{Mm,Nn} x_{Mm} x_{Nn} +O(sT)
\nonumber\\
\frac{y_{J\sigma j}-y}{sT}&=
q^{-1}_{Jj}+s(2 q^{-1}_{Jj} q^{-1}_{Mm}
-q^{-1}_{Jm} q^{-1}_{Mj}) x_{Mm}
+\frac{s^2}{2}[\det(q^{-1}) \epsilon_{jmn}\epsilon_{JMN}
+q^{-1}_{Jj}(q^{-1}_{Mm} q^{-1}_{Nn}
-q^{-1}_{Mn} q^{-1}_{Nm})
\nonumber\\
& \qqquad+2 q^{-1}_{Mm}
(q^{-1}_{Jj} q^{-1}_{Nn}-q^{-1}_{Jn} q^{-1}_{Nj})] x_{Mm} x_{Nn}
\nonumber\\
&=: 
C_{J\sigma j}+s C_{J\sigma j}^{Mm} x_{Mm}
+s^2 C_{J\sigma j}^{Mm,Nn} x_{Mm} x_{Nn}.
\end{align} 
We can therefore simplify (\ref{fluc4.77}) to 
\begin{align} \label{fluc4.79} 
\lambda^r_{J\sigma j}(x+p/T)
&= O(t^{[3r/2-1]\alpha} sT)+
2|\det(q)|^{r/2} t^{[3r/2-1]\alpha}\{
[f^{(1)}_{r/4}(1) C_{J\sigma j}]
\nonumber\\
&+ s[f^{(1)}_{r/4}(1) C_{J\sigma j}^{Mm}
+2f^{(2)}_{r/4}(1) C_{J\sigma j} C^{Mm}] x_{Mm}
\nonumber\\
&+ s^2 [f^{(1)}_{r/4}(1) C_{J\sigma j}^{Mm,Nn}
+2 f^{(2)}_{r/4}(1) C_{J\sigma j} C^{Mm,Nn}
+3 f^{(3)}_{r/4}(1) C_{J\sigma j} C^{Mm} C^{Nn}] x_{Mm} x_{Nn} \}
\nonumber\\
&=: O(t^{[3r/2-1]\alpha} sT)+
2|\det(q)|^{r/2} t^{[3r/2-1]\alpha}\{
D_{J\sigma j}(r)+ s D_{J\sigma j}^{Mm}(r) x_{Mm} 
\nonumber\\
&
+s^2 D_{J\sigma j}^{Mm,Nn}(r) x_{Mm} x_{Nn}\}.
\end{align}
Putting everything together now yields the following  
theorem.
\begin{Theorem} \label{th4.2}
For the classical limit and lowest order quantum corrections of 
expectation values of monomials of the operators $\widehat{q}_{J\sigma j}(r)$
for topologically cubic graphs we have 
\begin{align} \label{fluc4.81}
&\frac{\scpr{\psi^t_{\{g_{J\sigma j}\}}}
{\prod_{k=1}^N\widehat{q}_{J_k\sigma_k j_k}(r_k)
\psi^t_{\{g_{J\sigma j}\}}}}{||\psi^t_{\{g_{J\sigma j}\}}||^2} 
=(2|\det(q)|^{r/2} t^{[3r/2-1]\alpha})^N
\times\nonumber\\
&\qqquad\times  \{ [\prod_{k=1}^N D_{J_k\sigma_k j_k}(r)]
+ \frac{s^2}{4}\sum_{M,m}[\sum_{l=1}^N 
D_{J_l\sigma_l j_l}^{Mm,Mm}(r) \prod_{k\not=l} D_{J_k\sigma_k j_k}(r)) 
\nonumber\\
&\qqquad+ \sum_{1\le i< l\le N} D_{J_i\sigma_i j_i}^{Mm}(r) 
D_{J_l\sigma_l j_l}^{Mm}(r)
\prod_{k\not=l,i}  D_{J_k\sigma_k j_k}(r)] \}
\end{align}
where the constants are given by 
\begin{align*} 
C^{Mm}&=2 q^{-1}_{Mm}, 
\nonumber\\
C^{Mm,Nn}&=2 q^{-1}_{Mm} q^{-1}_{Nn}-q^{-1}_{Mn} q^{-1}_{Nm},
\nonumber\\
C_{J\sigma j}&= q^{-1}_{Jj},
\nonumber\\
C_{J\sigma j}^{Mm}&=
(2 q^{-1}_{Jj} q^{-1}_{Mm}-q^{-1}_{Jm} q^{-1}_{Mj}), 
\nonumber\\
C_{J\sigma j}^{Mm,Nn}&=
\frac{1}{2}[\det(q^{-1}) \epsilon_{jmn}\epsilon_{JMN}
+q^{-1}_{Jj}(q^{-1}_{Mm} q^{-1}_{Nn}-q^{-1}_{Mn} q^{-1}_{Nm})
+2 q^{-1}_{Mm}(q^{-1}_{Jj} q^{-1}_{Nn}-q^{-1}_{Jn} q^{-1}_{Nj})], 
\nonumber\\
D_{J\sigma j}(r) &= f^{(1)}_{r/4}(1) C_{J\sigma j},
\nonumber\\
D_{J\sigma j}^{Mm}(r) &=
f^{(1)}_{r/4}(1) C_{J\sigma j}^{Mm}+2f^{(2)}_{r/4}(1) C_{J\sigma j} C^{Mm},
\nonumber\\
D_{J\sigma j}^{Mm,Nn}(r) &=
f^{(1)}_{r/4}(1) C_{J\sigma j}^{Mm,Nn}
+2 f^{(2)}_{r/4}(1) C_{J\sigma j} C^{Mm,Nn}
+3 f^{(3)}_{r/4}(1) C_{J\sigma j} C^{Mm} C^{Nn},
\end{align*}
and the $f^{(k)}_{r/4}(1)=(r/4,k)$
are simply the binomial coefficients.\newline 
The first correction is small as long as $\alpha<1/2$. The error as 
compared to the first quantum correction of order  
$O(t^{(N[3r/2-1]\alpha} s^2)$ is a constant of order unity times
$t^{(N[3r/2-1]\alpha} sT$ and thus small as long as $0<\alpha$.
\end{Theorem}
So far we did not look at the classical limit and the 
first quantum corrections of (powers 
of) the volume operator itself but it is clear that it can be analyzed 
by similar methods, in fact, the analysis is even much simpler because 
we just need to expand $\lambda^r(x+p/T)$ in powers of $s$ without dividing
by $t$ and thus will have to do an expansion in terms of $y-1$ 
of one order less than for $\lambda^r_{J \sigma j}(x+p/T)$. Clearly the 
classical order will be 
that of $|\det(p)|^{r/2}=|\det(q)|^{r/2} t^{3r\alpha/2}=O(t^{3r\alpha/2})$
and the first quantum correction will be of order 
$O(t^{3r\alpha/2} s^2)$. We thus have, in expanding up to second order in 
$y-1$, where $y=\det(1+s q^{-1}x)^2$ as before
\begin{align} \label{fluc4.82}
\lambda^r(x+p/T)&= |\det(q)|^{r/2} t^{3r\alpha/2}
\bigg\{ 1+s f^{(1)}_{r/4}(1) C^{Mm} x_{Mm}\\
&\qquad+s^2 [f^{(2)}_{r/4}(1) C^{Mm,Nn}+f^{(1)}_{r/4}(1) C^{Mm} C^{Nn}]
x_{Mm} x_{Nn} \bigg\}+O(t^{3r\alpha/2} s^3).
\end{align}
Thus we obtain an analogue of theorem \ref{th4.2} above:
\begin{Theorem} \label{th4.3}
For the classical limit and lowest order quantum corrections of 
expectation values of powers of the volume operators 
$\widehat{\sqcup}_v^r$ for topologically cubic graphs we have 
\begin{equation}\label{fluc4.84}
\frac{\scpr{\psi^t_{\{g_{J\sigma j}\}}}
{\widehat{\sqcup}_v^r
\psi^t_{\{g_{J\sigma j}\}}}}{||\psi^t_{\{g_{J\sigma j}\}}||^2} 
=|det(q)|^{r/2} t^{3r\alpha/2}\{1+
\frac{s^2}{4}\sum_{M,m} [f^{(2)}_{r/4}(1) C^{Mm,Nn}+f^{(1)}_{r/4}(1) C^{Mm} 
C^{Nn}\}.
\end{equation}
The first correction is small as long 
as $\alpha<1/2$. The error as compared to the 
first quantum correction of order  
$O(t^{(N[3r/2-1]\alpha} s^2)$ is a constant of order unity times
$t^{(N[3r/2-1]\alpha} s^3$ and thus small as long as $0<\alpha$.
\end{Theorem}
%------------------------------------------------------------------------
\subsection{Application to the Hamiltonians}
%------------------------------------------------------------------------
\label{sea.3}
So far our considerations were completely general and model independent
and we see that our coherent states indeed predict small quantum 
predictions as long as $0<\alpha<1/2$ and $\ell_P/L\ll 1$ with 
controllable error. 
However, now
we will specialize to the case of scalar, electromagnetic and fermionic
fields coupled to gravity and compute the expectation values of the
relevant gravitational operators.

We recall from our companion paper \cite{ST01} that on a cubic graph,
the effective Hamiltonians for the scalar and the electromagnetic
field are  
\begin{align}
H^{\text{eff}}_{\text{KG}}&=\frac{1}{2Q_{\text{KG}}}\sum_v
  \left(\expec{\widehat{F}_{\text{kin}}(v)}\widehat{\pi}_v^2
  -\sum_{I\sigma I'\sigma'}
  \expec{\widehat{F}_{\text{der}}^{I\sigma I'\sigma'}(v)}
  \left(\partial_{e^\sigma_I}^+\ln U_v\right)
  \Big(\partial_{e^{\sigma'}_{I'}}^+\ln U_v\Big)
  -K^2\expec{\widehat{V}_v}(\ln U_v)^2\right),\\
H^{\text{eff}}_{\text{EM}}&=\frac{1}{2Q_{\text{EM}}}
  \sum_v\sum_{I\sigma I'\sigma'} 
  \bigg(\expec{\widehat{F}^{I\sigma I'\sigma'}_{\text{el}}(v)}
  Y_{I\sigma}Y_{I'\sigma'}\nonumber\\\label{ham1}
  &\qqquad\qqquad -\expec{\widehat{F}_{\text{mag}}^{I\sigma I'\sigma'}(v)}
  [\sum_{\sigma_1,\sigma_2} 
  \ln(H_{\beta^I_{\sigma;\sigma_1,\sigma_2}(v)})]\;
  [\sum_{\sigma'_1,\sigma'_2} 
\ln(H_{\beta^J_{\sigma';{\sigma'}_1{\sigma'}_2}(v)})]\bigg),
\end{align}
where $\expec{\,\cdot\,}$ denotes the expectation value in a
semiclassical state for the gravitational sector, and the
geometric operators are given by 
\begin{align*}
\widehat{F}_{\text{kin}}(v)&=
\frac{1}{\ell_P^{12}}
\bigg[\frac{1}{8}\sum_{\sigma_1,\sigma_2,\sigma_3} 
\frac{\sigma_1\sigma_2\sigma_3}{3!} \epsilon_{ijk}\epsilon^{IJK}
\hat{Q}^i_{I,\sigma_1}(v,\frac{1}{2})\hat{Q}^j_{J,\sigma_2}(v,\frac{1}{2})
\hat{Q}^k_{K,\sigma_3}(v,\frac{1}{2})\bigg]^\dagger
\bigg[\ldots\bigg],\\
\widehat{F}_{\text{der}}^{I\sigma I'\sigma'}&=
\frac{1}{4}\frac{1}{\ell_P^{8}}\sum_j
\bigg[\frac{\epsilon^{IJK}}{8} \epsilon_{jkl}
\sum_{\sigma_2,\sigma_3}
\hat{Q}^k_{J\sigma_2}(v,\frac{3}{4})
\hat{Q}^l_{K\sigma_3}(v,\frac{3}{4})\bigg]
\bigg[\ldots\, _j\, \ldots \bigg],\\
\widehat{F}^{I\sigma
  I'\sigma'}_{\text{el}}(v)&=\frac{1}{4}\frac{1}{\ell_P^4}
  \sum_j\hat{Q}^j_{I\sigma}(v,\frac{1}{2})
  \hat{Q}^j_{I'\sigma'}(v,\frac{1}{2}),\\
\widehat{F}^{I\sigma I'\sigma'}_{\text{mag}}(v)&=
  \frac{1}{64}\frac{1}{\ell_P^4}
  \sum_j\hat{Q}^j_{I\sigma}(v,\frac{1}{2})
  \hat{Q}^j_{I'\sigma'}(v,\frac{1}{2}).
\end{align*}
The matter fields are represented as 
\begin{equation*}
  \widehat{\phi}_v=\frac{\ln U_v}{i},\qquad \widehat{\pi}_v=i\hbar
  Q_{KG} Y_v,\qquad
  \widehat{E}_{I\sigma}(v)=i\hbar Q_{\text{EM}} Y_{I\sigma},\qquad
  \widehat{B}_{I\sigma;\sigma_1,\sigma_2}(v)
  =\frac{\ln(H_{\beta^I_{\sigma;\sigma_1,\sigma_2}(v)})}{i}, 
\end{equation*}
where the $Y$ are invariant derivatives on $\uone$, $U_v$ is a $\uone$ 
point-holonomy and $H_{\beta}$ a $\uone$ holonomy around a
\textit{minimal loop}.

The Hamiltonian for a fermionic field is given by 
\ba 
&& \hat{H}_{D,\gamma}=-\frac{m_P}{2\ell_P^3}\sum_{v,v'\in V(\gamma)}
[\hat{\theta}_B(v')\hat{\theta}^\dagger_A(v)
-\hat{\theta}'_B(v')\hat{\theta}^{\prime\dagger}_A(v)]
\times \nonumber\\
&\times&
\{ \frac{1}{8} \epsilon_{ijk} \epsilon^{IJK}
\sum_{\sigma_1,\sigma_2,\sigma_3}
\hat{Q}^i_{I\sigma_1}(v,\frac{1}{2})\hat{Q}^j_{J\sigma_2}(v,\frac{1}{2})
[\tau^k 
(h^{\sigma_3}_K(v)\delta_{v',f(e^{\sigma_3}_K(v))}-\delta_{v',v})]_{AB}\}
\nonumber\\
&&-
\{ \frac{1}{8} \epsilon_{ijk} \epsilon^{IJK}
\sum_{\sigma_1',\sigma_2',\sigma_3'}
[([h^{\sigma_3'}_K(v')]^{-1}\delta_{v,f(e^{\sigma_3'}_K(v'))}-
\delta_{v,v'})\tau^k]_{AB}
\hat{Q}^i_{I\sigma_1'}(v',\frac{1}{2})\hat{Q}^j_{J\sigma_2'}(v',\frac{1}{2})\}
\}
\nonumber\\\label{dirac}
&&
-i\hbar K_0\sum_{v,v'\in V(\gamma)}
\delta_{AB}\delta_{v,v'} [\hat{\theta}'_B(v')\hat{\theta}^\dagger_A(v)
-\hat{\theta}_B(v')\hat{\theta}^{\prime\dagger}_A(v)].
\ea
We strongly recommend to take a look at \cite{ST01} where the above
Hamiltonians are derived and all the ingredients are defined and 
discussed in detail!\\ 

We now proceed to compute the expectation values of the geometric
operators in a coherent state. To this end, we 
will use the formulae given in theorems \ref{th4.2} and \ref{th4.3} 
with the appropriate values of $r,N,J_k,\sigma_k,j_k$ inserted, and 
and perform the additional computations necessary. 
%-----------------------------------------------------------------
\subsubsection{The Kinetic Term}
%-----------------------------------------------------------------
For $F_{\text{kin}}$ we have to use theorem \ref{th4.2} with
$N=6$. Employing the relation \eqref{trans} between $\widehat{q}$ and
$\widehat{Q}$ we find
\begin{align} \label{fluc4.85}
\expec{\widehat{F}_{\text{kin}}}
&=\frac{1}{\ell_P^{12}}\frac{1}{(3!)^2}\left(\frac{2ta^{3r}}{r}\right)^6
(2|\det(q)|^{1/4} t^{[3/4-1]\alpha})^6\;
\epsilon^{J_1 J_2 J_3}\epsilon_{j_1 j_2 j_3}
\epsilon^{J_4 J_5 J_6}\epsilon_{j_4 j_5 j_6} 
\times \nonumber\\
&\times
  \{ [\prod_{k=1}^6 D_{J_k\sigma_k j_k}(1/2)]
+ \frac{s^2}{4}\sum_{M,m}[\sum_{l=1}^6
D_{J_l\sigma_l j_l}^{Mm,Mm}(1/2) \prod_{k\not=l} D_{J_k\sigma_k j_k}(1/2)) 
\nonumber\\
&+ \sum_{1\le i< l\le 6} 
D_{J_i\sigma_i j_i}^{Mm}(1/2) D_{J_l\sigma_l j_l}^{Mm}(1/2)
\prod_{k\not=l,i}  D_{J_k\sigma_k j_k}(1/2)] \}.
\end{align}
For $r=1/2$ we have 
\begin{equation}\label{fluc4.87}
a_1\doteq f^{(1)}_{1/8}(1)=\frac{1}{8},\;
a_2\doteq f^{(2)}_{1/8}(1)=-\frac{1}{8}\frac{7}{16}=-\frac{7}{128},\;
a_3\doteq f^{(3)}_{1/8}(1)=\frac{7}{128}\frac{15}{24}=\frac{35}{1024},
\end{equation}
and consequently
\begin{align} \label{fluc4.89}
\sum_{M,m} D_{J\sigma j}^{Mm,Mm}(1/2)&= [a_1+3 a_3] q^{-1}_{Jj} \tr (q^{-2})
-\frac{a_1}{2} q^{-3}_{Jj}\\
\sum_{Mm} D_{J_1\sigma_1 j_1}^{Mm}(1/2) D_{J_2\sigma_2 j_2}^{Mm}(1/2)
&= 4[a_1+a_2]^2 q^{-1}_{J_1 j_1} q^{-1}_{J_2 j_2}\tr (q^{-2})\nonumber\\
\label{fluc4.89b}
&\qquad-2 a_1[a_1+a_2] (q^{-1}_{J_1 j_1} q^{-3}_{J_2 j_2} 
+q^{-1}_{J_2 j_2} q^{-3}_{J_1 j_1})
+a_1^2 q^{-2}_{J_1 J_2} q^{-2}_{j_1 j_2}. 
\end{align}
Now we have to deal with the contractions in \eqref{fluc4.85}. It is easy
to see that  
\begin{align} \label{fluc4.91}
&
\epsilon^{J_1 J_2 J_3}\epsilon_{j_1 j_2 j_3}
\epsilon^{J_4 J_5 J_6}\epsilon_{j_4 j_5 j_6}
\prod_{k=1}^6 q^{-1}_{J_k j_k}]=\frac{36}{\det(q)^2},
\nonumber\\
&
\epsilon^{J_1 J_2 J_3}\epsilon_{j_1 j_2 j_3}
\epsilon^{J_4 J_5 J_6}\epsilon_{j_4 j_5 j_6}
q^{-3}_{J_l j_l}) \prod_{k\not=l} q^{-1}_{J_k j_k} 
=\frac{12\tr (q^{-2})}{\det(q)^2},
\nonumber\\
&
\epsilon^{J_1 J_2 J_3}\epsilon_{j_1 j_2 j_3}
\epsilon^{J_4 J_5 J_6}\epsilon_{j_4 j_5 j_6}
q^{-2}_{J_i J_l} q^{-2}_{j_i j_l})
\prod_{k\not=l,i}  q^{-1}_{J_k j_k} =0
\mbox{ if } l,i\in\{1,2,3\} \mbox{ or }
l,i\in\{4,5,6\}, 
\nonumber\\
&
\epsilon^{J_1 J_2 J_3}\epsilon_{j_1 j_2 j_3}
\epsilon^{J_4 J_5 J_6}\epsilon_{j_4 j_5 j_6}
q^{-2}_{J_i J_l} q^{-2}_{j_i j_l})
\prod_{k\not=l,i}  q^{-1}_{J_k j_k} 
=\frac{4\tr (q^{-2})}{\det(q)^2} \mbox{ otherwise}.  
\end{align}
Using the above together with \eqref{fluc4.89} in \eqref{fluc4.85} yields
\begin{align} \label{fluc4.92}
 \expec{\widehat{F}_{\text{kin}}(v)}&=
\frac{a^9}{\ell_P^{12}}\frac{4^6}{(3!)^2}t^6
\frac{(2|\det(q)|^{1/4} t^{[3/4-1]\alpha})^6}{\det(q)^2}\;
\{ 36 [a_1^6] 
+ \frac{s^2}{4}\tr (q^{-2})[ 6 a_1^5 (36 [a_1+3 a_3] -12 \frac{a_1}{2}) 
\nonumber\\
&\qquad+a_1^4 (15(4[a_1+a_2]^2 (36) -2 a_1[a_1+a_2] (12+12))+9 a_1^2]\}
\nonumber\\
&=\frac{a^9t^6}{\ell_P^{12}}\frac{1}{\sqrt{\det p}}
\{1
+\frac{t}{4}\tr (p^{-2})[ (5+24 a_3) 
+15(4[1+8a_2]^2 -\frac{4}{3}[1+8 a_2])+\frac{1}{4}]\} \nonumber\\
&=\frac{a^9t^6}{\ell_P^{12}}\frac{1}{\sqrt{\det p}}
\{1+t\frac{1707}{512}\tr (p^{-2})\}. 
\end{align}
Let us finally transform back to the dimensionfull quantity
$P=a^2p$. We get 
\begin{equation*}
\expec{\widehat{F}_{\text{kin}}}(v)=
\frac{1}{\sqrt{\det P(v)}}\left[1+ 
\frac{\ell_P^4}{t}\frac{1707}{512}\tr P^{-2}(v)\right].
\end{equation*}
%-----------------------------------------------------------------------
\subsubsection{The Derivative Term}
%-----------------------------------------------------------------------
The derivative term 
$F_{\text{der}}$ requires $N=4$. From theorem \ref{th4.2} we 
find
\begin{align} \label{fluc4.93}
\expec{F_{\text{der}}^{J\sigma J'\sigma'}}&=\frac{\sigma\sigma'}{4}
\frac{1}{4}\left(\frac{8}{3}\right)^4
\frac{t^4a^9}{\ell_P^8}
(2|\det(q)|^{3/8} t^{[9/8-1]\alpha})^4\;\sum_j 
\epsilon^{J J_1 J_1}\epsilon_{j j_1 j_2}
\epsilon^{J' J_3 J_4}\epsilon_{j j_3 j_4}
\frac{1}{16}\sum_{\sigma_1\ldots\sigma_4}\\ 
&\times
\{ [\prod_{k=1}^4 D_{J_k\sigma_k j_k}(3/4)]
+ \frac{s^2}{4}\sum_{M,m}[\sum_{l=1}^4
D_{J_l\sigma_l j_l}^{Mm,Mm}(3/4) \prod_{k\not=l} D_{J_k\sigma_k j_k}(3/4)) 
\nonumber\\
&+ \sum_{1\le i< l\le 4} 
D_{J_i\sigma_i j_i}^{Mm}(3/4) D_{J_l\sigma_l j_l}^{Mm}(3/4)
\prod_{k\not=l,i}  D_{J_k\sigma_k j_k}(3/4)] \}
\end{align}
For $r=3/4$ we have
\begin{equation}\label{fluc4.94}
a_1\doteq f^{(1)}_{3/16}(1)=\frac{3}{16},\;
a_2\doteq f^{(2)}_{3/16}(1)=-\frac{3}{16}\frac{29}{32}=-\frac{3\cdot 29}{2^9},\;
a_3\doteq f^{(3)}_{3/16}(1)=\frac{3\cdot 29}{2^9}\frac{45}{48}=
\frac{3^2\cdot 5\cdot 29}{2^{13}}.
\end{equation}
Furthermore the reader may verify that 
\begin{align} \label{fluc4.96}
& \sum_j 
\epsilon^{J J_1 J_2 }\epsilon_{j j_1 j_2}
\epsilon^{J' J_3 J_4}\epsilon_{j j_3 j_4}
\prod_{k=1}^4 q^{-1}_{J_k j_k}]=\frac{4 q^2_{JJ'}}{\det(q)^2},
\nonumber\\
&\sum_j 
\epsilon^{J J_1 J_2 }\epsilon_{j j_1 j_2}
\epsilon^{J' J_3 J_4}\epsilon_{j j_3 j_4}
q^{-3}_{J_l j_l}) \prod_{k\not=l} q^{-1}_{J_k j_k} 
=\frac{2[q^2_{JJ'}\tr (q^{-2})-\delta_{JJ'}]}{\det(q)^2},
\nonumber\\
&\sum_j 
\epsilon^{J J_1 J_2 }\epsilon_{j j_1 j_2}
\epsilon^{J' J_3 J_4}\epsilon_{j j_3 j_4}
q^{-2}_{J_i J_l} q^{-2}_{j_i j_l}
\prod_{k\not=l,i}  q^{-1}_{J_k j_k} =0
\mbox{ if } l,i\in\{1,2\} \mbox{ or }
l,i\in\{3,4\}, 
\nonumber\\
&\sum_j 
\epsilon^{J J_1 J_2 }\epsilon_{j j_1 j_2}
\epsilon^{J' J_3 J_4}\epsilon_{j j_3 j_4}
q^{-2}_{J_i J_l} q^{-2}_{j_i j_l}
\prod_{k\not=l,i}  q^{-1}_{J_k j_k} 
=\frac{q^2_{JJ'}\tr (q^{-2})+\delta_{JJ'}}{\det(q)^2} 
\mbox{ otherwise}. 
\end{align}
Thus we can finish with a tedious but straightforward computation:
\begin{align} \label{fluc4.97}
\expec{\widehat{F}_{\text{der}}^{J\sigma J'\sigma'}}&=\frac{\sigma\sigma'}{4}
\frac{1}{4}\left(\frac{8}{3}\right)^4
\frac{t^4a^9}{\ell_P^8}
(2|\det(q)|^{3/8} t^{[9/8-1]\alpha})^4\times\\
&\qquad\times\bigg\{ [a_1^4 \frac{4 q^2_{JJ'}}{\det(q)^2}]
+ \frac{s^2}{4}[ 4 a_1^3 
([a_1+3 a_3] \frac{4 q^2_{JJ'}}{\det(q)^2} \tr (q^{-2})
-\frac{a_1}{2}\frac{2[q^2_{JJ'}\tr (q^{-2})-\delta_{JJ'}]}{\det(q)^2})
\nonumber\\
&\qqquad+ a_1^2 
(4[a_1+a_2]^2 6 \frac{4 q^2_{JJ'}}{\det(q)^2} \tr (q^{-2})
-2 a_1[a_1+a_2] 12 
\frac{2[q^2_{JJ'}\tr (q^{-2})-\delta_{JJ'}]}{\det(q)^2}
\nonumber\\
&\qqquad+4 a_1^2 
\frac{q^2_{JJ'}\tr (q^{-2})+\delta_{JJ'}}{\det(q)^2})]
\bigg\}
\nonumber\\
&=\frac{\sigma\sigma'}{4}
\frac{t^4a^9}{\ell_P^8}
\frac{1}{\sqrt{|\det(p)|}}\;
\times\nonumber\\
&\qquad\times  \Big\{p^2_{JJ'}
+ \frac{t}{4}[ 
p^2_{JJ'}\tr (p^{-2})
[4(1+16 a_3)-\frac{1}{4}+\frac{8}{3} (3+16 a_2)^2-4 (3+16 a_2)+1]
\nonumber\\
&\qqquad+\delta_{JJ'}[\frac{1}{4}+4(3+16 a_2)+1]] \Big\}
\nonumber\\
&=\frac{\sigma\sigma'}{4}\frac{t^4a^9}{\ell_P^8}
\frac{1}{\sqrt{|\det(p)|}}\;
\Big\{p^2_{JJ'}
+t\Big[\frac{1173}{128} p^2_{JJ'}\tr (p^{-2})
+\frac{19}{32}\delta_{JJ'}\Big] \Big\}.
\end{align}
Again as a last step we transform to the dimensionfull quantity
$P=a^2p$:
\begin{equation*}
\expec{\widehat{F}^{J\sigma J'\sigma'}_{\text{der}}}=
\frac{\sigma\sigma'}{4}
\frac{1}{\sqrt{|\det(P)|}}\;
\Big\{P^2_{JJ'}
+\frac{\ell_P^4}{t}\Big[\frac{1173}{128} P^2_{JJ'}\tr (P^{-2})
+\frac{19}{32}\delta_{JJ'}\Big] \Big\}.
\end{equation*}
%----------------------------------------------------------------
\subsubsection{The Mass Term:}
%----------------------------------------------------------------
We now consider the mass term. Its basic
building block is the volume operator itself, so we can apply 
theorem \ref{th4.3} with $r=1$. In the by now familiar way we find
\begin{align} \label{fluc4.98}
\expec{\widehat{V}_v}
&=a^3|det(q)|^{1/2} t^{3\alpha/2}\{1+
\frac{s^2}{4}\sum_{M,m} [f^{(2)}_{1/4}(1) C^{Mm,Nn}+f^{(1)}_{1/4}(1) C^{Mm} 
C^{Nn}\}
\nonumber\\
&=a^3|det(q)|^{1/2} t^{3\alpha/2}\{1+
\frac{s^2}{4}\tr (q^{-2}) [\frac{1}{4}-4\frac{3}{32}] \}
\nonumber\\
&=a^3|det(p)|^{1/2}\{1-\frac{t}{32}\tr (p^{-2})\}\nonumber\\
&=\sqrt{\det P(v)}\left[1+\frac{\ell_P^7}{\sqrt{t}}\frac{1}{32}\tr
  P^{-2}(v)\right].
\end{align}
%-------------------------------------------------------
\subsubsection{The Maxwell Hamiltonian:}
%-------------------------------------------------------
The operators $\widehat{F}_{\text{el}}$ and $\widehat{F}_{\text{mag}}$ 
differ by their c-number
coefficients, but the gravitational operator at the heart of both is
the same, corresponding to $N=2$ and $r=1/2$. In both cases we
have to compute 
$\expec{\widehat{q}_{J_1j}(1/2)\widehat{q}_{J_2j}(1/2)}$.\newline
Let us use the definitions of $a_1,a_2,a_3$ given in \eqref{fluc4.87} and
equations \eqref{fluc4.89}, \eqref{fluc4.89b}. We find
\begin{align} \label{b4.25}
&<\widehat{q}_{J_1 j}(1/2)\widehat{q}_{J_2j}(1/2)>
= \delta_{j_1 j_2} (2|\det(q)|^{1/4} t^{[3/4-1]\alpha})^2\times
\nonumber\\
&\qqquad \times\bigg\{ [\prod_{k=1}^2 D_{J_k\sigma_k j_k}(1/2)] 
+ \frac{s^2}{4}\sum_{M,m}\Big[\sum_{l=1}^2 
D_{J_l\sigma_l j_l}^{Mm,Mm}(1/2) \prod_{k\not=l} D_{J_k\sigma_k
  j_k}(1/2)\nonumber\\ 
&\qqquad+\sum_{1\le i< l\le 2} D_{J_i\sigma_i j_i}^{Mm}(1/2) 
D_{J_l\sigma_l j_l}^{Mm}(1/2)
\prod_{k\not=l,i}  D_{J_k\sigma_k j_k}(1/2)\Big] \bigg\}
\nonumber\\
&= (2 a_1 |\det(q)|^{1/4} t^{[3/4-1]\alpha})^2
\Big\{q^{-2}_{J_1 J_2} 
+ \frac{s^2}{4}[2([1+3 \frac{a_3}{a_1}] q^{-2}_{J_1 J_2} \tr (q^{-2})
-\frac{1}{2} q^{-4}_{J_1 J_2})
\nonumber\\
&\qqquad+
4[1+\frac{a_2}{a_1}]^2 q^{-2}_{J_1 J_2}\tr (q^{-2})
-4 [1+\frac{a_2}{a_1}] q^{-4}_{J_1 J_2} 
+q^{-2}_{J_1 J_2} \tr (q^{-2}) 
] \Big\}
\nonumber\\
&= (|\det(q)|^{1/4} t^{[3/4-1]\alpha}/4)^2
\{q^{-2}_{J_1 J_2} 
+ \frac{s^2}{4}[
q^{-2}_{J_1 J_2} \tr (q^{-2})
(7+3 \frac{35}{2^7}-\frac{7}{2}+\frac{3^2\cdot 5^2}{2^6})
-q^{-4}_{J_1 J_2}(5-\frac{7}{4})]
\}
\nonumber\\
&= \frac{\sqrt{|\det(p)|}}{16} 
\{p^{-2}_{J_1 J_2} 
+t[\frac{763}{512} q^{-2}_{J_1 J_2} \tr (p^{-2})
-\frac{13}{16} p^{-4}_{J_1 J_2}]
\}.
\end{align}
We can now employ this result to give the explicit expressions for 
$\expec{\widehat{F}_{\text{el}}}$ and
$\expec{\widehat{F}_{\text{el}}}$. 
Upon using the above expectation value, we find that 
\begin{align*}
\expec{\widehat{F}^{I\sigma I'\sigma'}_{\text{el}}}&=
\frac{\sigma\sigma'}{4}\big[\sqrt{\det P(v)}P^{-2}_{II'}
+\frac{\ell_P^4}{t}\Big(\frac{763}{512}P^{-2}_{II'}\tr P^{-2}
-\frac{13}{16}P^{-4}_{II'}\Big)\Big],\\ 
\expec{\widehat{F}^{I\sigma I'\sigma'}_{\text{mag}}}&=
\frac{1}{64}\bigg[\sqrt{\det P(v)}P^{-2}_{II'}
+\frac{\ell_P^4}{t}\Big(\frac{763}{512}P^{-2}_{II'}\tr P^{-2}
 -\frac{13}{16}P^{-4}_{II'}\Big)\bigg]. 
\end{align*}
%---------------------------------------------------------------------
\subsubsection{The Fermionic Hamiltonian}
%---------------------------------------------------------------------
Due to explicit dependence of (\ref{dirac}) on $h'_e$ the 
expectation values computed so far are not quite sufficient in order
to compute the full expectation value of the fermionic Hamiltonian. 
Fortunately,
the Abelian nature of $U(1)^3$ allows for a simple transcription
of theorem \ref{th4.1} to this slightly more complicated situation.
Notice that at this point coherent states are, for the first time, 
essential, because weave states, being spin-network states would result
in zero expectation values.
\begin{Theorem} \label{th4.2a}
For the classical limit and lowest order quantum corrections of 
expectation values of monomials of the operators $\hat{q}_{J\sigma j}(r)$
times a holonomy operator 
for topologically cubic graphs we have 
\ba \label{4.17}
&& \frac{<\psi^t_{\{g_{J\sigma j}\}},
\hat{h}_{J_0\sigma_0 j_0}^\mu\prod_{k=1}^N\hat{q}_{J_k\sigma_k j_k}(r_k)
\psi^t_{\{g_{J\sigma j}\}}>}{||\psi^t_{\{g_{J\sigma j}\}}||^2} 
=e^{-t/4} h_{J_0\sigma_0 j_0}^\mu
(2|\det(q)|^{r/2} t^{[3r/2-1]\alpha})^N
\times\nonumber\\
&\times & \{ [\prod_{k=1}^N D_{J_k\sigma_k j_k}(r)]
+ \frac{s^2}{4}\sum_{M,m}[\sum_{l=1}^N 
D_{J_l\sigma_l j_l}^{Mm,Mm}(r) \prod_{k\not=l} D_{J_k\sigma_k j_k}(r)) 
\nonumber\\
&+& \sum_{1\le i< l\le N} D_{J_i\sigma_i j_i}^{Mm}(r) 
D_{J_l\sigma_l j_l}^{Mm}(r)
\prod_{k\not=l,i}  D_{J_k\sigma_k j_k}(r)] 
\}_{p\to p+\mu\sigma_0\delta_{J_0 j_0}t/4}
\nonumber\\
&& \frac{<\psi^t_{\{g_{J\sigma j}\}},
\prod_{k=1}^N\hat{q}_{J_k\sigma_k j_k}(r_k)\hat{h}_{J_0\sigma_0 j_0}^\mu
\psi^t_{\{g_{J\sigma j}\}}>}{||\psi^t_{\{g_{J\sigma j}\}}||^2} 
=e^{-t/4} h_{J_0\sigma_0 j_0}^\mu
(2|\det(q)|^{r/2} t^{[3r/2-1]\alpha})^N
\times\nonumber\\
&\times & \{ [\prod_{k=1}^N D_{J_k\sigma_k j_k}(r)]
+ \frac{s^2}{4}\sum_{M,m}[\sum_{l=1}^N 
D_{J_l\sigma_l j_l}^{Mm,Mm}(r) \prod_{k\not=l} D_{J_k\sigma_k j_k}(r)) 
\nonumber\\
&+& \sum_{1\le i< l\le N} D_{J_i\sigma_i j_i}^{Mm}(r) 
D_{J_l\sigma_l j_l}^{Mm}(r)
\prod_{k\not=l,i}  D_{J_k\sigma_k j_k}(r)] 
\}_{p\to p-\mu\sigma_0\delta_{J_0 j_0}t/4}
\ea
where the constants 
$D_{J\sigma j}(r),D_{J\sigma j}^{Mm}(r),D_{J\sigma j}^{Mm,Nn}(r)$ are 
defined in theorem \ref{th4.2} while $f^{(k)}_{r/4}(1)=(r/4,k)$
is simply the binomial coefficients. The first correction is small as long 
as $\alpha<1/2$. The error as compared to the 
first quantum correction of order  
$O(t^{(N[3r/2-1]\alpha} s^2)$ is a constant of order unity times
$t^{(N[3r/2-1]\alpha} sT$ and thus small as long as $0<\alpha$.
\end{Theorem}
Of course, in computing the quantum correction in terms of $p$ or $q$
rather than $p'=p\pm\mu\sigma_0\delta_{J_0 j_0}t/4$ or $q'=p' t^{-\alpha}$
up to order $t$ or $s^2$ respectively one is supposed to insert this
substitution into (\ref{4.17}) and to drop all higher order terms. \\
Proof of Theorem \ref{th4.2a}:\\
We begin with the operator identity \cite{Thiemann:2000bw}
\be \label{4.18}
\hat{h}_{J_0\sigma_0 j_0}^\mu=e^{-t/2} e^{-\mu\hat{p}_{J_0\sigma_0 j_0}}
\hat{g}_{J_0\sigma_0 j_0}^\mu
\ee
and exploit that our coherent states are eigenstates for 
$\hat{g}_{J_0\sigma_0 j_0}^\mu$ with eigenvalue
$g_{J_0\sigma_0 j_0}^\mu$. Moreover, our coherent states are expanded 
in terms of momentum operator eigenfunctions on which 
the operators $e^{-\mu\hat{p}_{J_0\sigma_0 j_0}}$ and 
$\hat{q}_{J_k\sigma_k j_k}(r_k)$ are simultaneously diagonal. It follows 
that
\ba \label{4.19}
&& 
<\hat{h}_{J_0\sigma_0 j_0}^\mu \prod_{k=1}^N\hat{q}_{J_k\sigma_k j_k}(r_k)>
=e^{-t/2} h_{J_0\sigma_0 j_0}^\mu  e^{-\mu p_{J_0\sigma_0 j_0}}
<e^{\mu \hat{p}_{J_0\sigma_0 j_0}}
\prod_{k=1}^N\hat{q}_{J_k\sigma_k j_k}(r_k)>
\nonumber\\
&& 
<\prod_{k=1}^N\hat{q}_{J_k\sigma_k j_k}(r_k)\hat{h}_{J_0\sigma_0 j_0}^\mu>
=e^{-t/2} h_{J_0\sigma_0 j_0}^\mu  e^{\mu p_{J_0\sigma_0 j_0}}
<e^{-\mu \hat{p}_{J_0\sigma_0 j_0}}
\prod_{k=1}^N\hat{q}_{J_k\sigma_k j_k}(r_k)>.
\ea
It is therefore sufficient to consider the expectation values
\ba \label{4.20}
&&
<e^{\nu \hat{p}_{J_0\sigma_0 j_0}}
\prod_{k=1}^N\hat{q}_{J_k\sigma_k j_k}(r_k)>
\nonumber\\ 
&=& 
\frac{
\sum_{n_{J\sigma j}} e^{\sum_{J\sigma j}[-t n_{J\sigma j}^2+
2n_{J\sigma j}(p_{J\sigma j}+\nu t\delta_{J\sigma j;J_0 \sigma_0 j_0}/2)]}
\prod_{k=1}^N\lambda^{r_k}_{J_k j_k}(n_{J\sigma j})
}{||\psi^t_{\{g_{J\sigma j}\}}||^2}
\nonumber\\
&=& 
\frac{
\sum_{n_{J\sigma j}} \int d^{18}x e^{\sum_{J\sigma j}[-x_{J\sigma j}^2+
2x_{J\sigma j}(p_{J\sigma j}-i\pi n_{J\sigma j}
+\nu t\delta_{J\sigma j;J_0 \sigma_0 j_0}/2)/T]}
\prod_{k=1}^N\lambda^{r_k}_{J_k j_k}(x_{J\sigma j}/T)
}{t^9\;||\psi^t_{\{g_{J\sigma j}\}}||^2}
\ea
where in the second
step we have again performed a Poisson transformation with periodicity 
parameter $T=\sqrt{t}$ (see the companion paper for more details).

As in \cite{ST01} we introduce the coordinates 
$x^\pm_{Jj}=(x_{J,+,j}\pm x_{J,-,j})/2$ and similar for $p^\pm_{Jj},
n^\pm_{Jj}$ with $x_{Jj}:=x^-_{Jj},p_{Jj}:=p^-_{Jj},n_{Jj}:=n^-_{Jj}$.
Then 
one can split the eighteen dimensional integral into two nine dimensional
ones with the result
\ba \label{4.21}
<.>&=&
\frac{2^9}{t^9\;||\psi^t_{\{g_{J\sigma j}\}}||^2}
\sum_{n_{J\sigma j}} 
[\int d^9x^+ e^{2\sum_{Jj}[-(x_{Jj}^+)^2+
2x_{Jj}^+ (p_{Jj}^+ -i\pi n_{J j}^+
+\nu t\delta_{J j;J_0 j_0}/4)/T]}]\;
\times\nonumber\\
&&\times
[\int d^9x e^{2\sum_{Jj}[-x_{Jj}^2+
2x_{Jj} (p_{Jj} -i\pi n_{J j}
+\nu \sigma_0 t\delta_{J j;J_0 j_0}/4)/T]}
\prod_{k=1}^N\lambda^{r_k}_{J_k j_k}(x_{J j}/T)].
\ea
Then, using the norm of our coherent states and dropping the 
$O(t^\infty)$ terms which come from the ones with $\sum_{J\sigma j}
n_{J\sigma j}^2>0$ we find 
\ba \label{4.22}
<.>&=&
\sqrt{\frac{2}{\pi}}^9 
e^{\frac{2}{t}\sum_{Jj}[(p^+ +\nu\delta_{J_0 j_0}t/4)_{Jj}^2
+(p+\nu\sigma_0 \delta_{J_0 j_0}t/4)_{Jj}^2-(p^+)_{Jj}^2-p_{Jj}^2]}
\times\nonumber\\
&& \times
[\int d^9x e^{-2\sum_{Jj}x_{Jj}^2}
\prod_{k=1}^N
\lambda^{r_k}_{J_k j_k}(x+\frac{p+\nu \sigma_0 t\delta_{J_0 j_0}/4}{T})]
\nonumber\\ &=&
\sqrt{\frac{2}{\pi}}^9 e^{\nu p_{J_0\sigma_0 j_0}} e^{t/4}
[\int d^9x e^{-2\sum_{Jj}x_{Jj}^2}
\prod_{k=1}^N
\lambda^{r_k}_{J_k j_k}(x+\frac{p+\nu \sigma_0 t\delta_{J_0 j_0}/4}{T})].
\ea
Combining (\ref{4.19}), (\ref{4.22}) we see that compared to 
(\ref{fluc4.81}) and the prefactor of $e^{-t/4} h_{J_0 \sigma_0 j_0}$
the remaining integral in (\ref{4.22}) is the one for the 
expectation value of the operator monomial
$\prod_{k=1}^N\hat{q}_{J_k\sigma_k j_k}(r_k)$ just that we have to 
evaluate it at $p+\nu \sigma_0 t\delta_{J_0 j_0}/4$ instead of at
$p$.\\
$\Box$\\
We are now ready to apply theorems \ref{th4.1} and \ref{th4.2a} to the 
case at hand. For $r=1/2$ we have 
\be \label{4.14}
a_1:=f^{(1)}_{1/8}(1)=\frac{1}{8},\;
a_2:=f^{(2)}_{1/8}(1)=-\frac{1}{8}\frac{7}{16}=-\frac{7}{128},\;
a_3:=f^{(3)}_{1/8}(1)=\frac{7}{128}\frac{15}{24}=\frac{35}{1024}
\ee
and 
\begin{align} \label{4.15}
D_{J\sigma j}(1/2)&= a_1 q^{-1}_{Jj}
\nonumber\\
D_{J\sigma j}^{Mm}(1/2)&=
a_1 (2 q^{-1}_{Jj} q^{-1}_{Mm}-q^{-1}_{Jm} q^{-1}_{Mj})
+2 a_2 q^{-1}_{Jj} q^{-1}_{Mm}
\nonumber\\
&= 2[a_1+a_2] q^{-1}_{Jj} q^{-1}_{Mm}-a_1 q^{-1}_{Jm} q^{-1}_{Mj}
\nonumber\\
D_{J\sigma j}^{Mm,Nn}(1/2)&=
\frac{a_1}{2}[\det(q^{-1}) \epsilon_{jmn}\epsilon_{JMN}
+q^{-1}_{Jj}(q^{-1}_{Mm} q^{-1}_{Nn}-q^{-1}_{Mn} q^{-1}_{Nm})
\nonumber\\
&\qqquad +2 q^{-1}_{Mm}(q^{-1}_{Jj} q^{-1}_{Nn}-q^{-1}_{Jn} q^{-1}_{Nj})] 
\nonumber\\
&\qqquad +2 a_2 q^{-1}_{Jj}(2 q^{-1}_{Mm} q^{-1}_{Nn}-q^{-1}_{Mn} q^{-1}_{Nm})
+3 a_3 q^{-1}_{Jj} q^{-1}_{Mm} q^{-1}_{Nn}
\nonumber\\
&=
\frac{a_1}{2}\det(q^{-1}) \epsilon_{jmn}\epsilon_{JMN}
+[\frac{3 a_1}{2}+2 a_2+3 a_3] q^{-1}_{Jj} q^{-1}_{Mm} q^{-1}_{Nn}
\nonumber\\
&\qqquad -[\frac{a_1}{2}+ 2 a_2]q^{-1}_{Jj} q^{-1}_{Mn} q^{-1}_{Nm}
-\frac{a_1}{2} q^{-1}_{Jn} q^{-1}_{Mm} q^{-1}_{Nj}.
\end{align}
It follows that
\begin{align} \label{4.16}
\sum_{M,m} D_{J\sigma j}^{Mm,Mm}(1/2) 
&=[a_1+3 a_3] q^{-1}_{Jj} q^{-1}_{Mm} q^{-1}_{Mm}
-\frac{a_1}{2} q^{-1}_{Jm} q^{-1}_{Mm} q^{-1}_{Mj}
\nonumber\\
&= [a_1+3 a_3] q^{-1}_{Jj} \mbox{tr}(q^{-2})
-\frac{a_1}{2} q^{-3}_{Jj} 
\nonumber\\
\sum_{Mm} D_{J_1\sigma_1 j_1}^{Mm}(1/2) D_{J_2\sigma_2 j_2}^{Mm}(1/2)
&=
(2[a_1+a_2] q^{-1}_{J_1 j_1} q^{-1}_{Mm}-a_1 q^{-1}_{J_1 m} q^{-1}_{Mj_1})
(2[a_1+a_2] q^{-1}_{J_2 j_2} q^{-1}_{Mm}-a_1 q^{-1}_{J_2 m} q^{-1}_{Mj_2})
\nonumber\\
&=
4[a_1+a_2]^2 q^{-1}_{J_1 j_1} q^{-1}_{J_2 j_2}\mbox{tr}(q^{-2})\nonumber\\
&\qqquad-2 a_1[a_1+a_2] (q^{-1}_{J_1 j_1} q^{-3}_{J_2 j_2} 
+q^{-1}_{J_2 j_2} q^{-3}_{J_1 j_1})
+a_1^2 q^{-2}_{J_1 J_2} q^{-2}_{j_1 j_2}. 
\end{align}
The relevant quantity for the Dirac Hamiltonian is
\ba \label{4.23}
\hat{F}_6 &:=& -\frac{i\hbar}{2} \sum_{v,v'\in V(\gamma)}  
\{
\nonumber\\
&& 
\epsilon^{JMN}\epsilon_{jmn}\frac{4^4}{2}
  [\hat{\theta}_B(v')(\hat{\theta}_A(v))^\dagger
  -\hat{\theta}'_B(v')(\hat{\theta}'_A(v))^\dagger]
  \sum_\sigma
\times \nonumber\\  
& \times & \{  
  <[\delta_{v',e_J^\sigma(v,1)} 
  (\sigma_j[1_2+\frac{\hat{h}_{J \sigma k}(v)-1}{2i}\tau_k])_{AB}
  -\delta_{v',e_J^\sigma(v,0)}\delta_{AB}]
\hat{q}_{M m}(v,1/2)\hat{q}_{N n}(v,1/2)>
\nonumber\\
&& -
<\hat{q}_{M m}(v',1/2)\hat{q}_{N n}(v',1/2)
  [\delta_{v,e_J^\sigma(v',1)} 
  ([1_2+\frac{\hat{h}_{J \sigma k}(v')^{-1}-1}{2i}\tau_k]\sigma_j)_{AB}
  -\delta_{v,e_J^\sigma(v',0)}\delta_{AB}]>
  \}
\nonumber\\
&+& 2 k_0 \delta_{AB}\delta_{v,v'}[\hat{\theta}'_B(v')
(\hat{\theta}_A(v)^\dagger-\hat{\theta}_B(v')(\hat{\theta}'_A(v))^\dagger]
\}
\ea
or more explicitly
\ba \label{4.24}
&&\hat{F}_6
-\frac{i\hbar}{2} \sum_{v,v'\in V(\gamma)}  
[-2 k_0 \delta_{AB}\delta_{v,v'}[\hat{\theta}'_B(v')
(\hat{\theta}_A(v)^\dagger-\hat{\theta}_B(v')(\hat{\theta}'_A(v))^\dagger]
\nonumber\\
&=& -\frac{i\hbar}{2} \sum_{v,v'\in V(\gamma)}  
\epsilon^{JMN}\epsilon_{jmn}\frac{4^4}{2}
  [\hat{\theta}_B(v')(\hat{\theta}_A(v))^\dagger
  -\hat{\theta}'_B(v')(\hat{\theta}'_A(v))^\dagger]
  \sum_\sigma
\times \nonumber\\  
& \times & 
 \{  
  [
  (\delta_{v',e_J^\sigma(v,1)} 
  (\sigma_j[1_2+\frac{(-1)}{2i}\tau_k])_{AB}
  -\delta_{v',e_J^\sigma(v,0)}\delta_{AB})
  <\hat{q}_{M m}(v,1/2)\hat{q}_{N n}(v,1/2)>
\nonumber\\
&&  + (\delta_{v',e_J^\sigma(v,1)}(\sigma_j\frac{1}{2i}\tau_k])_{AB}
  <\hat{h}_{J \sigma k}(v)\hat{q}_{M m}(v,1/2)\hat{q}_{N n}(v,1/2)>
  ]
\nonumber\\
&& -
  [
  (\delta_{v,e_J^\sigma(v',1)} 
  ([1_2+\frac{(-1)}{2i}\tau_k]\sigma_j)_{AB}
  -\delta_{v,e_J^\sigma(v',0)}\delta_{AB})
  <\hat{q}_{M m}(v',1/2)\hat{q}_{N n}(v',1/2)>
\nonumber\\  
&& +\delta_{v,e_J^\sigma(v',1)} (\frac{1}{2i}\tau_k\sigma_j)_{AB}
  <\hat{q}_{M m}(v',1/2)\hat{q}_{N n}(v',1/2)\hat{h}_{J \sigma k}(v')^{-1}>
  ]
  \}.
\ea
Let us write 
$q=p(v)t^{-\alpha},\;q'=p(v')t^{-\alpha},\;q_1=p_1(v)t^{-\alpha},
q'_1=p_1(v')t^{-\alpha}$ where 
$p_1(v)=p(v)+\sigma t\delta_{J k}/4,\;
p_1(v')=p(v')-\sigma t\delta_{J k}/4$. Then, using theorem
\ref{th4.2a}, we may write (\ref{4.24}) in the reduced form
\ba \label{4.25}
&&\hat{H}^{eff}_\gamma
-\frac{i\hbar}{2} \sum_{v,v'\in V(\gamma)}  
[-2 k_0 \delta_{AB}\delta_{v,v'}[\hat{\theta}'_B(v')
(\hat{\theta}_A(v)^\dagger-\hat{\theta}_B(v')(\hat{\theta}'_A(v))^\dagger]
\nonumber\\
&=& -\frac{i\hbar}{2} \sum_{v,v'\in V(\gamma)}  
\epsilon^{JMN}\epsilon_{jmn}\frac{4^4}{2}
  [\hat{\theta}_B(v')(\hat{\theta}_A(v))^\dagger
  -\hat{\theta}'_B(v')(\hat{\theta}'_A(v))^\dagger]
  \sum_\sigma
\times \nonumber\\  
& \times & 
 \{  
  [
  (\delta_{v',e_J^\sigma(v,1)} 
  (\sigma_j[1_2+\frac{(-1)}{2i}\tau_k])_{AB}
  -\delta_{v',e_J^\sigma(v,0)}\delta_{AB})
  <\hat{q}_{M m}(1/2)\hat{q}_{N n}(1/2)>_q
\nonumber\\ &&
  +e^{-t/4} (\delta_{v',e_J^\sigma(v,1)}
    (\sigma_j\frac{h_{J \sigma k}(v)}{2i}\tau_k])_{AB}
  <\hat{q}_{M m}(1/2)\hat{q}_{N n}(1/2)>_{q_1}
  ]
\nonumber\\
&& -
  [
  (\delta_{v,e_J^\sigma(v',1)} 
  ([1_2+\frac{(-1)}{2i}\tau_k]\sigma_j)_{AB}
  -\delta_{v,e_J^\sigma(v',0)}\delta_{AB})
  <\hat{q}_{M m}(1/2)\hat{q}_{N n}(1/2)>_{q'}
\nonumber\\&&
  +e^{-t/4}\delta_{v,e_J^\sigma(v',1)} 
  (\frac{h_{J \sigma k}(v')^{-1}}{2i}\tau_k\sigma_j)_{AB}
  <\hat{q}_{M m}(1/2)\hat{q}_{N n}(1/2)>_{q_1'}
  ]
  \}.
\ea
It remains to apply theorem \ref{th4.2a}. We have explicitly for arbitrary
invertible $q$
\ba \label{4.26}
&&\epsilon^{JJ_1 J_2}\epsilon^{j j_1 j_2}
<\hat{q}_{J_1 j_1}(1/2)\hat{q}_{J_2 j_1}(1/2)> 
\equiv \epsilon^{JJ_1 J_2}\epsilon^{j j_1 j_2}
<\hat{q}_{J_1\sigma_1 j_1}(1/2)\hat{q}_{J_2\sigma_2 j_2}(1/2)>
\nonumber\\
&=& \epsilon^{JJ_1 J_2}\epsilon^{j j_1 j_2} 
(2|\det(q)|^{1/4} t^{[3/4-1]\alpha})^2
\times\nonumber\\
&\times & \{ [\prod_{k=1}^2 D_{J_k\sigma_k j_k}(1/2)]
+ \frac{s^2}{4}\sum_{M,m}[\sum_{l=1}^2 
D_{J_l\sigma_l j_l}^{Mm,Mm}(1/2) \prod_{k\not=l} D_{J_k\sigma_k j_k}(1/2)) 
\nonumber\\
&+& \sum_{1\le i< l\le 2} D_{J_i\sigma_i j_i}^{Mm}(1/2) 
D_{J_l\sigma_l j_l}^{Mm}(1/2)
\prod_{k\not=l,i}  D_{J_k\sigma_k j_k}(1/2)] \}
\nonumber\\
&=&  \epsilon^{JJ_1 J_2}\epsilon^{j j_1 j_2}
(2|\det(q)|^{1/4} t^{[3/4-1]\alpha})^2
\times\nonumber\\
&\times & \{ a_1^2 q^{-1}_{J_1 j_1} q^{-1}_{J_2 j_2}
+ \frac{s^2}{4}[a_1
(
q^{-1}_{J_1 j_1} 
([a_1+3 a_3] q^{-1}_{J_2 j_2} \mbox{tr}(q^{-2})-\frac{a_1}{2} 
q^{-3}_{J_2 j_2})   
+q^{-1}_{J_2 j_2} 
([a_1+3 a_3] q^{-1}_{J_1 j_1} \mbox{tr}(q^{-2})-\frac{a_1}{2} 
q^{-3}_{J_1 j_1})   
)
\nonumber\\
&+& 
4[a_1+a_2]^2 q^{-1}_{J_1 j_1} q^{-1}_{J_2 j_2}\mbox{tr}(q^{-2})
-2 a_1[a_1+a_2] (q^{-1}_{J_1 j_1} q^{-3}_{J_2 j_2} 
+q^{-1}_{J_2 j_2} q^{-3}_{J_1 j_1})
+a_1^2 q^{-2}_{J_1 J_2} q^{-2}_{j_1 j_2} 
] \}.
\ea
We have
\ba \label{4.27}
\epsilon^{JJ_1 J_2}\epsilon^{j j_1 j_2}
q^{-1}_{J_1 j_1} q^{-1}_{J_2 j_2} &=& 
2\det(q^{-1}) q_{Jj}
\nonumber\\
\epsilon^{JJ_1 J_2}\epsilon^{j j_1 j_2}
q^{-1}_{J_1 j_1} q^{-3}_{J_2 j_2} &=& 
\det(q^{-1})(q_{Jj}\mbox{tr}(q^{-2}-q^{-1}_{Jj})
\nonumber\\
\epsilon^{JJ_1 J_2}\epsilon^{j j_1 j_2}
q^{-2}_{j_1 j_2} &=& 0.
\ea
Thus we can finish (\ref{4.26}) with
\ba \label{4.28}
&&\epsilon^{JJ_1 J_2}\epsilon^{j j_1 j_2}
<\hat{q}_{J_1 j_1}(1/2)\hat{q}_{J_2 j_1}(1/2)> 
\nonumber\\
&=&  
(2|\det(q)|^{1/4} t^{[3/4-1]\alpha})^2 \det(q^{-1})
\times\nonumber\\
&\times & \{ 2 a_1^2 q_{Jj}
+ \frac{s^2}{4}[2a_1(2[a_1+3 a_3] q_{J j} \mbox{tr}(q^{-2})-\frac{a_1}{2} 
[q_{Jj}\mbox{tr}(q^{-2})-q^{-1}_{Jj}])
\nonumber\\
&+& 
8[a_1+a_2]^2 q_{J j}\mbox{tr}(q^{-2})
-4 a_1[a_1+a_2] 
[q_{Jj}\mbox{tr}(q^{-2})-q^{-1}_{Jj}]
] \}
\nonumber\\
&=&  
4 t^{-\alpha/2}/\sqrt{\det(q)}
\{ 2 a_1^2 q_{Jj}
\nonumber\\
&&
+ \frac{s^2}{4}[
(4a_1 [a_1+3 a_3]+8[a_1+a_2]^2) q_{J j} \mbox{tr}(q^{-2})
-(a_1^2+4 a_1[a_1+a_2]) (q_{Jj}\mbox{tr}(q^{-2})-q^{-1}_{Jj})]
\}
\nonumber\\
&=&  
\frac{t^{-\alpha/2}}{16\sqrt{\det(q)}}
\{ 2 q_{Jj}
\nonumber\\
&&
+ \frac{s^2}{4}[
(4 [1+24 a_3]+8[1+8 a_2]^2) q_{J j} \mbox{tr}(q^{-2})
-(1+4[1+8a_2]) (q_{Jj}\mbox{tr}(q^{-2})-q^{-1}_{Jj})]
\}
\nonumber\\
&=&  
\frac{t^{-\alpha/2}}{16\sqrt{\det(q)}}
\{ 2 q_{Jj}+ 
\frac{s^2}{4}[
(4+\frac{3^2\cdot 5}{2^5}]+\frac{3^4}{2^5}) q_{J j} \mbox{tr}(q^{-2})
-\frac{13}{4} (q_{Jj}\mbox{tr}(q^{-2})-q^{-1}_{Jj})]
\}
\nonumber\\
&=&  
\frac{1}{8\sqrt{\det(p)}}
\{p_{Jj}+ 
\frac{t}{2^7}[75 p_{J j} \mbox{tr}(p^{-2})+52 p^{-1}_{Jj})]
\}
\nonumber\\
&=&  
\frac{t^{-\alpha/2}}{8\sqrt{\det(q)}}
\{q_{Jj}+ 
\frac{s^2}{2^7}[75 q_{J j} \mbox{tr}(q^{-2})+52 q^{-1}_{Jj})]
\}.
\ea
Notice that the classical limit is {\it precisely} the correct one
while the relative first quantum correction is given by
approximately $1.0 s^2\delta_{Jj}$ for flat initial data.

Now we should compute the additional corrections arising
when expanding
\be \label{4.29}
e^{-t/4}\epsilon^{JJ_1 J_2}\epsilon^{j j_1 j_2}
<\hat{q}_{J_1 j_1}(1/2)\hat{q}_{J_2 j_1}(1/2)>_{q\to q
+t^{1-\alpha}\nu\sigma_0\delta_{J_0 j_0}/4}
\ee
at $q$ up to order $s^2$. However, it is clear that the the additional
correction in $e^{-t/4}-1=s^2O(t^{2\alpha})$ and the one from 
$\delta q=s^2 O(t^{\alpha})$ are both of higher order in $s$ so that
we can drop the factors of $e^{-t/4}$ and the substitutions 
$q\to q_1,q'\to q'_1$ in (\ref{4.25}) which therefore can be written, up to
order $s^2$ as 
\ba \label{4.30}
&&\hat{H}^{eff}_\gamma
-\frac{i\hbar}{2} \sum_{v,v'\in V(\gamma)}  
[-2 k_0 \delta_{AB}\delta_{v,v'}[\hat{\theta}'_B(v')
(\hat{\theta}_A(v)^\dagger-\hat{\theta}_B(v')(\hat{\theta}'_A(v))^\dagger]
\nonumber\\
&=& -\frac{i\hbar}{2} \sum_{v,v'\in V(\gamma)}  
  \epsilon^{JMN}\epsilon^{jmn}
  [\hat{\theta}_B(v')(\hat{\theta}_A(v))^\dagger
  -\hat{\theta}'_B(v')(\hat{\theta}'_A(v))^\dagger]
  \sum_\sigma
\times \nonumber \\ 
& \times & 
 \{  
  [(\delta_{v',e_J^\sigma(v,1)} 
  (\sigma_j[1_2+\frac{h_{J \sigma k}(v)-1}{2i}\tau_k])_{AB}
  -\delta_{v',e_J^\sigma(v,0)}\delta_{AB})
  <\hat{q}_{M m}(1/2)\hat{q}_{N n}(1/2)>_{q=q(v)}
  ]
\nonumber\\
&& -
  [
  (\delta_{v,e_J^\sigma(v',1)} 
  ([1_2+\frac{h_{J \sigma k}(v')^{-1}-1}{2i}\tau_k]\sigma_j)_{AB}
  -\delta_{v,e_J^\sigma(v',0)}\delta_{AB})
  <\hat{q}_{M m}(1/2)\hat{q}_{N n}(1/2)>_{q=q(v')}
  ]
  \}
\nonumber\\
&=& -\frac{i\hbar}{2} \sum_{v,v'\in V(\gamma)}  
  [\hat{\theta}_B(v')(\hat{\theta}_A(v))^\dagger
  -\hat{\theta}'_B(v')(\hat{\theta}'_A(v))^\dagger]
  \sum_\sigma
\times \\  
& \times & 
 \{  
  [(\delta_{v',e_J^\sigma(v,1)} 
  (\sigma_j[1_2+\frac{h_{J \sigma k}(v)-1}{2i}\tau_k])_{AB}
  -\delta_{v',e_J^\sigma(v,0)}\delta_{AB})
  \frac{1}{8\sqrt{\det(p)}}
\times\nonumber\\ &&\times
  (p_{Jj}+\frac{t}{2^7}[75 p_{J j} \mbox{tr}(p^{-2})+52 p^{-1}_{Jj})])(v)
  ]
\nonumber\\
&& -
  [
  (\delta_{v,e_J^\sigma(v',1)} 
  ([1_2+\frac{h_{J \sigma k}(v')^{-1}-1}{2i}\tau_k]\sigma_j)_{AB}
  -\delta_{v,e_J^\sigma(v',0)}\delta_{AB})
  \frac{1}{8\sqrt{\det(p)}}
\times\nonumber\\ &&\times
  (p_{Jj}+\frac{t}{2^7}[75 p_{J j} \mbox{tr}(p^{-2})+52 p^{-1}_{Jj})])(v')
  ]
  \}.
\nonumber
\ea
%----------------------------------------------------------------------
\section{Towards Dispersion Relations}
\label{s5}
%----------------------------------------------------------------------
In the present section, we will bring together some of the results of
the companion paper and the previous section: We will compute
corrections to the standard dispersion relations for the scalar and
the electromagnetic field resulting from its coupling to QGR.  
The necessary calculations are performed in section \ref{se5.1} for the 
scalar and in \ref{se5.2} for the electromagnetic field. 
Similar computations can be performed for the fermions but they give no 
new insights so that we leave this to the interested reader.
We have set up the problem in such a way that the calculations for an 
arbitrary background metric but for a start we confine ourselves to the 
flat one. In section 
\ref{se5.3} we will discuss the results and compare them to those
obtained in \cite{Gambini:1998it,Alfaro:1999wd}. 
In our companion paper we have given some conceptual discussion of the 
issues involved in obtaining dispersion relation from QGR, so we will
mainly focus on the concrete calculations. 

In \cite{ST01} we have obtained Hamiltonian operators for the matter
fields of the form 
\begin{equation*}
\widehat{H}^{\text{eff}}_\gamma =
\frac{1}{2}\sum_{v,v',l,l'} \widehat{p}_l(v) P^{ll'}(v,v')\widehat{p}_{l'}(v') 
+\widehat{q}_l(v) Q^{ll'}(v,v') \widehat{q}_{l'}(v'),  
\end{equation*}
where the coefficients $P,Q$ are the expectation values of specific
operators on the gravitational Hilbert space. We have computed these
expectation values in the preceeding section. 

Note that these Hamiltonians are normal ordered with respect to the
annihilation and creation operators defined in \cite{ST01}. Thus, the 
expectation value of these Hamiltonians in a
coherent state peaked at a specific classical field configuration will 
yield \textit{precisely} its classical value.
Therefore, in discussing the dispersion relations, we will assume the
matter quantum fields to be in a coherent state and can  
effectively
work with the classical fields $p,q$. 
A similar argument can be given for the fully quantized Hamiltonians 
of \cite{ST01}, only that one has to consider a coherent
state for the \textit{combined} system of quantum matter \textit{and}
quantum gravity displayed in section 4 of \cite{ST01} as well.   

Summing up, in the following we will investigate Hamiltonians of the form 
\begin{equation}
\label{eq5.4}
\expec{\widehat{H}^{\text{eff}}_\gamma} =
\frac{1}{2}\sum_{v,v',l,l'} p_l(v) P^{ll'}(v,v')p_{l'}(v') 
+q_l(v) Q^{ll'}(v,v') q_{l'}(v').  
\end{equation}
The coefficients $P,Q$ can in principle be taken to be expectation values  
in a coherent state for the gravitational field peaked at an arbitrary 
point of the classical phase space. However, since we are interested
in dispersion relations, a notion that by definition describes the
propagation of fields in flat space, we will restrict considerations
to the case of GCS approximating flat Euclidean space (denoted by
$\Psi_{\text{flat}}$ in the following. 
Also, when considering application to situations such as the $\gamma$- 
ray burst effect, the curvature radius is always huge compared to
Planck length and does therefore not lead to any new quantum effects 
but just to classical redshifts which can easily be accounted for.  

Let us choose the canonical Euclidean coordinate system as global 
coordinates on $\Sigma$.    
In the $\uone^3$-setting, we can model the flat space situation by
choosing the classical values 
\begin{equation*}
 A_a^I(x)=0,\qquad E^a_I(x)=\delta^a_I\qquad \text{ for all } x\in\Sigma  
\end{equation*}
with respect to our global coordinates. Therefore all holonomies are
trivial and for the fluxes we find
\begin{equation*}
  p^e_i(v)=\frac{1}{a^2}\int_{S_e}dn^i.
\end{equation*}
We will also use the dimensionfull quantity $P^e_i(v)=a^2p^e_i(v)$. 

Let us come back to the discussion of \eqref{eq5.4}: Since the
coefficients in these Hamiltonians vary from vertex to vertex, the
equations of motion induced by \eqref{eq5.4} are still highly
complicated and an exact analytical treatment is beyond the scope of the
present paper. Moreover, the solutions to the equations of motion will not
have the character of plane waves, so the notion of a dispersion
relation is ill defined anyway.

In \cite{ST01} we argued that in the limit of low energies or,
equivalently, large wavelength the field propagation induced by
\eqref{eq5.4} \textit{can} be described by a dispersion relation: The
graph $\gamma$, the GCS is based on, breaks Euclidean
invariance. However on large scales this invariance is approximately
restored. 

As we can not easily \eqref{eq5.4} 
compute the solutions to the equations of motion of \eqref{eq5.4} and 
show that they reduce to approximate plane waves with a specific
dispersion relation in the low energy limit,  
the question is how one can nevertheless obtain  the dispersion relation
governing the propagation for low energies.

In \cite{ST01} we have sketched a tentative answer, which we will work out 
in the present section for the examples of the scalar and the
electromagnetic field. Let us review
the basic idea of the procedure before we spell out the details: 
We are going to replace \eqref{eq5.4} by a simpler Hamiltonian which
\begin{itemize}
\item is a good approximation of \eqref{eq5.4} for slowly varying
  $q$ and $p$ and  
\item is simple enough such that the EOM can be solved exactly. 
\end{itemize}
The resulting theory will be an approximation
for low energies, the detailed information 
contained in the full Hamiltonian
\eqref{eq5.4} which is only relevant for processes of very high energy 
gets integrated out. 
This idea underlies also the works \cite{Gambini:1998it} and 
\cite{Alfaro:1999wd} and, at a rather
simple level, is the basis for the recovery of continuum elasticity
theory from the atomic description in solid state physics 
(see for example \cite{Ashcroft:1976}). 

We will now turn to the scalar field Hamiltonian and explain the
steps we will take to implement the above idea in detail. The Maxwell
Hamiltonian will be treated along the same lines in section \ref{se5.2}. 
%-----------------------------------------------------------------------
\subsection{Dispersion Relation for the Scalar Field}
%-----------------------------------------------------------------------
\label{se5.1}
The basic field variables underlying the quantization of the scalar
field in \cite{ST01},
\begin{equation}
\label{eq5.16}
  \phi_v=\phi(\evec{x}(v)),\qquad\text{ and }\qquad \pi_v=\int_{R_v}\pi,
\end{equation}
were represented by the operators $-i\ln U(v)$, $Y_v$. $R_v$ is the
cell containing $v$ 
in a polyhedral decomposition of $\Sigma$ dual to $\gamma$. 
According to what we 
have said in the introduction to this chapter, in the considerations
to follow we will replace these operators by their classical
counterparts \eqref{eq5.16} upon assuming the quantum fields to be in
a coherent state.     

Using the results of section \ref{s4}, the Hamiltonian for the scalar
field  we are considering can be written as 
\begin{equation}
\label{eq5.18}
  H^{\text{eff}}_{\text{KG}}
  =\frac{1}{2Q_{\text{KG}}} 
\left(F_{\text{kin}}(\pi)
+F_{\text{der}}(\phi)+K^2F_{\text{m}}\right),
\end{equation}
where
\begin{align*} 
F_{\text{m}}
&=\sum_v\sqrt{\det P(v)}\left[1+\frac{\ell_P^7}{\sqrt{t}}\frac{1}{32}\tr
  P^{-2}(v)\right]\phi_v^2,\\ 
F_{\text{der}}&=\frac{1}{4}\sum_v\sum_{I\sigma
  I'\sigma'}\bigg[
\frac{\sigma\sigma'P^2_{II'}(v)}{\sqrt{\det P(v)}}\\
&\qqquad+\frac{\ell_P^4}{t}
\frac{\sigma\sigma'}{\sqrt{\det P(v)}}
\left(\frac{1173}{128}\tr(P^{-2})P^2_{II'}(v)
+\frac{19}{32}\delta_{II'}\right)\bigg]
\partial_{e_{\sigma I}}^{+}\phi_v\partial_{e_{\sigma' I'}}^{+}\phi_v,\\
F_{\text{kin}}
&=\sum_v\frac{1}{\sqrt{\det P(v)}}\left[1+ 
\frac{\ell_P^4}{t}\frac{1707}{512}\tr P^{-2}(v)\right]\pi_v^2.
\end{align*}
Now we will express the field quantities $\phi_v,\pi_v$ by the 
basic fields $\phi(\evec{x}),\pi(\evec{x})$, using an approximation 
which is good in the case $\phi(\evec{x}),\pi(\evec{x})$ vary only
very little on the scale $\epsilon$ of the graph. The idea is to
isolate the rough structural properties of \eqref{eq5.18} that
lead to corrections as compared to the standard dispersion relations 
and to discard the microscopic details that will only yield higher order
corrections which are not visible in the long wavelength regime.  

To this end, we Taylor expand the field variables $\phi_v,\pi_v$
around the location $\evec{x}(v)$ of the vertex $v$, i.e. we make the
replacements 
\begin{align*}
\phi_v&\longrightarrow\phi(\evec{x}(v)),\\  
\pi_v&\longrightarrow \pi(\evec{x}(v))
\text{Vol}(R_v)+a^{(a)}(v)\partial_a\pi(\evec{x}(v))+\ldots,\\
\partial^+_{e_I}\phi_v&\longrightarrow b_I^{(a)}
\partial_a\phi(\evec{x}(v))
+b_I^{(a)}b_I^{(a')}\partial_a\partial_{a'}\phi(\evec{x}(v))+\ldots.
\end{align*}
and truncate the right hand sides at the desired order. Note that in
the above formulae we have introduced the geometric quantities
\begin{equation*}
b_I^{(a)}(v)\doteq x^a(f(e_I(v)))-x^a(v),\qquad
a^{(a)}(v)\doteq\int_{R_v}x^a\,d^3x,  
\end{equation*}
and let us furthermore define
\begin{equation*}
\widetilde{b}_I^{(a)}(v)\doteq
\frac{1}{2}\left(x^a[f(e_I(v))]-x^a[f(e_I^-(v))]\right) 
\end{equation*}
which we will have opportunity to use below. 
Also, it is perhaps worthwhile to remind the reader at this point 
that all edges are taken to be outgoing from $v$. 

Then we replace the coefficients of the continuum fields by graph
averages and the sums by integrals. As argued in \cite{ST01}, this is
a good approximation, as long as $\phi$ and $\pi$ are slowly varying
on the graph scale $\epsilon$. Let us detail this step for the example 
of the mass term. We write
\begin{align*}
  F_{\text{m}}
&=\sum_v \text{Vol}(R_v)\frac{\sqrt{\det P(v)}}{\text{Vol}(R_v)}
\left[1+\frac{\ell_P^7}{\sqrt{t}}\frac{1}{32}\tr
  P^{-2}(v)\right]\phi_v^2\\
&\approx\sum_v \phi_v^2 \text{Vol}(R_v)
\left(\aver{\frac{\sqrt{\det P(\,\cdot\,)}}{\text{Vol}(R_{\,\cdot\,})}}
+\frac{\ell_P^7}{\sqrt{t}}\frac{1}{32}
\aver{\frac{\sqrt{\det P(\,\cdot\,)}\tr
  P^{-2}(\,\cdot\,)}{\text{Vol}(R_{\,\cdot\,})}}\right)\\
&\approx \int_\Sigma \phi(x)\, d^3x\,\,
\left(\aver{\frac{\sqrt{\det P(\,\cdot\,)}}{\text{Vol}(R_{\,\cdot\,})}}
+\frac{\ell_P^7}{\sqrt{t}}\frac{1}{32}
\aver{\frac{\sqrt{\det P(\,\cdot\,)}\tr
  P^{-2}(\,\cdot\,)}{\text{Vol}(R_{\,\cdot\,})}}\right) 
\end{align*}
where $\langle\langle\,\cdot\,\rangle\rangle$ denotes the
\textit{graph average}
\begin{equation*}
  \aver{C(\,\cdot\,)}\doteq \frac{1}{N}\sum_v C(v)
\end{equation*}
for vertex dependent quantities $C(v)$. $N$ denotes the number of vertices
of the graph. In case the graph has a countably infinite number of
vertices, the above definition has to be replaced by an appropriate
limit of finite sums.   

Analogously, we make the replacements 
\begin{align*}
F_{\text{kin}}&\longrightarrow
\int_\Sigma \left({A_0}{}+{A_1}{}\right)\pi^2(\evec{x})
+{A_0^{(a)(a')}}{}\partial_a\pi(\evec{x})\partial_{a'}\pi(\evec{x})
+\ldots\,d^3x,\\ 
F_{\text{der}}&\longrightarrow
\int_\Sigma \left({B_0^{(a)(a')}}{}+{B_1^{(a)(a')}}{}\right)
\partial_a\phi(\evec{x})\partial_{a'}\phi(\evec{x})\\
&\qquad\qqquad
+\frac{1}{4}{B_0^{(ab)(a'b')}}{}
\partial_a\partial_b\phi(\evec{x})\partial_{a'}\partial_{b'}\phi(\evec{x})
+\frac{1}{3}{B_0^{(abc)(a')}}{}
\partial_a\partial_b\partial_c\phi(\evec{x})\partial_{a'}\phi(\evec{x})
+\ldots\,d^3x,\\
{F}_{\text{m}}&\longrightarrow
\int_\Sigma \left({C_0}{}+{C_1}{}\right)\phi^2(\evec{x})+\ldots\,d^3x
\end{align*}
where the coefficients in the kinetic term are defined to be 
\begin{align*}
A_0&=\aver{\frac{V^2_v}{\sqrt{\det P(v)}}},\\
A_1&=\frac{1707}{512}\frac{\ell_P^4}{t}\aver{\frac{V^2\tr P(v)}{\sqrt{\det
    P(v)}}},\\
A_1^{(a)(a')}&=\aver{\frac{a^{(a)}(v)a^{(a')}(v)}{\sqrt{\det P(v)}}},
\end{align*}
the ones in the derivative term as 
\begin{align*}
B_0^{(a)(a')}&=\sum_{I,I'}\aver{  
\sqrt{\det P(v)}P^2_{II'}
\widetilde{b}_I^a\widetilde{b}_{I'}^{a'}(v)},\\
B_1^{(a)(a')}&=\frac{\ell_P^4}{t}\sum_{I,I'}\aver{
\sqrt{\det P(v)}
\left(\frac{1173}{128}\tr(P^{-2})P^2_{II'}(v)
+\frac{19}{32}\delta_{II'}\right)
\widetilde{b}_{I}^{a}\widetilde{b}_{I'}^{a'}},\\
B_0^{(abc)(a')}&=\sum_{I,I'}\aver{  
\sqrt{\det P(v)}P^2_{II'}
\widetilde{b}_I^a\widetilde{b}_I^b\widetilde{b}_I^c\widetilde{b}_{I'}^{a'}(v)},\\
B_0^{(ab)(a'b')}&=\sum_{I,I'}\aver{  
\sqrt{\det P(v)} P^2_{II'}
\widetilde{b}_I^a\widetilde{b}_I^b
\widetilde{b}_{I'}^{a'}\widetilde{b}_{I'}^{b'}(v)}.
\end{align*}
and finally the coefficients in the mass term by
\begin{align*}
C_0&\doteq \aver{\sqrt{\det P(v)}},\\ 
C_1&\doteq \frac{1}{32}\frac{\ell_P^7}{\sqrt{t}}
\aver{\sqrt{\det P(v)}\tr P^{-2}(v)}.
\end{align*}
Note that we have just written down the leading order terms and the
first order corrections, where a ``first order correction'' is either
of 
\begin{enumerate}
\item A term that is \textit{next to leading order} in the Taylor
  expansion and \textit{leading} order with respect to the
  fluctuation calculation.
\item A term that is \textit{leading order} in the Taylor expansion  
and next to leading order in the fluctuation calculation. 
\end{enumerate}
Terms that are leading order in the fluctuation calculation carry a
superscript $0$ while that are first order corrections in the
fluctuation calculation are marked by a superscript 1.  
Finally note that we have dropped terms that end up to be a total 
derivative and therefore do not contribute to the Hamiltonian.  

Before we write down the resulting dispersion relation we invoke our
restriction to random processes which imply Euclidean invariance 
on large scales of the resulting random graphs. That is we assume 
\begin{equation*}
{A_0^{(a)(a')}}{}\sim\delta^{aa'},\qquad
{B_0^{(a)(a')}}{}\sim\delta^{aa'},\qquad
{B_1^{(a)(a')}}{}\sim\delta^{aa'}. 
\end{equation*}
For the tensors of fourth rank, the situation is slightly more
complicated: $\delta^{ab}\delta^{cd}$, $\delta^{ac}\delta^{db}$ and 
$\delta^{ad}\delta^{bc}$ span the space of rotationally invariant
tensors of fourth rank. But contraction of any of them with 
$k_{a}k_{b}k_{c}k_{d}$ 
is equal to $\betr{k}^4$.

We can now write down the dispersion relation for the Hamiltonian
resulting from the above replacements  
\begin{equation}
\label{eq5.15}
\begin{split}
\omega^2(\evec{k})=&K^2\left[{A_0}{}{C_0}{}+
{A_0}{}{C_1}{}+{A_1}{}{C_0}{}\right]\\
&+\betr{k}^2\left[{A_0}{}{B_0}+{A_0}{}{B_1^{(1)(1)}}{}
    +{A_1}{}{B_0^{(1)(1)}}{}
    +K^2{A_0^{(1)(1)}}{}{C_0}{}\right]\\
&+\betr{k}^4\left[\frac{1}{4}{A_0}{}{B_0^{(11)(11)}}{}
    -\frac{1}{3}{A_0}{}{B_0^{(111)(1)}}{}
    +{A_0^{(1)(1)}}{}{B_0^{(1)(1)}}{}\right]\\
&+\ldots. 
\end{split}  
\end{equation}
We will discuss the physical content of \eqref{eq5.15} in section
\ref{se5.3}. Before that, we give a similar calculation for the
electromagnetic field. 
%--------------------------------------------------------------------------
\subsection{The Electromagnetic Field}
\begin{figure}
\centerline{\epsfig{file=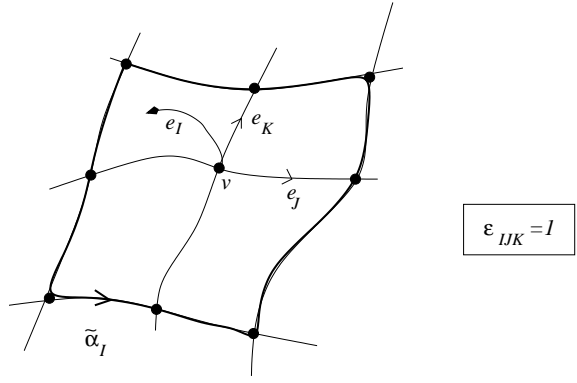, height=5cm}}
\caption{\label{loop}The loop $\widetilde{\alpha}_I(v)$}
\end{figure}

%--------------------------------------------------------------------------
\label{se5.2}
This section is devoted to the calculation of an (approximate)
dispersion relation for the electromagnetic field. The treatment is
completely analogous to the one given for the scalar field in the last
section, so we can be rather brief, here. 
Again, we introduce the continuum fields $A_a(\evec{x})$, $E^a(\evec{x})$
underlying the regularization and quantization of the Hamiltonian
performed in \cite{ST01} and also the classical quantities 
\begin{equation}
\label{eq5.10}
\begin{split}
  E_e=\int_{S_e}E &\text{ which was represented by } Y_e,\\
  A_e=\int_e A&\text{ which was represented by } -i \ln H_e
\end{split}    
\end{equation}
(subject to the subtleties associated with the logarithm spelled out in 
detail in \cite{ST01})
in the quantum Hamiltonian \eqref{ham1}. When we replace the  
gravitational operators in the Hamiltonian by their expectation values 
obtained in the last section and the operators for the matter fields
by their classical counterparts \eqref{eq5.10} we get 
\begin{equation*}
\expec{\widehat{H}_{M,\gamma}}_{\Psi_{\text{flat}}}= \frac{1}{2Q_{\text{M}}}
\left(F_{\text{el}}(E)+F_{\text{mag}}(B)\right)
\end{equation*}
where 
\begin{align*}
F_{\text{el}}(E)&=\sum_v\sum_{I\sigma I'\sigma'}\bigg[
\sqrt{\det P(v)}P^{-2}_{II'}
+\frac{\ell_P^4}{t}\Big(\frac{763}{512}P^{-2}_{II'}\tr P^{-2}
-\frac{13}{16}P^{-4}_{II'}\Big)\bigg]
\frac{\sigma\sigma'}{4}E_{e_{\sigma I}}(v)E_{e_{\sigma'I'}}(v),\\ 
F_{\text{mag}}(A)&=\frac{1}{16}\sum_v\sum_{II'}
\bigg[\sqrt{\det P(v)}P^{-2}_{II'}
+\frac{\ell_P^4}{t}\Big(\frac{763}{512}P^{-2}_{II'}\tr P^{-2}
 -\frac{13}{16}P^{-4}_{II'}\Big)\bigg]
 A_{\widetilde{\alpha}_I}A_{\widetilde{\alpha}_{I'}}, 
\end{align*}
where $\widetilde{\alpha}(v)$ is the loop around the vertex $v$ ``in the 
$I$-plane'' as depicted in figure \ref{loop}. 
Now we Taylor-expand the $A_\alpha$, $E_e$. To this end we   
introduce some geometric quantities:
\begin{gather*}
s^e_{a}(v)\doteq \int_{S^e_v}n_a(\evec{y})\, dy,\qquad
s^e_{ab}(v)\doteq \int_{S^e_v}n_a(\evec{y})(\evec{y}- \evec{x}(v))_b\, dy,\\
s^e_{abc}(v)\doteq \int_{S^e_v}n_a(\evec{y})(\evec{y}- \evec{x}(v))_b
(\evec{y}- \evec{x}(v))_c\, dy, 
\end{gather*}
where $n$ denotes the normal to the surface of integration. Moreover 
\begin{equation*}
  \widetilde{s}^I_a(v)\doteq \frac{1}{2}
\left(s^{e_I}_{b}(v)-s^{e^-_I}_{c}(v)\right),\qquad
\widetilde{s}^I_{ab}(v)\doteq \frac{1}{2}\left(s^{e_I}_{ab}(v)-s^{e^-_{I}}_{ab}(v)\right),\dots.
\end{equation*}
Now we can make the replacement for the electric field: 
\begin{equation*}
  E_e\vla
  s^e_{a}(v)E^a(\evec{x}(v))+s^e_{ab}(v)\partial^bE^a(\evec{x}(v))+\ldots.  
\end{equation*}
We proceed in a similar fashion for the connection: 
\begin{equation}
b_\alpha^{ab}(v)\doteq \int_0^1\dot \alpha^a(s)(\evec{\alpha}(s)-\evec{x}(v))^b,\qquad
b_\alpha^{abc}(v)\doteq \int_0^1\dot \alpha^a(s)
(\evec{\alpha}(s)-\evec{x}(v))^b(\evec{\alpha}(s)-\evec{x}(v))^c
\end{equation}
whence we replace 
\begin{equation*}
A_\alpha\vla b_\alpha^{ab}(v)\partial_bA_a(\evec{x}(v))
+\frac{1}{2}b_\alpha^{abc}(v)\partial_b\partial_cA_a(\evec{x}(v))+\ldots.
\end{equation*}
Inserting this into the expressions for $F_{\text{el}}$ and $F_{\text{mag}}$
and subsequently replacing the resulting coefficients by graph
averages results in the total replacement
\begin{equation}
\label{eq5.11}
\begin{split}
F_{\text{el}}&\vla\int_\Sigma \left(S^{(0)}_{(a)(a')}
+S^{(1)}_{(a)(a')}\right)E^a(\evec{x})E^{a'}(\evec{x})
+2 S^{(0)}_{(a)(a'b')}E^a(\evec{x})\partial^{b'}E^{a'}(\evec{x})+\ldots,\\
F_{\text{mag}}&\vla\left(B_{0}^{(ab)(a'b')}+B_{1}^{(ab)(a'b')}\right) 
\partial_{b}A_{a}(\evec{x})\partial_{b'}A_{a'}(\evec{x})
+B_{0}^{(ab)(a'b'c')}\partial_{b}A_{a}(\evec{x})\partial_{b'}
\partial_{c'}A_{a'}(\evec{x})+\ldots, 
\end{split}
\end{equation}
where 
\begin{align*}
S^{(0)}_{(a)(a')}&=\sum_{II'}\aver{
\sqrt{\det P}P^{-2}_{II'}(\blank)\widetilde{s}^I_a(\blank)\widetilde{s}^{I'}_{a'}(\blank)},\\
S^{(1)}_{(a)(a')}&=\frac{\ell_P^4}{t}\sum_{II'}\aver{
\Big(\frac{763}{512}P^{-2}_{II'}\tr P^{-2}(\blank)
-\frac{13}{16}P^{-4}_{II'}(\blank)\Big)
\widetilde{s}^I_a(\blank)\widetilde{s}^{I'}_{a'}(\blank)},\\
S^{(0)}_{(a)(a'b')}&=\sum_{II'}\aver{
\sqrt{\det P}P^{-2}_{II'}(\blank)\widetilde{s}^I_a(\blank)
\widetilde{s}^{I'}_{a'b'}(\blank)},
\end{align*}
and analogously 
\begin{align*}
B_{0}^{(ab)(a'a')}&=\sum_{II'}\aver{
\frac{\sqrt{\det P}}P^{-2}_{II'}(\blank)
b^{ab}_{\widetilde{\alpha}_I}(\blank)b^{a'b'}_{\widetilde{\alpha}_{I'}}(\blank)},\\
B_{1}^{(ab)(a'b')}&=
\frac{\ell_P^4}{t}\sum_{II'}\aver{\frac{1}{V_v}
\Big(\frac{763}{512}P^{-2}_{II'}\tr P^{-2}(\blank)
-\frac{13}{16}P^{-4}_{II'}(\blank)\Big)
b^{ab}_{\widetilde{\alpha}_I}(\blank)b^{a'b'}_{\widetilde{\alpha}_{I'}}(\blank))},\\
B_{0}^{(ab)(a'b'c')}&=\sum_{II'}\aver{
\frac{\sqrt{\det P}}{V_v}P^{-2}_{II'}(\blank)
b^{ab}_{\widetilde{\alpha}_I}(\blank)b^{a'b'c'}_{\widetilde{\alpha}_{I'}}(\blank)}.
\end{align*}
Now we can make the replacements \eqref{eq5.11} and obtain a
Hamiltonian for the continuum fields $A_a(\evec{x})$, $E^a(\evec{x})$.
A straightforward calculation yields the resulting equations of motion:
\begin{equation}
\label{eq5.12}
\begin{split}
\ddot{A}^d=&
\partial_b\partial_{b'}A^{a'}S_{(d)(a)}
\left(B^{(db)(a'b')}-B^{(a'b)(db')}\right)\\
&+\partial_b\partial_{b'}\partial^{c'}A^{a'}\Big[
B^{(db)(a'b'c')}-B^{(a'b)(db'c')}
+\left(S_{(d)(ac')}-S_{(a)(dc')}\right)
\left(B^{(ab)(a'b')}-B^{(a'b)(ab')}\right)\Big]\\
&+\ldots 
\end{split}
\end{equation}
where we have used shorthands 
$S_{(a)(a')}\doteq S^{(0)}_{(a)(a')}+S^{(1)}_{(a)(a')}$ and similarly 
for the other tensors $S_{()()},B^{()()}$.

Before we spell out the resulting dispersion relation, we use the
rotation invariance of the graph on large scales: It is
clear that 
\begin{equation*}
  S_{(a)(a')}\sim\delta_{aa'},\qquad S_{(a)(a'b')}\sim \epsilon_{aa'b'}.
\end{equation*}
For the tensors of higher rank, the situation is slightly more
complicated: For rank four, the space of invariant tensors is three
dimensional, the space of rank five tensors is ten dimensional. But if we 
take into consideration the symmetries of the terms, these tensors get 
contracted with, there is only one invariant tensor left in each
case. We define 
\begin{align*}
c_1^{(0/1)}&\doteq\frac{1}{3}\sum_i{S^{(0/1)}_{(a)(a)}}{},&
c_3^{(0/1)}&\doteq\frac{1}{6}\left(\sum_{ab}{B_{0/1}^{(ab)(ab)}}{}
-\sum_a{B_{0/1}^{(aa)(aa)}}{}\right),\\
c_2&\doteq\frac{1}{6}\sum_{abc}\epsilon^{abc}{S^{(0)}_{(a)(bc)}}{},&
c_5&\doteq\frac{1}{6}\sum_{abc}\epsilon_{bac}
{B_{0}^{(bc)(acc)}}{}.
\end{align*}
A straightforward calculation shows that the equations of motion 
\eqref{eq5.12} simplify to 
\begin{equation}
\label{eq5.13}
  \ddot{\evec{A}}(t,\evec{x})=\left(c_1^{(0)}c_3^{(0)}+c_1^{(0)}c_3^{(1)}
+c_1^{(1)}c_3^{(0)}\right)\Delta\evec{A}(t,\evec{x})
+\left(c_2c_3^{(0)}-c_1^{(0)}c_5\right)\Delta\rotat
\evec{A}(t,\evec{x}).
\end{equation}
Note that in the last equation we have just kept terms of leading
order and first order corrections, in the sense that we have explained 
in the previous section. Also we have eliminated a term containing
$\diver A$ by choosing the appropriate gauge. 

Equation \eqref{eq5.13} leads to a chiral modification of the
dispersion relation for electromagnetic waves:  
Let a unit vector
$\evec{e}_3$ be given and choose $\evec{e}_1$,$\evec{e}_2$ such that the
$\evec{e}_i$ form a righthanded orthonormal triple. Then a circularly 
polarized wave of helicity $\pm$, propagating in the direction given 
by $e_3$ can be written as 
\begin{equation*}
\evec{A}_k(t,\evec{x})=A_0\left[
\evec{e}_1\cos\left(\omega_\pm(k)t-k\,\evec{e}_3\cdot\evec{x}\right)
\pm \evec{e}_2\sin\left(\omega_\pm(k)t-k\,\evec{e}_3\cdot\evec{x} \right)\right].
\end{equation*}
This is a solution to the wave equation \eqref{eq5.13} provided that 
\begin{equation}
\label{eq5.14}
\omega_\pm(k)=\betr{k}\sqrt{\left(c_1^{(0)}c_3^{(0)}+c_1^{(0)}c_3^{(1)}
+c_1^{(1)}c_3^{(0)}\right)\pm \left(c_2c_3^{(0)}-c_1^{(0)}c_5\right)k}.
\end{equation}
Thus we have found a chiral modification to the dispersion relation. 
Note that this chiral modification is similar but not completely
analogous to the birefringence occurring for light propagation in some
crystals. The latter effect is not isotropic, it also depends on the
direction of propagation relative to the symmetry axes of the
crystal, whereas the chiral effect found here is isotropic. This can
be seen from the fact that nothing in the above formulae depends on
the direction of the vector $\evec{e}_3$.
We can now proceed to a discussion of results. 
%----------------------------------------------------------------------
\subsection{Discussion}
%----------------------------------------------------------------------
\label{se5.3}
Let us start the discussion of the results of the last section by 
considering the physical units and orders of magnitude of the
various terms appearing.  
We will use $F_{\text{der}}$, the derivative term in the scalar
field Hamiltonian, as an example -- similar
considerations apply to the other terms.

The classical term corresponding to ${B_0^{(a)(a')}}+{B_1^{(a)(a')}}$ is 
$\sqrt{\det q}q^{aa'}$. The latter is
dimensionless, since $q$ is. ${B_0^{(a)(a')}}$ has the structure
\begin{equation}
\label{eq5.21}
 B_0^{(\blank)(\blank)} \sim \frac{1}{\text{Vol}}\frac{P^2}{\sqrt{\det P}}bb 
\end{equation}
where $\text{Vol}$ is a volume. 
Since $[P]= \text{meter}^2$, $[P^2/\sqrt{\det P}]=\text{meter}$. $b$
is also a length, unit-wise, so $B_0^{(a)(a')}{}$ is indeed dimensionless.
${B_1^{(a)(a')}}{}$ has the structure
\begin{equation}
\label{eq5.22}
{B_1^{(\blank)(\blank)}}{}\sim  \frac{\ell_P^4}{t\text{Vol}}
\frac{P^2\tr P^{-2}-1}{\sqrt{\det P}}bb,
\end{equation}
so it is again dimensionless as it should be. 
The structure of ${B^{(ab)(a'b')}_0}{}$ is
\begin{equation}
\label{eq5.23}
{B_0^{(\blank\blank)(\blank\blank)}}{}\sim \frac{1}{\text{Vol}}\frac{P^2}{\sqrt{\det P}}bbbb,
\end{equation}
so its unit is $\text{meter}^2$ which is the correct one for a term
proportional to $\betr{k}^4$ in the dispersion relation.\newline
As for orders of magnitude, we remark the following. Assume  
$q_{ab}=O(1)$ in the chosen coordinate system. Then 
\begin{equation}
\label{eq5.24}
P=O(\epsilon^2),\qquad \text{Vol}=O(\epsilon^3)\quad\text{ and }\quad
b=O(\epsilon).  
\end{equation}
Using \eqref{eq5.21} it follows that ${B_0^{(a)(a')}}{}=O(1)$, so the
leading order term has the right order of magnitude.\newline
As for the order of magnitude of ${B_1^{(a)(a')}}{}$, we use 
\eqref{eq5.22} and \eqref{eq5.24} to conclude that  
\begin{equation*}
{B_1^{^{(\blank)(\blank)}}}{}=O\left(\frac{1}{t}\frac{\ell_P^4}{\epsilon^4}\right)
=O\left(\left(\frac{\ell_P}{L}\right)^{2-4\alpha}\right)=O(t^{1-2\alpha})
\end{equation*}
which is very small since $\alpha<1/2$.\newline
Consider finally ${B_0^{(ab)(a'b')}}{}$: From \eqref{eq5.23} and \eqref{eq5.24}
we see that ${B_0^{(ab)(a'b')}}{}=O(\epsilon^2)$.\newline 
As for the other terms in the dispersion relation, similar results can 
be seen to hold: The leading order term has same unit and order of
magnitude as the corresponding classical term and the ratio of leading 
order to first order correction is of order $t^{1-2\alpha}$. 

We will now discuss the structure of the dispersion relations
\eqref{eq5.15}, \eqref{eq5.14}. The coefficients appearing are given as
graph averages of certain local geometric quantities of the random
graph. Let us call these graph averages \textit{moments} of the random 
graph prescription (RGP for short).  
So, in order to get numerical statements from the 
results of the last section, one has to fix the scale $L$, an RGP 
and compute the relevant moments. 
Such a computation might be hard to perform analytically, but 
a computer can easily determine the moments occurring in
\eqref{eq5.15},\eqref{eq5.14} for a given RGP, so this calculation
does not present a principal difficulty. 

The more serious issue here is that there are certainly many RGPs, 
all leading to different graph averages and hence 
different predictions, and it is a priori not clear how one can single out 
the ``right'' one. We note however that for not too pathological RGPs,
the graph averages should be approximately equal so that at least the
size of the different terms in the dispersion relations 
is not too sensitive on the choice of the RGP. 
Moreover, again for a not too pathological RGP, 
the moments showing up in the dispersion relations should be
related. To give an example, a plausible assumption is that 
\begin{equation*}
  \aver{\sqrt{\det P}}{} \approx 
  \left(\aver{\frac{1}{\sqrt{\det P}}}{}\right)^{-1} 
\end{equation*}
and that their difference would not depend very strongly on the
chosen prescription.
Thus there will be approximate relations between the different
coefficients in the dispersion relations which are not affected by the 
choice of a specific RGP. 
 
Moreover we note that the leading order terms in the coefficients 
depend on the RGP. This might at first seem to
be a problem as well, since it means that we will have to tune the
RGP in such a way that the leading order terms assume their classical 
values. On the other hand, this can be seen as a
blessing: Fixing the leading order term means to fix one moment of 
the RGP. Via the relations conjectured above, this will also approximately 
fix other moments, independently of the specific RGP assumed, 
and thereby to a certain extent the higher order corrections.

Investigations in this direction are worthwhile but beyond the scope of 
the present work. Let us for the rest of this section assume that a
prescription is fixed and the relevant graph averages have been computed. 

Next we observe that two different sorts of corrections appear in the 
dispersion relations: The first sort of correction is simply a correction
to the leading order term coming out of the fluctuation calculation of 
section \ref{s4}. Its relative magnitude was found to be
$t^{1-2\alpha}$. We will call this sort of correction
a \textit{fluctuation correction}.\newline
The other sort of correction is a term containing a higher power of 
$\betr{k}$ as compared to the standard dispersion relation. 
We will call this kind of correction a \textit{lattice correction}. 
We have demonstrated for the example of ${B^{(ab)(a'b')}_0}{}$ 
that the terms
proportional to $\betr{k}^4$ are of the order $\epsilon^2$, therefore
the relative magnitude of the lattice corrections is of the order 
\begin{equation*}
O\left(\frac{\epsilon^2}{\lambda^2}\right)= \frac{L^2}{\lambda^2}\;\;
O\left(t^\alpha\right)\le O(t^\alpha). 
\end{equation*}
Similarly the terms proportional to $\betr{k}^3$ in the dispersion
relation for the electromagnetic field are of the order 
$t^\alpha L/\lambda$. 

When comparing our results for the electromagnetic field with the ones of
\cite{Gambini:1998it,Alfaro:1999wd}
we find the following: The result of Pullin and Gambini \cite{Gambini:1998it} 
does not contain any fluctuation corrections. This is however not result
of the calculation but rather assumed from the beginning. As for the
lattice corrections, they find a chiral modification to the dispersion 
relation as we do here. The relative magnitude of the correction is
however $\ell_P/\lambda$.\newline
Alfaro et al. \cite{Alfaro:1999wd} also do not have fluctuation
corrections by assumption. They find the helicity dependent 
correction of \cite{Gambini:1998it} and the present work, again of the 
order $\ell_P/\lambda$. They also get higher order corrections the
precise structure of which depends on a parameter which is not
fixed.

Thus our results agree with that of
\cite{Gambini:1998it,Alfaro:1999wd} as far as the structure of the
dispersion relation is concerned. We additionally have fluctuation
corrections and, most importantly, the corrections found {\it do not 
scale with an integer power of $\ell_P$, contrary to their finding}.
This signals a warning to assumptions made in \cite{Amelino-Camelia:1998pp}
to take into account only corrections which are of the order 
$(\ell_P/L)^n$ where $n$ is an integer. Notice that the fluctuation 
correction and the lattice correction are equal at $\alpha=\frac{1}{3}$.
Thus the leading correction is always of the order of at least $t^{1/3}$.     

Finally we should make a few remarks concerning a possible detection of 
the corrections in experiments. The fluctuation corrections will not  
show up in an experiment testing for a frequency dependence of the
velocity $c$ of light, since they merely correspond to a frequency
independent shift of $c$. Also, these corrections are certainly not
measurable by measuring the flight-time of photons since their velocity
would already be the ``bare'' leading order term plus the fluctuation
correction.    
Fluctuation corrections may however be measurable by comparing
flight-times of photons in different geometries, since the corrections
will change 
when the calculations presented in this chapter are repeated with LQC
approximating a non-flat spacetime. To discuss how this could be done
in practice is however beyond the scope of the present work.\newline 
Whether the lattice corrections are big enough to be detectable in
the data from current or planned $\gamma$-ray burst observations 
crucially depends on the values of $\alpha$ and $L$. 
For the value $\alpha=1/3$ which renders fluctuation and lattice
corrections equal in magnitude (and which is close to the lower bound 
value $2/5$ derived in \cite{ST01}), and $L$ of the order of a 
$\gamma$-ray wavelength, a rough estimate shows that {\it the lattice
corrections would indeed be detectable in the foreseeable future}.  

So, to conclude this chapter, we should repeat that not too deep a 
significance  
should be attached to the precise values of the coefficients in the 
dispersion relations obtained: There are still some ambiguities present 
in the GCS which we will discuss below: The quantization of the 
Hamiltonians, 
in the procedure to obtain the dispersion relations from the expectation values and, as a consequence, in the coefficients themselves. Also the replacement
$\sutwo\rightarrow\uone$ will certainly affect the precise numerical
outcome. Most significantly, so far we have little control on what will 
happen to the size of quantum corrections when our kinematical coherent 
states are replaced by physical ones. Within our kinematical scheme
the structure of the corrections, as well as the orders of magnitude 
$t^\alpha$, $t^{1-2\alpha}$ of the two sorts of corrections are 
robust, however. Thus we are possibly in trouble because such corrections
seem to lie in the detectable regime. If such corrections are not found,
then presumably it is not justified to use kinematical coherent states.

Finally, the approximate relations between the different graph 
averages will make the predictions of a more
complete calculation much less dependent on the random graph
prescription chosen, then one might at first fear. Similar remarks apply
if, as advocated for example in \cite{Bombelli:2000ua}, instead of
working with a fixed random graph, one averages over many of them. 
(Notice that also in that case, averaging procedures are not unique). 
In order to remove those ambiguities
one should probably set up a variational principle in order to optimize 
a family of semiclassical states according to a given set of observables.
%----------------------------------------------------------------------
\section{Summary and Outlook}
%----------------------------------------------------------------------
\label{s6}
In the this work we have presented a calculation of dispersion
relations for the scalar and the electromagnetic field coupled to
quantum general relativity. These dispersion relations bear
corrections to the standard ones, due to the discreteness of the
states of the geometry and to the bound on the uncertainty 
product of configuration and momentum variables in QGR. 
The calculations rest on the quantization of the matter parts of the
Hamilton constraint given in \cite{ST01} and the coherent states for
QGR constructed and analyzed in \cite{Thiemann:2000bw,Thiemann:2000ca,Thiemann:2000bx,Sahlmann:2001nv}

Corrections to dispersion relations due to QGR were also computed  
in \cite{Gambini:1998it,Alfaro:1999wd} and the present work partly
rests on the ideas implicit
and explicit in these pioneering works. The form of the correction
term in the dispersion relation for the electromagnetic field 
found in the present work agrees with that of
\cite{Gambini:1998it,Alfaro:1999wd}. This is
not too big a surprise since there is no other rotation invariant 
term in $\evec{k}$ of the same order. 
However, we find important differences in 
the order of magnitude of the effects, as compared to
\cite{Gambini:1998it,Alfaro:1999wd}.  
Moreover, the results of the present work are more specific,
since a definite class of semiclassical states, the coherent states
for QGR are employed in the calculation.  

Rather than making precise numerical predictions, the aim of the
present work is to demonstrate the steps necessary in such a
calculation, to highlight the issues that remain to be clarified and 
to give a robust estimate of the size of the effects. 

In this spirit, we have simplified the calculation of the expectation
values in \ref{s4} by replacing the full gauge group by its
I\"on\"u-Wigner limit $\uone^3$. This replacement will certainly
affect the precise numerical outcome but not the order of magnitude of 
the correction.  Also we have not specified a
prescription for obtaining random graphs, but only assumed general
properties that such a procedure will have. Most importantly, the effect
of using kinematical rather than physical coherent states is presently not
well understood.

The main achievements of the present work can be summarized as
follows:

The calculation given in section \ref{s4} shows how expectation
values of complicated operators in coherent states for quantum general
relativity can be computed and there is no principal difficulty in 
repeating such a calculation for the full gauge group $\sutwo$. 

Perhaps even more important are the order of magnitude estimates of
the resulting effects obtained in this work: They depend on 
very few parameters and will continue to hold true when more general 
complexifier coherent states \cite{Thiemann:2002vj} are used. The main choices that enter are:
\begin{itemize}
\item A complexifier $C$ has to be chosen for the construction of the
  coherent states. (Of course, there are more general semiclassical
states than coherent ones).
\item A class of observables has to be chosen 
  that should be approximated well by the coherent states.     
\item A (random, averaged) graph has to be chosen.
\end{itemize}
The other parameters are fixed by the above choices: 
The requirement that
$C/\hbar$ is dimensionless forces the parameter $t$ in the definition 
of the resulting coherent states to be $(\ell_P/a)^n$ where $n$ is some 
positive number and $a$ a length scale which is not yet fixed. 

The nature of these observables (do they involve one-, two- or
three-dimensional integrations? etc.) determines a) a length scale
$L$ and b) the exponent $\beta$ in the expression for the classical
error $(\epsilon/L)^{2\beta}$. 

The length scale $a$ gets fixed to be $L$ by requiring fluctuations of
configuration and momentum degrees of freedom to be
equal. Finally, the typical edge-length $\epsilon$ of the random 
graph is found to be a weighted geometric mean by requiring the
fluctuations to be minimal. Thus, at least within the vast class of 
complexifier coherent states, the structure of the ambiguities and their 
principal effects on the orders of the magnitudes of the 
quantum corrections {\it can be neatly classified}!

Many things remain to be done before one can really obtain reliable
predictions of observable effects from quantum general relativity:  

The procedure used to obtain dispersion relations from the discrete
classical Hamiltonians has to be further analyzed, and rigorously
justified at least in models which can be solved analytically. 
The influence of the choice of a random graph should be investigated,
and concrete procedures have to be implemented. A more distant goal is 
to also analyze possible back reaction effects of the matter on the
gravitational field. These were neglected in \cite{ST01} and in this work
since it would require to solve the combined matter -- geometry 
Hamiltonian constraint and force us to work with physical coherent states.

Thus, although we certainly did not carry out a first principle 
calculation, we hope to have made a modest contribution to an 
understanding what the principal problems are and how such a computation
could possibly be carried out in principle. Also, we hope to have 
demonstrated that QGR is still far from making reliable semiclassical
predictions until one is convinced of the physical relevance of a definite 
scheme. However, it should have become clear that once such a scheme has 
been identified, QGR {\it is} able to provide {\it precise} numerical
predictions. In any case, at least for the limited purpose of showing
that some version of the quantum Hamiltonian constraint is correct (for
which kinematical coherent states are unavoidable), the results of the
present two papers should be relevant.
\\
\\
\\ 
{\bfseries \Large Acknowledgements}
\\
\\
It is a pleasure for us to thank Abhay Ashtekar, Luca Bombelli, 
Arundhati Dasgupta, Rodolfo Gambini, Jurek Lewandowski, 
Fotini Markopoulou Kalamara, Hugo
Morales-Tecotl, Jorge Pullin,
and Oliver Winkler for numerous valuable discussions.  
We also thank the Center for Gravitational Physics of The
Pennsylvania State University, where part of this work was completed, for 
warm hospitality. T.T. was supported in part by NSF grant PHY 0090091 to The
Pennsylvania State University.
H.S. also gratefully acknowledges the splendid hospitality at the 
Universidad Autonoma Metropolitana Iztapalapa, Mexico City, and at 
the University of Mississippi, as well as the financial support by the
Studienstiftung des Deutschen Volkes. 
\bibliographystyle{JHEP-2}
\bibliography{lqg2}         %  die bibliographie
\end{document}